\documentclass[aps,prb,onecolumn,supergroupaddress,notitlepage]{revtex4-1}

\usepackage[dvipdfmx]{graphicx}
\usepackage[usenames]{color}
\usepackage[pdfusetitle, bookmarks=false,colorlinks=true,citecolor=blue,urlcolor=blue,linkcolor=black]{hyperref}
\usepackage{comment}
\usepackage{amsmath,amssymb,amsfonts}
\usepackage{bm,braket,array}
\usepackage{mathrsfs} 

\def\im{{\rm Im }}
\def\mod{{\rm mod }}
\def\Tr{{\rm Tr }}

\def\ch{{\rm ch \, }}

\def\p{\partial}
\def\lto{\longrightarrow}
\def\wt{\widetilde}

\def\Hom{{\rm Hom}}

\def\spin{{\rm Spin}}
\def\spinc{{\rm Spin}^c}
\def\pinp{{\rm Pin}^+}
\def\pinm{{\rm Pin}^-}
\def\pinc{{\rm Pin}^c}

\def\ua{\uparrow}
\def\da{\downarrow}

\newcommand{\C}{\mathbb{C}}

\newcommand{\R}{\mathbb{R}}
\newcommand{\Z}{\mathbb{Z}}

\newcommand{\cM}{\mathcal{M}}

\newcommand{\bx}{\bm{x}}
\newcommand{\bk}{\bm{k}}

\def\widebar{\accentset{{\cc@style\underline{\mskip10mu}}}} 
\def\wideubar{\underaccent{{\cc@style\underline{\mskip10mu}}}} 

\makeatletter
\renewcommand{\theequation}{
\arabic{section}.\arabic{equation}}
\@addtoreset{equation}{section}
\makeatother

\usepackage{float}

\newcommand{\xv}{\textbf{x}}
\newcommand{\kv}{\textbf{k}}
\newcommand{\iv}{\textbf{i}}
\newcommand{\jv}{\textbf{j}}
\newcommand{\ev}{\textbf{e}}

\begin{document}
\title{Many-body topological invariants in fermionic symmetry protected
  topological phases:
  Cases of point group symmetries}
\author{Ken Shiozaki}
\email{shiozaki@illinois.edu}
\author{Hassan Shapourian}
\author{Shinsei Ryu}
\affiliation{Department of Physics, University of Illinois at Urbana Champaign, Urbana, IL 61801, USA}
\date{\today}

\begin{abstract}
We propose 
the definitions of many-body topological invariants 
to detect symmetry-protected topological phases protected by point group symmetry, 
using partial point group transformations 
on a given short-range entangled quantum ground state.
Here, partial point group transformations $g_D$ are defined by point group transformations 
restricted to a spatial subregion $D$, 
which is closed under the point group transformations 
and sufficiently larger than the bulk correlation length $\xi$. 
By analytical and numerical calculations,
we find that the ground state expectation value of 
the partial point group transformations behaves generically as 
$\braket{GS | g_D | GS} 
\sim 
\exp \Big[ i \theta+ \gamma - \alpha \frac{{\rm Area}(\partial D)}{\xi^{d-1}} \Big]$.
Here, 
${\rm Area}(\partial D)$ is the area of the boundary of the subregion $D$, and $\alpha$ is a dimensionless constant. 
The complex phase of the expectation value $\theta$ is 
quantized and serves as the topological invariant, 
and $\gamma$ is a scale-independent topological contribution to the amplitude. 
The examples we consider include the $\Z_8$ and $\Z_{16}$ invariants of
topological superconductors protected by  
inversion symmetry in $(1+1)$ and $(3+1)$ dimensions, respectively,
and 
the lens space topological invariants
in $(2+1)$-dimensional fermionic topological phases. 
Connections to topological quantum field theories and cobordism classification 
of symmetry-protected topological phases are discussed. 
\end{abstract}


\maketitle

\tableofcontents

\newpage 
\section{Introduction}

\subsection{Topological phases in the presence of symmetries and topological quantum field theories}

Topological phases of matter are gapped quantum phases 
which cannot be adiabatically connected to a trivial state,
i.e., tensor product state. 
Topological phases can be discussed in the presence of various symmetries,
such as  
time-reversal, charge conjugation, and/or space group symmetry.  
More specifically, it may occur that
a gapped quantum phase,
which can be adiabatically connected to a trivial state in the absence of symmetries, 
cannot be turned into a trivial phase
once a certain set of symmetries are enforced.
Such gapped quantum phases
are called symmetry-protected topological (SPT) phases. 
Topological insulators and superconductors
are celebrated examples of SPT phases of fermions.~\cite{ 
Schnyder2008, Kitaev2009, Ryu2010, Hasan2010, 
Qi2011, Cho2011, Fidkowski2011, Gu2012, Freed2013, 
Wang2014, Kapustin2015a, HsiehChoRyu2016, 
witten2015fermion, Metlitski2015, Freed2016, Bhardwaj2016} 
Other examples of bosonic SPT phases have been also widely discussed.~\cite{
Pollmann2010, Chen2011b, Schuch2011, 
Chen2013, Lu2012, 
Levin2012, SuleXiaoRyu2013, 
vishwanath2013physics, 
Kapustin2014symmetry, Freed2016} 
On the other hand, topologically ordered phases~\cite{
WenNiu1990, wen2004quantum, MooreRead1991, Kitaev2003, 
Levin2005, Kitaev2006, KongWen2014, 
Witten1989, TuraevViro1992, 
Dijkgraaf1990, Propitius1995}
are phases of matter which are topologically distinct from a trivial state even in the absence of symmetries. 
Topologically ordered phases can be enriched by the presence of symmetries;
They can exhibit a particular pattern of symmetry fractionalization,
which can be used to distinguish and characterize different topologically 
ordered phases with symmetries.
Topologically ordered phases of this kind are called symmetry-enriched.~\cite{
Wen2002, 
Mesaros2013, EssinHermele2013, 
HungWen2013, Lu2016, 
Barkeshli2014, Teo2015, 
Turaev2000, turaev2010homotopy}
Our main focus below will be SPT phases, although
some of our discussion should be readily applicable to
symmetry-enriched topological phases as well. 

%

For gapped phases of matter, 
it is reasonable to expect that their 
low-energy and long-wavelength physics is captured by
topological quantum field theories (TQFTs) of some sort. 
More specifically,
let us consider topological phases protected/enriched by 
a set of global symmetries, which form a symmetry group $\wt{G}$. 
It is convenient to decompose the symmetry group $\wt{G}$ into 
the part which consists of unitary on-site (or ``internal'') symmetries ($=G$),
and the part which consists of symmetry transformations 
which reverse the orientation of spacetime manifolds,
such as time-reversal and parity.
(Other spatial symmetries, such as point group symmetries, will be discussed momentarily, but for now, 
we focus on orientation-reversing symmetries, such as parity, or reflection in one direction.)
If the underlying system includes fermions, 
it is convenient to include the fermion number parity to the latter part. 

For the purpose of detecting (topological) properties of the gapped quantum phases,
it is well-advised to couple them to a background $G$ gauge field. 
One can then integrate over all matter degrees of freedom (matter fields $=\{\phi_i\}$) of 
the topological phases. 
This procedure leads to the partition function
\begin{align}
Z(X, \eta, A) = \int \prod_i D \phi_i e^{- S_X(\{ \phi_i \}, \eta, A)}. 
\label{path integral}
\end{align}
Here, $X$ is a closed $(d+1)$-dimensional spacetime manifold,
$\{ \phi_i\}$ includes all matter fields, 
$A$ is the background $G$ gauge field which couples to the matter fields. 
On the other hand, 
$\eta$ is a ``structure'' endowed to the manifold such as 
an orientation, 
a $\spin$ or $\spinc$ structure for real or complex fermions, respectively. 
\footnote{
Recall that for an orientable $(d+1)$-dimensional manifold $X$, 
the structure group $SO(d+1)$ on its tangent spaces
acts on frame fields (vielbein). 
For a relativistic fermion field on $X$,
$SO(d+1)$ is lifted to $Spin(d+1)$. 
The choice of signs that arises in this lifting defines a spin structure. 
It should be noted and stressed that 
fermions in condensed matter systems are not relativistic and not always sensitive to spin structures.
However, when some sort of topological media (topological phases) are realized,
effective relativistic fermions can emerge, which do depend on spin structures.
In short, spin may emerge from short-range entanglement in quantum ground states.
}
In short, the symmetry group $\wt{G}$ of a given gapped 
(topological) phase enters into the 
corresponding partition function as the input data $(X, \eta, A)$. 

For gapped phases of matter, 
when the correlation length of the system is much shorter than the system size,
the resulting partition functions are expected not to depend very sensitively on 
the details of the spacetime manifold $X$, and the background gauge field $A$. 
In particular, for topologically non-trivial gapped phases, 
the partition function $Z(X,\eta, A)$,
or equivalently the effective action $S_{eff}(X,\eta,A)=-\ln Z(X,\eta,A)$,
may have a topological term --
a $U(1)$ phase of the partition function
(the imaginary part of the Euclidean effective action),
which is insensitive to the metric on $X$ 
(i.e., it is invariant under diffeomorphisms)
as well as small variations of the background gauge field. 
In the limit of zero-correlation length, 
the partition function consists solely of a topological term. 

Put differently, 
the partition function $Z(X, \eta, A)$ is expected to define a TQFT, 
or more precisely, the so-called $G$-equivariant TQFT
(or $G$-equivariant Spin TQFT if we are interested in fermionic condensed matter systems).
\cite{turaev2010homotopy,Moore2006,Kapustin2015,gaiotto2015spin, 
WilliamsonWang2016hamiltonian,
Shiozaki-Ryu, 
KapustinTurzilloYou2016topological, 
BultinckWilliamsonHaegemanVerstraete2016fermionic,
KapustinTurzilloYou2016spin} 
For a given closed manifold $X$ with background structures specified by $(\eta, A)$, 
the TQFT yields a topological invariant.

When the gapped phase in question has no topological order as in an SPT phase, 
i.e., it has a unique ground state, 
\footnote{
In the terminology of TQFTs, 
the property that the gapped phase has a unique ground state 
for all spatial manifolds is called invertible.}
it was further proposed that the corresponding partition function depends only 
on the certain equivalence classes of the background 
manifolds (with structure $\eta$ and the background gauge field), 
the cobordism class of $(X, \eta, A)$.~\cite{Kapustin2014symmetry}
Here, two $(d+1)$-dimensional spacetime manifolds $X_{1,2}$ 
(with structure $\eta_{1,2}$ and background fields $A_{1,2}$)
are called cobordant when 
one can find a $(d+2)$-dimensional manifold $Y$
with an appropriate background gauge field that can interpolate 
$(X_1,\eta_1, A_1)$ and $(X_2,\eta_2, A_2)$, 
i.e., $\partial Y = X_1 \sqcup X_2$. 
This relation can be used to define equivalence classes of 
$(d+1)$-dimensional manifolds 
(with a given structure and background gauge field configurations). 
It should be noted that the claim that the partition function is cobordism invariant 
is more stringent than topological (diffeomorphism) invariance. 

One can further introduce an Abelian group structure to 
the equivalence classes of $(X, \eta, A)$ by taking 
the disjoint union as an operation. 
The resulting group is called the cobordism group and denoted by
$\Omega_{d+1}^{\rm str}(B G)$,
which is Abelian (e.g., $\mathbb{Z}_n$, $\mathbb{Z}_n\oplus\mathbb{Z}_m$).
Here, $B G$ is the classifying space of $G$.~\cite{stong2015notes, Gilkey}
When there is no symmetry, we simply put a single point as $BG$, $BG=pt$.

In topological theories in which the partition function is given by a pure phase, 
$Z(X,\eta,A)$ can be thought of as a homomorphism 
\begin{align}
Z : \Omega_{d+1}^{\rm str}(B G) \to U(1), \quad (X,\eta,A) \mapsto Z(X,\eta,A). 
\end{align}
Thus, 
a useful way to classify possible topological $U(1)$ 
phases of $Z(X, \eta, A)$ is by using 
the cobordism group
classification of manifolds with structures.
\cite{Kapustin2014symmetry, Kapustin2015a, Freed2016}
It was proposed that the torsion part of the cobordism group ${\rm Tor}\ \Omega_{d+1}^{\rm str}(B G)$ provides a 
possible classification of topological phases of matter. 

%
%
%
%
%

\subsection{SPT invariants in terms of ground state wave functions}


The purpose of the paper is to construct topological invariants,
i.e., 
quantities which take the same value for all points (Hamiltonians)
in a given gapped quantum phase
and can be used as ``order parameters'' of topological phases.
In particular, 
in the condensed matter context, 
we wish to define and compute 
the topological invariants
for a given Hamiltonian or a ground state, i.e.,
within the operator formalism. 
(We will mainly focus on ground states of fermionic SPT phases, which are unique. The topological invariants in this context 
are often called SPT invariants.)
The above consideration suggests that 
the topological $U(1)$ phase of 
the path integral evaluated on a suitable manifold with structures 
$[X,\eta,A]$ can be thought of as 
a meaningful topological invariant (SPT invariant).
%
Our task is then to find a way to ``simulate'' the path integral 
\eqref{path integral}
defined for the data $[X,\eta, A]$. 
(It should be stressed that we are here to construct 
{\it many-body} topological invariants,
as opposed to {\it single-particle} topological invariants,
which have been commonly discussed in the literature.)

Of particular importance for our purposes is 
the generator $[X,\eta,A]$ of the cobordism group $\Omega_{d+1}^{\rm str}(B G)$,
which we simply call the {\it generating manifold} in the following. 
It is on this manifold that evaluating the path integral (the partition function)
gives rise to a ``least possible'' or ``most fundamental'' 
topological $U(1)$ phase;
The topological $U(1)$ phases for other possible 
manifolds $[X,\eta,A]$ are given as an integer multiple 
of the topological $U(1)$ phase for the generating manifold.

Our proposal to define/construct the topological invariants can be summarized
by the following set up and operations $(i)$-$(v)$:
\begin{itemize}
\item[$(i)$] closed $d$-dimensional space manifolds $M$ on which the Hamiltonian is defined, 
\item[$(ii)$] orientation or spin structures $\eta_M$ on $M$, 
\item[$(iii)$] twisted boundary conditions (background flat gauge fields $A_M$) on $M$ by on-site unitary symmetry $G$, 
\item[$(iv)$] symmetry operations $\hat g$ on the many-body Hilbert space,  
\item[$(v)$] the {\it partial (point group) operation} 
$\hat g_D$ on a subregion $D$ 
of the space manifold $M$.
\end{itemize}

Let $\Ket{GS(M,A_M,\eta_M)}$ be the ground state of the Hamiltonian $H(\eta_M,A_M)$ 
with the spin structure $\eta_M$ and the twisted boundary condition $A_M$. 
From $(i-iv)$, we can extract 
a set of $U(1)$ phases $\{ e^{i \Phi(M,\eta_M,A_M,g)} \}$ by 
\begin{align}
\hat g \Ket{GS(M,\eta_M,A_M)} 
= e^{i \Phi(M,\eta_M,A_M,g)} \Ket{GS \big (M,g(\eta_M),g(A_M) \big)}, 
\end{align}
where $g(\eta_M)$ and $g(A_M)$ are 
the spin structure and 
twisted boundary condition 
mapped by symmetry transformation $g$.  
In particular,
when
$\Ket{GS(M,\eta_M,A_M)}$ 
and 
$\Ket{GS \big (M,g(\eta_M),g(A_M) \big)}$
are in the same Hilbert space,
we can extract the $U(1)$ phases by
\begin{align}
\Bra{GS \big (M,g(\eta_M),g(A_M) \big)}
\hat g \Ket{GS(M,\eta_M,A_M)} 
= e^{i \Phi(M,\eta_M,A_M,g)}. 
\label{wfn overlap}
\end{align}
The wave function overlap \eqref{wfn overlap},
and hence the $U(1)$ phase, 
can be readily interpreted as a spacetime path integral
on the spacetime
which takes the form of a mapping torus,  
\begin{align}
X = M \times_{f} S^1:= M \times [0,1]/\big( (x,0) \sim (f(x),1) \big), 
\end{align}
where $f : M \to M$ is a diffeomorphism on $X$
induced by the action of $\hat{g}$. 
For instance, it can be simply the identity $f = {\rm id}$ or the space inversion $f : x \to -x$. 
Another example in $(2+1)d$ is the case when $M$ is the 2-torus $T^2$ and $f$ is chosen as modular transformations acting on $T^2$. 
In Refs.\ \onlinecite{ZaletelMongPollmann2013, Hung2014, HeMoradiWen2014, Moradi2014, Williamson2014, You2015, HuangWei2015}, 
the action of the modular transformations on the ground state(s) 
is discussed,
as a method to extract
data of topological phases, 
i.e., 
representations of the mapping class group of the space manifold.

\subsection{Partial point group operation and spacetime path-integral}
\label{Partial point group operation and spacetime path-integral}

In some symmetry classes, 
the subset of the above ingredients $(i)$-$(iv)$ is sufficient to define the topological invariant for the generating manifold 
(see, for example, Refs.~\onlinecite{Resta1998, RyuHatsugai2006, Pollmann2012, Wen2014, Zaletel2014, Hung2014, Williamson2014, Yonekura2016}). 
For example, the generating manifold of $\spin$ cobordism group 
$\Omega^{\spin}_2(pt) = \Z_2$,
which is relevant for the topological classification of class D superconductors
in $(1+1)$ dimensions,
is the 2-torus $T^2$ with periodic boundary conditions in both time and space directions.
\cite{Moore2006, Kirby}
On the other hand,
there are other symmetry classes for which
the generating manifold cannot be realized as a mapping torus. 
For example, 
the generating manifold of $\pinm$ cobordism group $\Omega_2^{\pinm}(pt) =
\Z_{8}$,
which is relevant for the topological classification of class D superconductors with reflection symmetry with $R^2 = (-1)^F$ in $(1+1)$ 
dimensions,
is the real projective plane $\R P^2$.~\cite{Kirby, Kapustin2015a} 
(Here $(-1)^F$ is the fermion number parity.) 
Since $\R P^2$ is not a mapping torus,  
the topological invariant cannot be constructed by using $(i)$-$(iv)$.

In these cases, the last ingredient $(v)$,
partial symmetry operations which act on a given subregion $D$ of the total
space manifold $M$,
is necessary to construct topological invariants.
For previous studies using partial symmetry operations to detect properties of topological phases,
see for example, Refs.\
\onlinecite{RyuHatsugai2006, Pollmann2012, Tu2013, Shapourian-Shiozaki-Ryu, Shiozaki-Ryu}. 
In particular, our approach, which applies to fermionic SPT phases in arbitrary
dimensions, is partly motivated by Ref.\ \onlinecite{Pollmann2012}.
In Ref.\ \onlinecite{Pollmann2012},
Pollmann and Turner showed that the $\Z_2$ invariant for 
the inversion-symmetric Haldane chain (a bosonic SPT phase) can be detected by the partial 
inversion~\cite{Cen2009} on the ground state. 
In the path-integral picture, 
the Pollmann-Turner invariant is interpreted as the spacetime path-integral  
on the projective plane, $\mathbb{R}P^2$.~\cite{Shiozaki-Ryu} 
Tu et al.~\cite{Tu2013} also showed that the partial lattice translation  
contains useful information on (2+1)-dimensional topologically ordered phases
such as the central charge and topological spins.

Let us now give a more detailed definition of partial symmetry operations. 
Specifically, we consider a Hamiltonian which is invariant under $\wt G$,
$\hat g H \hat g^{-1} = H$,
$g \in \wt G$. 
The symmetry operation $\hat{g}$ acts on the underlying fermionic operators as
\begin{align}
\hat g \psi^{\dag}_i(\boldsymbol{x}) \hat g^{-1} 
= \psi^{\dag}_j(g \cdot \boldsymbol{x}) U_{ji}, 
\quad 
  \hat g \ket{0}_{\psi} = \ket{0}_{\psi},
  \quad g \in \wt G, 
\end{align}
where $\psi^{\dag}_i(\boldsymbol{x})$ is a fermion creation operator 
at $\boldsymbol{x}\in M$, 
and 
$i,j$ represents internal degrees of freedom, 
and $\ket{0}_{\psi}$ is the Fock vacuum of 
the $\psi_i(\boldsymbol{x})$ fermion. 
(We focus on fermionic systems here.  
A similar definition applies to bosonic topological phases). 
We now choose a subregion $D$ of $M$
which is closed under the action of the group $\wt G$,
i.e.,
$\boldsymbol{x} \in D \Leftrightarrow g\cdot 
\boldsymbol{x} \in D$ for any $g \in \wt G$. 
We define the partial transformation by restricting 
the transformation $\hat{g}$ to 
the subregion $D$ as 
\begin{align}
&
\hat g^{\ }_D \psi^{\dag}_i(\boldsymbol{x}) \hat g_D^{-1} 
= \left\{\begin{array}{ll}
\psi^{\dag}_j(g \cdot \boldsymbol{x}) U_{ji} & (\boldsymbol{x} \in D) \\
\psi^{\dag}_j(\boldsymbol{x}) & (\boldsymbol{x} \notin D) \\
\end{array}\right.  
,
\qquad
\hat g_D \ket{0}_{\psi} = \ket{0}_{\psi}, 
\end{align}
for $g \in \wt G$.
We take the subregion $D$ such that 
the length scale of $D$ is sufficiently larger than the correlation length $\xi$. 
For a given ground state $\ket{GS}$ on $M$, 
the topological invariant associated with the symmetry $\wt G$ is given by 
the expectation value $\braket{GS | \hat g_{D} | GS}$ of the partial symmetry 
transformation. 

\begin{figure}[!]
 \begin{center}
  \includegraphics[width=\linewidth, trim=0cm 0cm 0cm 0cm]{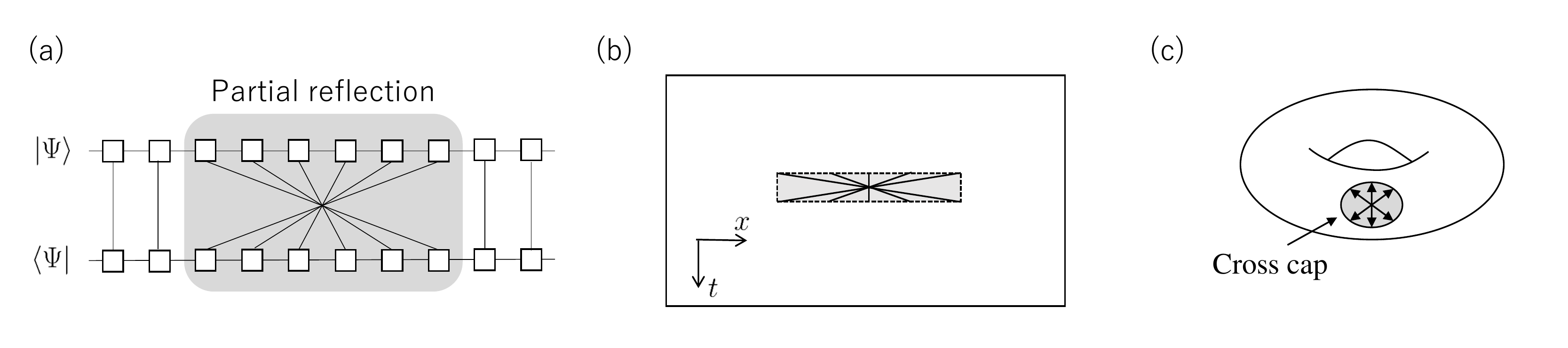}
 \end{center}
 \caption{
 [a] Fermionic matrix product representation of the partial reflection. 
 [b] Path integral representation of the partial reflection. 
 [c] One cross-cap on the torus. 
 }
 \label{Fig:Crosscap}
\end{figure}

Of particular importance is the case when $\hat{g}$ or $\hat{g}_D$ is 
a point group operation. 
In this case,  
$\braket{GS | \hat g_{D} | GS}$ 
can be interpreted, in the path-integral picture,
as a path-integral on the spacetime manifold which may not be obtained 
as a mapping torus. 
For example,
let us consider partial reflection 
$\hat{R}_I$
acting on 
a segment $I$ of the total  $(1+1)d$ system.
The action of partial reflection $\hat{R}_I$
on a ground state
(represented here by using a (fermionic) matrix product state)
is shown schematically in 
Fig.\ \ref{Fig:Crosscap} [a].
In the path-integral representation 
(Fig.\ \ref{Fig:Crosscap} [b]), 
$\braket{GS | \hat{R}_I | GS}$ 
can be interpreted as a path-integral on a manifold,
which is obtained from the original spacetime 
by first introducing a slit $[-\epsilon, \epsilon] \times [0,L]$ at time $t=0$
and then applying reflection on the slit.
This procedure is topologically equivalent to introducing a cross-cap in the spacetime torus
as shown in Fig.\ \ref{Fig:Crosscap} [c]. 
For example, 
as we will show 
in Sec.\ \ref{Sec:3} 
(see also Ref.\ \onlinecite{Shapourian-Shiozaki-Ryu}),
for 
$(1+1)d$ topological superconductors protected by 
reflection symmetry, 
the
ground state expectation value of $\hat{R}_I$ 
is given by 
\begin{align}
\braket{GS | \hat R_{I} | GS} 
\lto \frac{e^{\pm \frac{\pi i}{4}}}{\sqrt{2}}
\quad 
(L, N-L \gg \xi), 
\qquad
I = [0,L], 
\label{eq:GS_expectation_partial_reflection}
\end{align}
where   
$I = [0,L]$ is an interval on the circle with $N$ sites,
and $\xi$ is the correlation length. 
The $U(1)$ phase $e^{\pm \frac{\pi i}{4}}$ correctly reproduces
the known the $\Z_8$ topological classification of 
$(1+1)d$ topological superconductors protected by 
reflection symmetry.
Furthermore, we will confirm that the $U(1)$ phase agrees with the result from
the TQFT;
the corresponding (spin) TQFT path-integral on $\R P^2$ is given by (or computes) 
the $\Z_8$ Brown invariant of manifolds with $\pinm$ structure,~\cite{Kirby, Kapustin2015a}
\begin{align}
Z(\R P^2, \eta_{\pm}) = e^{\pm \frac{\pi i}{4}},
\label{eq:partition_function_RP2_pin-}
\end{align}
where $\eta_{\pm}$ represent two different $\pinm$ structures 
associated with 
the non-contractible loop on $\R P^2$.
(In the operator formalism, 
the different $\pinm$ structures correspond to the presence or absence of 
the fermion parity operator $(-1)^F$ in the partial reflection 
-- see Sec.\ \ref{Sec:3} for details.)

Technically, the spacetime manifold created by the partial reflection (\ref{eq:GS_expectation_partial_reflection})
[Fig.~\ref{Fig:Crosscap} (c)]
agrees with the real projective plane $\R P^2$ only up to a genus.
This is however not a problem if phases of our interest are SPT phases, i.e., if the underlying TQFT is invertible.
More precisely, instead of $Z(\R P^2,\eta_{\pm})$,
the expectation value of the partial reflection
is related to the partition function of a $\pinm$ TQFT on the connected sum
of $\R P^2$ and the 2-torus,
\begin{align}
\braket{GS_a|\hat R_I|GS_a}
  \sim
  Z\big((\R P^2,\eta_{\pm}) \# (S^1_{a} \times S^1_{ns}) \big), \qquad a=ns,r, 
\label{eq:partial_reflection_general_formula}
\end{align}	
where the subscript $a=r,ns$ specifies the boundary condition of the circle $S^1$; 
$r/ns$ represents the periodic/anti-periodic boundary condition, respectively;
$X \# Y$ means the connected sum of $X$ and $Y$.
Note that, by the invertiblity assumption,
only a single ground state $|GS_a\rangle$ appears
on the LHS of (\ref{eq:partial_reflection_general_formula}).
Since one can recast the RHS as
$Z\big( (\R P^2,\eta_{\pm}) \# (S^1_{a} \times S^1_{ns}) \big)
= Z(\R P^2, \eta_{\pm}) Z(S^1_a \times S^1_{ns}) / Z(S^2) =
Z(\R P^2, \eta_{\pm})$,
we conclude
$
\braket{GS_a|\hat R_I|GS_a}
\sim
Z(\R P^2, \eta_{\pm})$, 
i.e., the equivalence between
(\ref{eq:GS_expectation_partial_reflection}) and (\ref{eq:partition_function_RP2_pin-}). 
(In this discussion, as noted earlier, 
the equivalence between the expectation value of the partial reflection and
the TQFT partition function holds only up to the amplitude (modulus) --
see (\ref{eq:GS_expectation_partial_reflection});
As for the partition function $Z(S^1_a \times S^1_{ns})=1$, see
\eqref{eq:z2inv_1d_sc} for details.)
The above argument also shows that the boundary condition
imposed on the total space is not important.
The same argument can be applied to the partial point group
transformation in any space dimensions
as far as we consider SPT phases (invertible TQFT).


For generic partial symmetry transformations, 
from numerical and analytical calculations, 
we find the following behavior
\begin{align}
\braket{GS | \hat g_{D} | GS} 
= \exp \left[ i \theta+ \gamma - \alpha \frac{{\rm Area}(\partial D)}{\xi^{d-1}} + \cdots \right]
\label{Eq:PPGT}
\end{align}
as a leading contribution to the expectation value 
for ${\rm Area} (\partial D)/\xi^{d-1} \gg 1$. 
\footnote{
A logarithmic contribution 
$\log \big[ {\rm Area}(\p D)/\xi^{d-1} \big]$ 
may also appear 
as in our example in Eq.\ \eqref{result partial fermion parity 4d}.
}
Here, 
${\rm Area}(\partial D)$ is the area of the boundary of the subregion $D$,
and $\alpha$ is a dimensionless constant. 
The complex phase $\theta$ is well quantized for a sufficiently large region $D$ and represents 
a topological invariant.
In addition, 
$\gamma$ is a scale-independent part of the amplitude,
and can be thought of as yet another topological contribution to
the ground state expectation value of the partial symmetry operation. 
It is natural to expect that the topological $U(1)$ phase $e^{ i \theta}$ is the same as 
the TQFT partition function $Z(X,\eta,A)$.

In the following sections, we will show that this equivalence between 
the ground state expectation value of partial point group transformations
and TQFT partition functions 
holds quite generically, 
in any dimensions and for various point group symmetries. 
For examples, in $(2+1)$-dimensional topological phases, 
the ground state expectation of the partial rotation gives rise to 
the partition functions on the lens space. 
In $(3+1)$ dimensions, the partial inversion effectively induces a cross-cap and 
gives rise to the TQFT partition function on the 4-dimensional real projective 
space $\R P^4$.

It should be noted that
the cobordism classification of SPT phases 
using TQFTs, at least in its original form, 
does not discuss point group operations
other than those that reverse the orientation of 
the spacetime. 
For orientation-reversing symmetries, 
the corresponding partial transformations
give rise to generating manifolds, and hence
the desired SPT invariants
(at least for the examples studied in this paper).
On the other hand, for
partial transformations
for other point group symmetries, 
their connection to 
the cobordism classification of SPT phases using TQFTs
is less obvious. 
Even so,
since the low-energy TQFTs descriptions
of the form \eqref{path integral}
are often Lorentz invariant,
we expect that the orientation-preserving point group operations, 
at least within the low-energy TQFTs,  
can be implemented as on-site unitary operations. 
For example, $n$-fold discrete rotations ($C_n$ rotations)
can be implemented as $\mathbb{Z}_n$ discrete unitary on-site symmetries. 
If so, the effects of point group symmetries can be incorporated 
by introducing a proper background gauge field, 
e.g., by including the $\mathbb{Z}_n$ symmetries by taking $G=\mathbb{Z}_n$.  
(For similar claims, see 
Ref.~\onlinecite{Song-Huang-Fu-Hermele2016, ThorngrenElse2016gauging}.) 
We will verify this claim for selected examples, 
non-chiral topological insulators in $(2+1)d$ protected by $C_n$ symmetry.
Following the above discussion, the relevant cobordism group is
$\Omega^{\spinc}_3(B\mathbb{Z}_n)$. 
The generating manifold is the lens space $L(n,1)$. 
We will show that, by using partial $C_n$ rotations, 
we can simulate the path-integral on the generating manifold $L(n,1)$. 
Our calculations for explicit models
strongly suggest that the partial point group operations give 
rise to SPT invariants for SPT protected by 
the point group symmetries.

\subsection{Bulk-boundary correspondence and the reduced density matrix}

We confirm numerically the area law for the expectation value of
the partial symmetry operation, Eq.\ \eqref{Eq:PPGT}, 
in various examples.
Intuitively, the area law also follows naturally from 
the bulk-boundary correspondence,
which we will make use of to develop our analytical calculations.
Let us consider the reduced density matrix $\rho_D$
for the subregion $D$, 
\begin{align}
\rho_D = \Tr_{M \backslash D} \big[ \ket{GS} \bra{GS} \big] 
= \frac{e^{- H_E}}{\Tr_{D} \big[ e^{- H_E } \big] }, 
\end{align}
where $H_E$ is the entanglement Hamiltonian.
By the bulk-boundary correspondence of a SPT phase,  
we then expect that the entanglement Hamiltonian $H_E$ is given by, 
in the low energy subspace of the Hilbert space on the subregion $D$, 
the {\it physical} Hamiltonian $H_{\partial D}$ describing 
low-energy gapless boundary degrees of freedom
~\cite{RyuHatsugai2006,Li2008,Qi2011b}:
\begin{align}
	H_E 
 \sim
\frac{\xi}{v} 
	H_{\partial D}.
\label{eq:bulk_boundary_correspondence}
\end{align}
Here, the effective temperature in the reduced density matrix   
is given in terms of the bulk correlation length $\xi$,
and $v$ is the velocity of the gapless excitation. 
%
The partial point group transformation $\hat g_D$ induces a symmetry action 
$\hat g_{\p D}$ on the Hilbert space of the boundary theory;
The point group transformation on the gapless theory $H_{\p D}$ can be 
specified by explicitly expressing the boundary low-energy excitations of $D$ 
in terms of the bulk degrees of freedom $\psi_i(\boldsymbol{x})$.
Then, the ground state expectation (\ref{Eq:PPGT}) can be written in terms of $H_{\p D}$ as 
\begin{align}
\braket{GS | \hat g_{D} | GS} 
= \Tr_D \big[ \hat g_D \rho_D \big]
\sim
\frac{\Tr_{\p D} \big[ \hat g_{\p D} e^{- \frac{\xi}{v} H_{\p D}} \big]}{\Tr_{\p D} 
\big[ e^{- \frac{\xi}{v} H_{\p D} } \big] }. 
\end{align}
As an example, 
we will use this bulk-boundary correspondence 
to compute the expectation values of partial symmetry operations in
(2+1)-dimensional SPT phase
by using the corresponding (1+1)-dimensional CFTs.
We will also present similar calculations for higher dimensions
using free theories as the boundary theories. 
With these analytical calculations, together with numerics, 
we will confirm the formula (\ref{Eq:PPGT}).

\begin{table}[!]
\begin{center}
\caption{
\label{tab:list}
List of many-body topological invariants for fermionic SPT phases 
in the present paper. 
The first column specifies symmetry classes. 
``A'' and ``D'' represent Altland-Zirnbauer symmetry classes.~\cite{altland1997nonstandard}
$R$, $C$, $(-1)^F$, $C_n$, and $I$ are reflection, 
charge conjugation, fermion parity, $n$-fold rotation, 
and inversion symmetries, respectively. 
}
\begin{tabular}{| >{\centering\arraybackslash}m{2.3cm} | >{\centering\arraybackslash}m{0.8cm} | >{\centering\arraybackslash}m{2.8cm} | >{\centering\arraybackslash}m{3cm} | >{\centering\arraybackslash}m{3.4cm} |>{\centering\arraybackslash}m{3.7cm} |>{\centering\arraybackslash}m{1cm} | }
\hline
Symmetry class & Space dim. & Topological classification & Spacetime manifold & Topological invariant & Comment & Sec. \\
\hline
D & 1 & $\Omega^{\spin}_2(pt) = \Z_2$ & $T^2$ & 
$$\begin{array}{l}
\Braket{GS_r | (-1)^F | GS_r}
\end{array}$$
& $\Ket{GS_r}$ is the ground state with PBC. & 
\ref{sec:(1+1)D}
\\
\hline
$$\begin{array}{c}
{\rm D} + R, \\
R^2 = 1
\end{array}$$
& 1 & $\Omega^{\pinp}_2(pt) = \Z_2$ & Klein bottle & 
$$\begin{array}{c}
\Braket{GS_r | R | GS_r}
\end{array}$$
& Full reflection on the ground state with PBC 
& \ref{sec:(1+1)D+R+}\\
\hline
$$\begin{array}{c}
{\rm A} + C, \\
C^2 = 1
\end{array}$$
& 1 & 
$\Omega^{{\rm Spin}^{\tilde c+}}_2(pt) \ni \Z_2$
& $T^2$ & 
\begin{align*}
\frac{\Braket{GS(\pi) | C | GS(\pi)}}{\Braket{GS(0) | C | GS(0)}}
\end{align*}
& 
$\Ket{GS(\theta)}$ is the ground state with the twisted boundary condition by $\theta$. 
& \ref{sec:(1+1)A+C}\\
\hline
$$\begin{array}{c}
{\rm D} + R, \\
R^2 = (-1)^F 
\end{array}$$
& 1 & $\Omega^{\pinm}_2(pt) = \Z_8$ & $\R P^2$ & 
$$\begin{array}{l}
\Braket{GS | R_I | GS}
\end{array}$$
& Partial reflection 
&\ref{sec:(1+1)D+R-}\\
\hline
A + $R$ & 1 & 
$\Omega^{\pinc}_2(pt) = \Z_4$
& $\R P^2$ & 
$$\begin{array}{l}
\Braket{GS | R_I | GS}
\end{array}$$
& Bond center partial reflection. 
A $U(1)$ phase associated with $R$ is chosen so that $R^2 = (-1)^F$. 
&\ref{sec:(1+1)A+R}\\
\hline
$$\begin{array}{lc}
{\rm D} + R, \\
R^2 = 1
\end{array}$$
& 2 & $\Omega^{\pinp}_3(pt) = \Z_2$ & Klein bottle $\times S^1$ & 
$$\begin{array}{l}
\prod_{\eta} \Braket{GS(\eta) | R | GS(\eta)}
\end{array}$$
& $\eta$ runs over $\{(r,r)$, $(r,ns)$, $(ns,r)$, $(ns,ns)\}$ sectors. 
&\ref{sec:(2+1)D+R+}\\
\hline 
$$\begin{array}{c}
{\rm chiral} \\
D + C_n, \\
(C_n)^n = (-1)^F
\end{array}$$
& 2 & Spin TQFT with background framing & Lens space $L(n,1)$ & 
$$\begin{array}{l}
\Braket{GS | C_{n,D} | GS}
\end{array}$$
& Partial $n$-fold rotation 
&\ref{$(2+1)d$ $(p_x - i p_y)$ chiral superconductor with rotation symmetry}\\
\hline
$$\begin{array}{c}
{\rm nonchiral} \\
D + C_2, \\
(C_2)^2 = (-1)^F
\end{array}$$
& 2 & $\Omega^{\spin}_3(B \Z_2) = \Z_8$ & $\R P^3$ & 
$$\begin{array}{l}
\Braket{GS | C_{2,D} | GS}
\end{array}$$
& Partial 2-fold rotation 
& \ref{Sec:4-3}\\
\hline
$$\begin{array}{c}
{\rm nonchiral} \\
A + C_n
\end{array}$$
& 2 & $\Omega^{\spinc}_3(B \Z_n)$ & Lens space $L(n,1)$ & 
$$\begin{array}{l}
\Braket{GS | C_{n,D} | GS}
\end{array}$$
& Partial $n$-fold rotation 
& \ref{Sec:4-4-2}\\
\hline
$$\begin{array}{c}
{\rm D} + I, \\
I^2 = (-1)^F
\end{array}$$
& 3 & $\Omega^{\pinp}_4(pt) = \Z_{16}$ & $\R P^4$ & 
$$\begin{array}{l}
\Braket{GS | I_D | GS}
\end{array}$$
 & Partial inversion 
&\ref{Sec:5-1}\\
\hline
A + I & 3 & $\Omega^{\pinc}_4(pt) = \Z_{8} \oplus \Z_2$ & $\R P^4$ for subgroup $\Z_8$ & 
$$\begin{array}{l}
\Braket{GS | I_D | GS}
\end{array}$$
& Partial inversion. 
A $U(1)$ phase associated with $I$ is chosen so that $I^2 = (-1)^F$. 
&\ref{Sec:5-2}\\
\hline
\end{tabular}
\end{center}
\end{table}

%
%

\subsection{Outline}

In this paper, we construct and evaluate many-body topological (SPT)
invariants for various fermionic SPT phases. 
In particular, for the cases with point group symmetries and unitary symmetries. 
Table~\ref{tab:list} is the list of many-body topological invariants for fermionic SPT 
phases studied in the present paper.

Partly overlapping results were already reported in our previous paper, Ref.\ \onlinecite{Shapourian-Shiozaki-Ryu}.
In Ref.~\onlinecite{Shapourian-Shiozaki-Ryu},
we reported the construction of topological invariants,
taking as examples,  (1+1) and (3+1)-dimensional topological superconductors
protected by an orientation-reversing symmetry (an inversion or time-reversal).
(We shall recapitulate some essential points of
Ref.~\onlinecite{Shapourian-Shiozaki-Ryu} with regard to the partial reflection
in $(1+1)d$ SPT phases.)
The result for the case of time-reversal symmetry is one of the main differences,
between Ref.~\onlinecite{Shapourian-Shiozaki-Ryu} and the present work,
i.e., here we focus on point group symmetries and unitary symmetries.  
Because of the anti-unitary nature of time-reversal, the construction of 
many-body topological invariants is somewhat more complicated, and involves
the so-called partial transpose (or its proper extension to fermionic systems). 

Another difference between Ref.~\onlinecite{Shapourian-Shiozaki-Ryu} and this work
is that in Ref.\ \onlinecite{Shapourian-Shiozaki-Ryu}
the path-integral representation of many-body topological invariant
was developed,
which can be applied to a given Hamiltonian without the knowledge of the ground state wave function. 
There, the spacetime manifold is discretized by using the Suzuki-Trotter decomposition of the thermal density matrix.
We showed how to introduce a cross-cap in the spacetime path integral by using
orientation-reversing symmetries.
In the present paper, however,
we focus on the operator formalism, in which 
the input data for the many-body topological invariants
is a ground state wave function of SPT phases.

The rest of the paper is organized as follows:
\begin{itemize}
\item
  We start, in Sec.\ \ref{Sec:2}, by first introducing examples of topological invariants generated 
by mapping tori for fermionic topological phases. 
For these cases, partial transformations are not necessary to discuss their
topological invariants.

\item
In Sec.\ \ref{Sec:3}, 
partial reflection in $(1+1)$ dimensions is discussed. 
There, we illustrate that for some symmetry classes it is not sufficient to use
full symmetry transformations to topological phases. 
We show that 
the ground state expectation value of partial reflection faithfully captures the $\Z_8$ topological invariant for 
class D topological superconductors with reflection symmetry. 

\item
In Sec.\ \ref{Sec:4}, we discuss partial $C_n$ rotations in 
$(2+1)$-dimensional topological phases. 
By using the modular properties of CFTs, 
the ground state expectation values of partial rotations will be interpreted as 
partition functions of the $(2+1)$-dimensional TQFTs  on the lens space.
The relation between partial rotations and
the $\spinc$ cobordism classification for 
on-site $\Z_n$ symmetry will be discussed. 

\item
 In Sec.\ \ref{Sec:5}, we show that 
partial inversion can be used to construct 
the topological invariant
of $(3+1)$-dimensional topological phases protected by
inversion symmetry.
We also discuss the general formula for 
the ground state expectation of partial inversion on 
topological superconductor and insulator in any even spacetime dimension,
by using their boundary Dirac fermion theory on a sphere. 

\item
  Finally, we conclude in Sec.~\ref{Sec:6} with some discussion and outlooks. 
We also explain technical details in four appendices. 

\end{itemize}

\section{Symmetry transformation on twisted ground states}
\label{Sec:2}
In this section, 
we discuss fermionic SPT phases that can be detected by
``full'' (as opposed to partial) symmetry transformation on the 
twisted ground states, 
such as $(1+1)d$ class D topological superconductors. 
It is sufficient for detecting the topological invariants not to use the partial symmetry 
transformations, for example, $(1+1)d$ bosonic SPT phases with on-site 
unitary symmetry~\cite{Pollmann2012, Moore2006, Shiozaki-Ryu} 
and $(1+1)d$ class D superconductor~\cite{Moore2006}. 
In bosonic SPT phases with on-site symmetry, 
Hung-Wen~\cite{Hung2014} generalized this approach 
to those in higher spacetime dimensions, where they also discussed 
modular transformations on the space manifold. 
However, for fermionic topological phases, there are few literatures for this approach to 
detect topological invariants, thus, it might be useful to explain details.


\subsection{$(1+1)d$ topological superconductors ($\Omega^{\rm Spin}_2(pt)=\mathbb{Z}_2$)}
\label{sec:(1+1)D}

We start with the definition of the many-body $\Z_2$ topological invariant 
for $(1+1)d$ superconductors. 
The topological classification is given by the spin cobordism group 
\begin{align}
\Omega_2^{\rm Spin}(pt) = \Z_2. 
\end{align}
The generating manifold is the 2-torus $T^2$ with periodic boundary conditions for 
time and space directions.
\cite{Kirby}
In the operator formalism, 
the path integral on the generating manifold   
corresponds to 
the expectation value of the fermion parity 
with respect to the ground state on a closed space circle
with the periodic boundary condition.~\cite{Moore2006, read2000paired, ChoShiozakiRyuLudwig2016}

\subsubsection{The Kitaev Majorana chain and the many-body $\Z_2$ invariant}

Our construction of the SPT invariant can be best explained by 
taking an example -- the Kitaev Majorana chain. 
\cite{kitaev2001unpaired}
Let us consider a closed chain with $N$ sites. 
Let $f_j$ be complex fermions defined on the $j$-site. 
Then,
the Hamiltonian of the Kitaev chain is given by 
\begin{align}
H_{r/ns}& = \frac{1}{2} \sum_{j=1}^{N-1} \big[ -f^{\dag}_j f_{j+1} - f_j f_{j+1} + h.c. \big]
\pm \frac{1}{2}\big[ -f^{\dag}_N f_1 - f_N f_1 + h.c. \big]. 
\label{Eq:1D_D_Comp}
\end{align}
This Hamiltonian is fine-tuned
such that it is at a renormalization group fixed point with zero correlation length. 
The subscript $r$ and $ns$ represent the periodic boundary condition 
(the ``Ramond'' sector) $f_{N+1} = f_1$, 
and the anti-periodic boundary condition 
(the ``Neveu-Schwarz'' sector) 
$f_{N+1} = - f_1$, respectively. 
The fermion parity operator $(-1)^F$ is defined by 
\begin{align}
(-1)^F := (-1)^{\sum_{j=1}^N f^{\dag}_j f_j}. 
\end{align}
The anti-periodic boundary condition can be thought of as 
a symmetry twist by the fermion parity $(-1)^F$.
(This symmetry twist can be interpreted as an 
introduction of a topological defect, i.e.\ 
the location of the defect can move along the closed chain by a local unitary transformation. )

Introducing real fermion operators $c^L_j, c^R_j$ at the $j$-th site by 
\begin{align}
c_j^L = i(f_j - f_j^{\dag}), \quad 
c_j^R = f_j + f_j^{\dag}, 
\label{eq:1d_sc_def_real_fermions}
\end{align}
the Hamiltonian and the fermion number parity 
can be rewritten as
\begin{align}
H_{r/ns} &= \frac{i}{2} \sum_{j=1}^{N-1} c^R_j c^L_{j+1} \pm \frac{i}{2} c^R_N c^L_1, 
\label{Eq:1D_D_Real}
\qquad
(-1)^F =  (-i c^L_1 c^R_1) (-i c^L_2 c^R_2) \cdots (-i c^L_{N} c^R_{N}). 
\end{align}
It is also convenient to 
introduce complex fermions $g_j$ living on the bond $(j, j+1)$ as 
\begin{align}
g_j := \frac{c_j^R + i c^L_{j+1}}{2},  
\quad 
g^{\dag}_j := \frac{c_j^R -i c^L_{j+1}}{2}.
\end{align}
In terms of $g_j, g^{\dag}_j$, the Hamiltonian is written as 
\begin{align}
H_{r/ns} 
= \sum_{j=1}^{N-1} \big[ g^{\dag}_j g_j - \frac{1}{2} \big] 
\pm \big[ g^{\dag}_N g_N - \frac{1}{2} \big]. 
\end{align}
The ground states $\ket{GS_{r/ns}}$ of $H_{r/ns}$ are given by 
\begin{align}
\ket{GS_r} = \ket{0_g}, \quad  
\ket{GS_{ns}} = g^{\dag}_N \ket{0_g}, 
\end{align}
where $\ket{0_g}$ is the Fock vacuum of $g_j$ fermions. 
Explicitly, the ground states $\ket{GS_{r/ns}}$ can be written 
in terms of the $f_j$ fermions 
and their Fock vacuum $\ket{0_f}$ as 
\begin{align}
\ket{GS_{r/ns}} 
&\sim \left[ \prod_j (1+f^{\dag}_j) \mp \prod_j (1- f^{\dag}_j) \right] \ket{0_f} 
\nonumber \\
&\sim \sum_{n : {\rm odd/even}} \sum_{1 \leq p_1 < p_2 < \cdots < p_n \leq N} f^{\dag}_{p_1} f^{\dag}_{p_2} \cdots f^{\dag}_{p_n} \ket{0_f} 
\label{Eq:1D_D_GS}
\end{align}
up to a normalization. 
Here, we used the abbreviation 
$\prod_j (1\pm f^{\dag}_j) := (1\pm f^{\dag}_1) \cdots (1\pm f^{\dag}_N)$.

As advocated above, we consider the 
the fermion number parity of the ground state of $H_{r/ns}$. 
Noticing the following relation between 
the fermion parities defined for $f_j$ and $g_j$ fermions 
\begin{align}
(-1)^F &= - (-1)^G,
\nonumber \\ 
(-1)^G &:= (-1)^{\sum_{j=1}^N g^{\dag}_j g_j} 
= (-i c^R_1 c^L_2) (-i c^R_2 c^L_3) \cdots (-i c^R_N c^L_1), 
\end{align}
the fermion number parity of the ground states is computed as 
\begin{align}
(-1)^F \ket{GS_r} = - \ket{GS_r}, \quad 
(-1)^F \ket{GS_{ns}} = \ket{GS_{ns}}. 
\label{intermediate results}
\end{align}
The odd fermion parity of the ground state 
for the periodic boundary condition is the 
hallmark of 
$(1+1)d$ topological superconductors. 
While the above result 
\eqref{intermediate results}
(and hence \eqref{eq:z2inv_1d_sc} below) 
is derived for the specific model \eqref{Eq:1D_D_Comp},
the Kitaev chain in the limit of zero correlation length, 
the same result
holds for any gapped $\Z_2$ nontrivial superconductors with a unique ground state on a closed chain, 
since the fermion number parity $(-1)^F$ 
has a definite $\Z_2$ value on ground states $\ket{GS_{r/ns}}$. 
We show, as an example, the explicit form of the ground state with a exactly solvable model 
with a many-body interaction and finite correlation length~\cite{katsura2015exact} in Sec.~\ref{An example of interacting Majorana chain}.

To compare these results with the $(1+1)d$ spin TQFT, 
we recast our result in the form of partition functions. 
Since our theory is gapped, 
the partition functions in the zero temperature limit consist only of the ground states,  
\begin{align} \label{eq:z2inv_1d_sc} 
Z(T^2,(r,r)) &= {\rm Tr}_r [(-1)^F]  = \braket{GS_r | (-1)^F | GS_r} = -1, 
\nonumber \\
Z(T^2,(ns,r)) & = {\rm Tr}_r [1] = \braket{GS_r | GS_r} = 1, 
\nonumber \\
Z(T^2,(r,ns)) &= {\rm Tr}_{ns} [(-1)^F] = \braket{GS_{ns} | (-1)^F | GS_{ns}} = 1. 
\nonumber \\
Z(T^2,(ns,ns)) &= {\rm Tr}_{ns} [1] = \braket{GS_{ns} | GS_{ns}} = 1. 
\end{align}
Here, the notation $Z(T^2,(a,b))$ means the partition function on $T^2$ with 
$a$-($b$-)boundary condition for the time (space) direction. 
Recall that 
the periodic boundary condition for the time direction in 
the path integral picture 
corresponds to, in the operator picture, 
the insertion of the fermion number parity $(-1)^F$ operator
in the trace. 
Moreover, notice that these results are modular invariant: 
the modular transformations on $T^2$ permutes $(r,ns)$, $(ns,r)$ and $(ns,ns)$ sectors 
while the $(r,r)$ sector is unchanged.

To further support the claim that the fermion number parity 
of the ground states with twisted boundary conditions is 
an SPT invariant, 
we now compare the above partition functions (\ref{eq:z2inv_1d_sc}) of the Kitaev model 
with the Arf invariant of $\spin$ structures,~\cite{Kapustin2015a}
which is a $\Z_2$-valued function 
\begin{align}
{\rm Arf} : \spin(M) \to \Z_2 = \{0,1\}
\end{align}
on $\spin$ structures of a given oriented manifold $M$.  
See Appendix \ref{Arf and Brown invariant} for a review of the Arf invariant, 
where we illustrate how to compute the Arf invariants on a given oriented 2-manifold. 
The Arf invariants on $T^2$ are summarized in (\ref{eq:app_arf_torus}), 
which coincide with the torus partition functions (\ref{eq:z2inv_1d_sc}) of the Kitaev chain with twisted boundary conditions, says, 
\begin{align}
Z(T^2,\eta) = (-1)^{{\rm Arf}(\eta)}, 
\quad
\eta \in \spin(T^2). 
\end{align}

%
%
%


\subsubsection{The trivial phase}

Let us consider the Hamiltonian 
\begin{align}
H^{\rm triv}_{r/ns} = \sum_{j=1}^N \big[ f^{\dag}_j f_j - \frac{1}{2} \big] = \frac{i}{2} \sum_{j=1}^N c^L_j c^R_j,
\end{align}
which realizes a trivial superconductor.  
There is no difference between $r$ and $ns$ sectors since there is no hopping term. 
The ground state is just the Fock vacuum of $f$ fermions, 
\begin{align}
\ket{GS^{\rm triv}_{r/ns}} = \ket{0_f}.
\end{align}
The partition functions (SPT invariants) can be computed as
\begin{align}
Z^{\rm triv}(T^2,(r,r)) &= {\rm Tr}_r [(-1)^F]  = \braket{GS^{\rm triv}_r | (-1)^F | GS^{\rm triv}_r} = 1, 
\nonumber \\
Z^{\rm triv}(T^2,(ns,r)) &= {\rm Tr}_r [1] = \braket{GS^{\rm triv}_r | GS^{\rm triv}_r} = 1, 
\nonumber \\
Z^{\rm triv}(T^2,(r,ns)) &= {\rm Tr}_{ns} [(-1)^F] = \braket{GS^{\rm triv}_{ns} | (-1)^F | GS^{\rm triv}_{ns}} = 1. 
\nonumber \\
Z^{\rm triv}(T^2,(ns,ns)) &= {\rm Tr}_{ns} [1] = \braket{GS^{\rm triv}_{ns} | GS^{\rm triv}_{ns}} = 1. 
\end{align}
There is no spin structure dependence, which is consistent with
the triviality of this phase.

\subsubsection{An example of interacting Majorana chains}
\label{An example of interacting Majorana chain}

To demonstrate that 
the definition of the $\Z_2$ SPT invariant  
(\ref{eq:z2inv_1d_sc}) is applicable to interacting Majorana chains, 
we consider an example of the many-body ground states 
of an interacting Majorana chain 
discussed in Ref.~\onlinecite{katsura2015exact}. 
Let us consider the following Hamiltonian including 
an extended Hubbard interaction 
\begin{align}
H_{r/ns} 
&= \sum_{j=1}^{N-1} \big[ -t f^{\dag}_j f_{j+1} + \Delta f_j f_{j+1} + h.c. \big] 
\pm \big[ -t f^{\dag}_N f_1 + \Delta f_N f_1 + h.c. \big] \nonumber \\
&\quad 
- \frac{\mu}{2} 
\sum_{j=1}^{N}(f^{\dag}_j f_j + f^{\dag}_{j+1} f_{j+1}-1) 
+ U 
\sum_{j=1}^{N}\big( f^{\dag}_j f_j-\frac{1}{2} \big) \big( f^{\dag}_{j+1} f_{j+1}-\frac{1}{2} \big) 
\end{align}
with $\mu = 4 \sqrt{U^2 + tU + \frac{t^2-\Delta^2}{4}}$. 
The ground state is given by 
\begin{align}
\ket{GS_{r/ns}(\alpha)} 
\sim \left[ \prod_j (1+ \alpha f^{\dag}_j) \mp \prod_j (1- \alpha f^{\dag}_j) \right] \ket{0_f}, 
\end{align}
where $\alpha$ is defined by $\alpha = \sqrt{\cot \frac{\theta}{2}}, 
\theta = \tan^{-1}\big( 2 \Delta/\mu \big)$. 
Clearly, the R sector consists only of odd numbers of fermions and 
shows 
the nontrivial $\Z_2$ invariant 
$\Braket{GS_r(\alpha)|(-1)^F |GS_r(\alpha)} = -1$.

\subsection{$(1+1)d$ superconductors protected by reflection 
($\Omega^{\rm Pin^+}_2(pt)=\mathbb{Z}_2$)
}
\label{sec:(1+1)D+R+}
Next, we consider $(1+1)d$ topological superconductors 
protected by reflection $R$ with $R^2=1$.
These topological superconductors can be thought of as a CPT dual of class DIII
topological superconductors. 
The topological classification is given by the $\pinp$ cobordism group~\cite{
Kirby, Kapustin2015a}
\begin{align}
\Omega_2^{\pinp}(pt) = \Z_2, 
\label{eq:bord_1d_pinp}
\end{align}
and the generating manifold is the Klein bottle $KB$. 
Since the Klein bottle is realized in the mapping torus 
\begin{align}
S^1 \times_{R} S^1 = S^1 \times [0,1] / \big( (x,0) \sim (-x,1) \big), 
\end{align}
we expect that 
the many-body $\Z_2$ invariant can be 
computable by using full symmetry operations
on the twisted ground states.

\subsubsection{A nontrivial model and the many-body $\Z_2$ invariant}
We consider the reflection symmetric pair of Kitaev chains,
Eq.\ (\ref{Eq:1D_D_Comp}), 
\begin{align}
H_{r/ns} 
&= \frac{1}{2} \sum_{j=1}^{N-1} \big[ -f^{\dag}_{\uparrow,j} f_{\uparrow,j+1} - f_{\uparrow,j} f_{\uparrow,j+1} + h.c. \big] 
\pm \frac{1}{2}\big[ -f^{\dag}_{\uparrow, N} f_{\uparrow,1} - f_{\uparrow,N} f_{\uparrow,1} + h.c. \big] 
\nonumber \\
&\quad 
+ \frac{1}{2} \sum_{j=1}^{N-1} \big[ -f^{\dag}_{\downarrow,j} f_{\downarrow,j+1} + f_{\downarrow,j} f_{\downarrow,j+1} + h.c. \big] 
\pm \frac{1}{2}\big[ -f^{\dag}_{\downarrow, N} f_{\downarrow,1} + f_{\downarrow,N} f_{\downarrow,1} + h.c. \big]. 
\end{align}
We consider reflection that exchanges two flavors 
$\uparrow$ and $\downarrow$ as 
\begin{align}
R f^{\dag}_{\uparrow, j} R^{-1} = f^{\dag}_{\downarrow, N-j+1}, \quad 
R f^{\dag}_{\downarrow, j} R^{-1} = f^{\dag}_{\uparrow, N-j+1}, \quad 
R \ket{0_f} = \ket{0_f}, 
\end{align}
where $\ket{0_f}$ is the Fock vacuum of the $f_{\uparrow}$ and $f_{\downarrow}$ fermions. 
The ground states $\ket{GS_{r/ns}}$ of the R and NS sectors 
are given by 
\begin{align}
\ket{GS_{r/ns}} 
= 
\Big( \sum_{n: {\rm odd/even}} \sum_{1 \leq p_1 < \cdots < p_n \leq N} f^{\dag}_{\uparrow, p_1} \cdots f^{\dag}_{\uparrow, p_n} \Big) 
\Big( \sum_{m: {\rm odd/even}} i^{m} \sum_{1 \leq q_1 < \cdots < q_m \leq N} f^{\dag}_{\downarrow, q_1} \cdots f^{\dag}_{\downarrow, q_m} \Big)  
\ket{0_f}.  
\end{align}
Then, we have 
\begin{align}
R \ket{GS_{r/ns}} 
= \left\{\begin{array}{ll}
- \ket{GS_{r}} \\
\ket{GS_{ns}}
\end{array}\right.
.
\end{align}
Thus, the many-body $\Z_2$ invariant is given by 
the ground state expectation value of full reflection on the superconducting chain with the periodic boundary condition, 
\begin{align}
Z(KB,(ns,r)) = {\rm Tr}_r \big[ R \big] = \braket{GS_r | R | GS_r} = -1. 
\label{eq:1d_sc_r_z2inv}
\end{align}
This agrees with the cobordism classification (\ref{eq:bord_1d_pinp}).

\subsection{$(1+1)d$ insulators with charge conjugation symmetry
($\Omega_2^{\mathrm{Spin}^{\tilde{c}+}}(pt)$)}
\label{sec:(1+1)A+C}

Our next example concerns 
$(1+1)d$ topological insulators protected by 
charge conjugation with $C^2 = 1$, 
where the action of $C$ on the fermionic Fock space is defined as
\begin{align}
C \psi^{\dag}_i(x) C^{-1} = {\cal C}_{ij} \psi_j(x), 
\quad {\cal C} {\cal C}^* = 1, 
\quad C \ket{0} = \ket{\rm full}, 
\end{align}
where $\psi_i^{\dag}(x)$ is a complex fermion operator 
defined on a closed ring of length $L$, 
$\ket{0}$ and $\ket{\rm full}$ are 
the Fock vacuum and the fully occupied state of $\psi^{\dag}_i(x)$ fermions,
respectively. 
This symmetry is relevant to, for example, the fermionic Hubbard model 
at half filling. 
At the level of non-interacting fermions, 
the ensemble of single-particles Hamiltonians with 
this symmetry 
is equivalent to 
the ensemble of BdG Hamiltonians
(symmetry class D in Altland-Zirnbauer symmetry classes
\cite{altland1997nonstandard}), 
where $C$ plays the role of the 
particle-hole symmetry (constraint) for Nambu spinors. 
The topological classification 
for this symmetry class in $(1+1)d$ is $\mathbb{Z}_2$
in the absence of interactions, 
and this is expected to be so even in the presence of interactions. 
In the following,
we give the definition of the many-body $\Z_2$ invariant 
in terms of the ground state with twisted boundary conditions. 
The relevant spin structures here
for insulators with charge-conjugation symmetry
are called $\spin^{\tilde c+}$ structures.
This has to be distinguished 
from $\spin$ structures for 
charge neutral fermions in superconductors.

%


\subsubsection{The $\Z_2$ equivariant line bundle}

For our construction of the $\mathbb{Z}_2$ many-body topological invariant
we consider twsited spatial boundary conditions by symmetry of the problem. 
Thanks to the $U(1)$ particle number conservation, 
we can introduce the twisted boundary condition $\psi(x+L) = e^{i \theta} \psi(x)$. 
Let $\Ket{GS(\theta)}$ be the ground state of the Hamiltonian with the twist 
$\theta \in U(1)$. 
Since charge conjugation $C$ flips the $U(1)$ flux, 
we have a $\Z_2$-equivariant complex line bundle over the flux space $U(1)$: 
\begin{align}
C \Ket{GS(\theta)} = e^{i \phi(\theta)} \Ket{GS(-\theta)}, 
\quad 
e^{i \phi(\theta)} e^{i \phi(-\theta)} = 1. 
\label{eq:1d_A+C_C_action} 
\end{align}
This $\Z_2$-equivariant structure leads to the $\Z_2$ quantization of the Berry phase 
\begin{align}
\gamma 
= \exp \Big[ \oint_{-\pi}^{\pi} d \theta \braket{GS(\theta) | \partial_{\theta} |GS(\theta)} \Big]
= e^{i \phi(\pi) - i \phi(0)}
= \frac{\braket{GS(\pi) | C | GS(\pi)}}{\braket{GS(0) | C |GS(0)}}
\in  \{ \pm 1 \}.
\label{Eq:1D_Z2_C}
\end{align}
Here we used 
$\int_0^{\pi} d \theta \braket{GS(\theta) | \partial_{\theta} | GS(\theta)} 
= \int_0^{\pi} d \theta \braket{GS(\theta) | C^{-1} \partial_{\theta} C | GS(\theta)) } 
= i (\phi(\pi) - \phi(0)) - \int_{-\pi}^0 d \theta \braket{GS(\theta) | \partial_{\theta} GS(\theta)}$. 
This is a candidate of the many-body $\Z_2$ invariant. 
Our remaining task is to show that the existence of a model with 
nontrivial $\Z_2$ invariant.

\subsubsection{A nontrivial model and the $\Z_2$ invariant}
Let us consider the following two orbital model of complex fermions 
$a_j, b_j$ in a closed ring, 
\begin{align}
H = \sum_{j=1}^{N-1} b^{\dag}_j a_{j+1} + b^{\dag}_N a_{1} + h.c., 
\label{eq:1d_A_C}
\end{align}
This model is invariant under the charge conjugation symmetry defined by 
\begin{align}
C a_j C^{-1} = a^{\dag}_j, \quad 
C b_j C^{-1} = -b^{\dag}_j, \quad 
C \ket{0} = \ket{\rm full}. 
\end{align}
We introduce the $U(1)$-twisted Hamiltonian $H(\theta)$ by 
\begin{align}
H(\theta):= \sum_{j=1}^{N-1} b^{\dag}_j a_{j+1} + e^{-i \theta} b^{\dag}_N a_{1} + h.c., 
\quad 
C H(\theta) C^{-1} = H(- \theta). 
\end{align}
By introducing ``bond'' complex fermions 
$c_j, d_j (j=1, \dots N-1), c_N(\theta), d_N(\theta)$ as 
\begin{align}
&c_j := \frac{b_j + a_{j+1}}{\sqrt{2}}, \quad 
d_j := \frac{b_j - a_{j+1}}{\sqrt{2}}, \quad  (j=1, \dots  N-1), 
\nonumber \\
&c_N(\theta) := \frac{b_N + e^{-i \theta} a_1}{\sqrt{2}}, \quad 
d_N(\theta) := \frac{b_N - e^{-i \theta} a_1}{\sqrt{2}}, 
\end{align}
the Hamiltonian $H(\theta)$ can be written as  
\begin{align}
H(\theta) 
= \sum_{j =1}^{N-1} [c^{\dag}_j c_j - d^{\dag}_j d_j] 
+ c_N^{\dag}(\theta) c_N(\theta) - d^{\dag}_N(\theta) d_N(\theta). 
\end{align}
The ground state of $H(\theta)$ is obtained by filling $d_j$ and $d_N(\theta)$ 
fermions, 
\begin{align}
\ket{GS(\theta)} = d^{\dag}_1 \cdots d^{\dag}_{N-1} d^{\dag}_N(\theta) \ket{0}. 
\label{eq:1d_A_C_GS}
\end{align}
An explicit calculation shows
that the Berry phase for $|GS(\theta)\rangle$ is non-trivial,
$\gamma = -1$. 
This can be confirmed from the $\Z_2$-equivariant structure on 
the ground state $\Ket{GS(\theta)}$. 
By making use of 
\begin{align}
&
C d^{\dag}_j C^{-1} = - c_j \, (j=1, \dots, N-1), 
\quad 
C d^{\dag}_N(\theta) C^{-1} = - c_N(-\theta), 
\nonumber \\
&c_N(-\theta) b^{\dag}_N a^{\dag}_1 = - e^{i \theta} d^{\dag}_N(- \theta) + b^{\dag}_N a^{\dag}_1 c_N(- \theta), 
\end{align}
then we have the nontrivial $\Z_2$-equivariant line bundle over $U(1)$ as 
\begin{align}
C \ket{GS(\theta)} 
&\sim c_1 \cdots c_{N-1} c_N(\theta) \ket{\rm full} 
\nonumber \\
&\sim c_1 \cdots c_{N-1} c_N(\theta) a^{\dag}_1 \cdots a^{\dag}_N b^{\dag}_1 \cdots b^{\dag}_N \ket{0} 
\nonumber \\
&\sim e^{i \theta} \ket{GS(-\theta)}. 
\end{align}

\subsubsection{$\Z_2$ anomaly on the edge state}

Let us now have a further look at
the topological non-triviality of the model,
from the point of view of its boundaries. 
By the bulk-boundary correspondence,
the non-trivial bulk topological invariant ($\gamma=-1$)
is expected to manifests itself, in the presence of boundaries (edges),
as a quantum anomaly. 
Let us now consider a topological insulator
with the non-trivial bulk $\Z_2$ invariant 
with open boundary conditions,
and focus on the low-energy excitations of the one of the edges.
There is a fermion zero mode
created/annihilated by a complex fermion creation operator $a^{\dag}$/$a$. 
The charge conjugation (particle-hole) operator $C$ acts
on the fermion mode as
$C a^{\dag} C^{-1} = a$.
The edge theory defined in terms of $a^{\dag}, a$ is anomalous in the sense that
both the $U(1)$ charge conservation and charge conjugation $C$ symmetries
(the total symmetry group $=U(1) \rtimes C$)
cannot be imposed on a unique ground state. 
In fact, the ground state with the $U(1)$ symmetry
(having a definite particle number)
is either $\ket{0}$ or $a^{\dag}\ket{0}$.
However, $C$ exchanges $\ket{0}$ and $a^{\dag} \ket{0}$ (where we set $C \ket{0} \sim a^{\dag} \ket{0}$). 
It is also easy to see that this anomaly is $\mathbb{Z}_2$ in the sense that if
we stack two identical copies of the system, this anomaly disappears.

\subsubsection{Cut and glue construction and the $\Z_2$ invariant}

Once the physics at the edge of non-trivial $\mathbb{Z}_2$ topological
insulators are understood, 
one can use the cut and glue construction~\cite{Qi2011b}
to give a proof that the corresponding bulk $\Z_2$ invariant (\ref{Eq:1D_Z2_C}) is nontrivial.
Let us consider an interval $I=[0,N]$ of the $\Z_2$ nontrivial insulating chain.
The low-energy excitations can be described by
two sets of edge complex fermion creation/annihilation operators,
$\{a^{\dag}_1, a^{\ }_1\}$ and $\{b^{\dag}_N, b^{\ }_N\}$.
The action of $C$ on these fermion opeartors is given by 
$C a_1^{\dag} C^{-1}=a_1$ and $C b_N^{\dag} C^{-1} = -b_N$.
The unique ground state of the Hamiltonian on the closed chain can be approximated by the ground state of the following gluing Hamiltonian 
\begin{align}
H_{\rm glue}(\theta)=e^{-i \theta} b_N^{\dag} a_1 + h.c., 
\end{align}
where we have introduced the twisted boundary condition by the $U(1)$ symmetry.
A direct calculation shows that the $\Z_2$ invariant (\ref{Eq:1D_Z2_C}) is nontrivial, $\gamma = -1$.

\subsection{$(2+1)d$ superconductor with reflection symmetry 
($\Omega_3^{{\rm Pin}^+}(pt)=\mathbb{Z}_2$)}
\label{sec:(2+1)D+R+}

In this section, we give the many-body definition of 
the $\Z_2$ topological invariant 
for $(2+1)$-dimensional superconductors 
protected by reflection symmetry with $R^2 = 1$.
Topological superconductors in this symmetry class 
can be considered as a CPT dual of 
class DIII topological superconductors in $(2+1)d$. 
Here, 
the reflection acts on the fermion creation operator 
$\psi^{\dag}_i(x,y)$ as
\begin{align}
R \psi^{\dag}_i(x,y) R^{-1} = \psi^{\dag}_j(-x,y) {\cal R}_{ji}, 
\quad 
{\cal R}^2 = 1, 
\end{align}
where $(x,y)$ is the spatial coordinate 
and $i,j$ are the flavor (orbital, spin, etc.) indices. 
The topological classification is given by the $\pinp$ cobordism group 
\begin{align}
\Omega^{\pinp}_3(pt) = \Z_2. 
\end{align}
The generating manifold of the cobordism group $\Z_2$ is $KB \times S^1$
($KB=\,$ the Klein bottle),~\cite{Kirby}
which is a mapping torus.
In other words, 
in the operator formalism, the many-body $\Z_2$ invariant can be constructed  
by considering a fully symmetry action on twisted ground states. 

\subsubsection{The $\Z_2$ equivariant line bundle}
Let us consider a superconductor on 2-torus $T^2$. 
There are four distinct spin structures: $\{(r,r),(r,ns),(ns,r),(ns,ns)\}$. 
Since the reflection transformation $R$ preserves these spin structures, 
we have the following $\Z_2$ equivariant bundle over four points: 
\begin{align}
R \ket{GS(\eta)} = (-1)^{\nu(\eta)} \ket{GS(\eta)}, \qquad 
\eta \in \{(r,r),(r,ns),(ns,r),(ns,ns)\}. 
\end{align}
To remove ``weak indices'' 
\eqref{eq:1d_sc_r_z2inv}
arising from $(1+1)d$ superconductors 
with reflection symmetry, we define the ``strong'' $\Z_2$ invariant $(-1)^{\nu}$ specific 
to $(2+1)d$ superconductors by 
\begin{align}
(-1)^{\nu} = \prod_{\eta \in \{(r,r),(r,ns),(ns,r),(ns,ns)\}} (-1)^{\nu(\eta)}. 
\end{align}
In the following, we show an example of a $\Z_2$ nontrivial model.

\subsubsection{A nontrivial model and the many-body $\mathbb{Z}_2$ invariant}
A model Hamiltonian is a reflection symmetric pair of $(p_x+i p_y)$ and  $(p_x - i p_y)$ 
superconductors  
\begin{align}
H
&= \sum_{\bk} \psi^{\dag}_{\uparrow}(\bk) (m-\cos k_x - \cos k_y) \psi_{\uparrow}(\bk) 
+ \frac{1}{2} \sum_{\bk} (\sin k_x + i \sin k_y) \psi^{\dag}_{\uparrow}(\bk) \psi^{\dag}_{\uparrow}(-\bk) + h.c.
\nonumber \\
&+ \sum_{\bk} \psi^{\dag}_{\downarrow}(\bk) (m-\cos k_x - \cos k_y) \psi_{\downarrow}(\bk) 
+ \frac{1}{2} \sum_{\bk} (\sin k_x - i \sin k_y) \psi^{\dag}_{\downarrow}(\bk) \psi^{\dag}_{\downarrow}(-\bk) + h.c. 
\end{align}
The reflection $R$ acts on the fermion fields as
\begin{align}
R \psi^{\dag}_{\uparrow}(k_x,k_y) R^{-1} = \psi^{\dag}_{\downarrow}(k_x,-k_y), 
\quad 
R \psi^{\dag}_{\downarrow}(k_x,k_y) R^{-1} = \psi^{\dag}_{\uparrow}(k_x,-k_y), 
\quad 
R \ket{0} = \ket{0}, 
\end{align}
where $\ket{0}$ is the Fock vacuum of
the $\psi_{\uparrow, \downarrow}$ fermions. 
The ground state for each spin structure is given by
the following BCS form:
\cite{read2000paired}
\begin{itemize}
\item For the R-R sector, 
the set of allowed momenta are  
$(k_x,k_y) = (\frac{2 \pi n_x}{L}, \frac{2 \pi n_y}{L})$, 
and the ground state is given by
\begin{align}
&\ket{\Psi(r,r)} = \Psi^{\dag}_{\rm UP}
\cdot 
\exp \left( \sum_{
\substack{k_x>0\\ 
\bk \neq (0,0), (0,\pi), (\pi,0), (\pi,\pi)} }
\left\{ g_+(\bk) \psi^{\dag}_{\uparrow}(\bk) \psi^{\dag}_{\uparrow}(-\bk) + g_-(\bk) \psi^{\dag}_{\downarrow}(\bk) \psi^{\dag}_{\downarrow}(-\bk) \right\} \right)  \ket{0}.   
\end{align}
\item 
For the NS-NS sector, 
the set of allowed momenta are
$(k_x,k_y) = (\frac{2 \pi}{L}(n_x+\frac{1}{2}), \frac{2 \pi}{L}(n_y+\frac{1}{2}) )$, 
and the ground state is given by
\begin{align}
&\ket{\Psi(ns,ns)} = \exp \left( \sum_{k_x > 0} \left\{ g_+(\bk) \psi^{\dag}_{\uparrow}(\bk) \psi^{\dag}_{\uparrow}(-\bk) + g_-(\bk) \psi^{\dag}_{\downarrow}(\bk) \psi^{\dag}_{\downarrow}(-\bk) \right\} \right) \ket{0}. 
\end{align}
\item
For the R-NS sector, 
the set of allowed momenta are
$(k_x,k_y) = (\frac{2 \pi n_x}{L}, \frac{2 \pi}{L} (n_y+\frac{1}{2}))$,
and the ground state is given by
\begin{align}
&\ket{\Psi(r,ns)} = \exp \left( \sum_{k_x > 0} \left\{ g_+(\bk) \psi^{\dag}_{\uparrow}(\bk) \psi^{\dag}_{\uparrow}(-\bk) + g_-(\bk) \psi^{\dag}_{\downarrow}(\bk) \psi^{\dag}_{\downarrow}(-\bk) \right\} \right) \ket{0}.
\end{align}
\item 
For the NS-R sector, 
the set of allowed momenta are
$(k_x,k_y) = (\frac{2 \pi}{L}(n_x+\frac{1}{2}), \frac{2 \pi n_y}{L} )$, 
and the ground state is given by
\begin{align}
&\ket{\Psi(ns,r)} = \exp \left( \sum_{k_x > 0} \left\{ g_+(\bk) \psi^{\dag}_{\uparrow}(\bk) \psi^{\dag}_{\uparrow}(-\bk) + g_-(\bk) \psi^{\dag}_{\downarrow}(\bk) \psi^{\dag}_{\downarrow}(-\bk) \right\} \right) \ket{0}.
\end{align}
\end{itemize}
Here,  $g_{\pm}(\boldsymbol{k})$ is given by
\begin{align}
g_{\pm}(\boldsymbol{k}) = \frac{\sqrt{\epsilon^2_k + \sin^2 k_x + \sin^2 k_y} - \epsilon_{\boldsymbol{k}}}{\sin k_x \mp i \sin k_y}, 
\quad 
\epsilon_{\boldsymbol{k}} = m - \cos k_x - \cos k_y, 
\end{align}
and $\Psi^{\dag}_{\rm UP}$ is the contribution from unpaired fermions 
and given by
\begin{align}
\Psi^{\dag}_{\rm UP} 
:= \left\{ \begin{array}{ll}
\psi^{\dag}_{\uparrow}(0,0) \psi^{\dag}_{\uparrow}(\pi,\pi) \psi^{\dag}_{\uparrow}(0,\pi) \psi^{\dag}_{\uparrow}(\pi,0) 
\psi^{\dag}_{\downarrow}(0,0) \psi^{\dag}_{\downarrow}(\pi,\pi) \psi^{\dag}_{\downarrow}(0,\pi) \psi^{\dag}_{\downarrow}(\pi,0)& (m<-2) \\
\psi^{\dag}_{\uparrow}(0,0) \psi^{\dag}_{\uparrow}(0,\pi) \psi^{\dag}_{\uparrow}(\pi,0) \psi^{\dag}_{\downarrow}(0,0) \psi^{\dag}_{\downarrow}(0,\pi) \psi^{\dag}_{\downarrow}(\pi,0)& (-2<m<0) \\
\psi^{\dag}_{\uparrow}(0,0) \psi^{\dag}_{\downarrow}(0,0) & (0<m<2) \\
1  & (2<m) \\
\end{array}\right. 
\end{align}
Since the condensate of Cooper pairs in the BCS ground states is reflection symmetric, 
a nontrivial phase arises from unpaired fermions: 
\begin{align}
R \ket{\Psi(r,r)}
&= \left\{ \begin{array}{ll}
- \ket{\Psi(r,r)} & (-2<m<0, 0<m<2) \\
\ket{\Psi(r,r)} & (m<-2, 2<m) \\
\end{array}\right. , 
\nonumber  \\
R \ket{\Psi(\eta)} &= \ket{\Psi(\eta)} \ (\eta \in \{(r,ns), (ns,r), (ns,ns)\}). 
\end{align}
Thus, 
the topological invariant is 
$(-1)^{\nu} = -1$ for topologically nontrivial phases $(-2<m<0, 0<m<2)$.

\section{Partial reflections}
\label{Sec:3}

We now discuss some $(1+1)d$ SPT phases,
of which the detection requires the real projective plane $\R P^2$ 
as a generating manifold of the relevant cobordism group. 
For example, $(1+1)d$ bosonic SPT phases with reflection symmetry is classified by 
the unoriented cobordism group $\Omega^{O}_2(pt) = \Z_2$ with the generating manifold 
$\R P^2$.~\cite{Kapustin2014symmetry}
Other fermionic examples are discussed in the subsequent sections. 
Topologically, $\R P^2$ is realized by introducing 
a cross-cap on the spacetime manifold $S^2$. 
In the operator formalism, introducing a cross-cap 
is intuitively equivalent to 
acting with a partial reflection $R_I$ on an interval $I$ 
of a closed ring $S^1$,
as shown in Fig.\ \ref{Fig:Crosscap}. 
In $(1+1)d$ bosonic SPT phases protected by reflection symmetry, 
Pollmann and Turner showed that the expectation value of the 
partial reflection $\braket{GS | R_I | GS}$ on the ground state $\ket{GS}$ 
faithfully captures the $\Z_2$ SPT invariant.~\cite{Pollmann2012} 
In this section, 
we extend this approach to fermionic SPT phases. 
Some contents of this section were also discussed in Ref.\ \onlinecite{Shapourian-Shiozaki-Ryu}.

\subsection{$(1+1)d$ topological superconductors 
protected by reflection 
($\Omega_2^{{\rm Pin}^-}(pt)=\mathbb{Z}_8$)}
\label{sec:(1+1)D+R-}

Let us consider $(1+1)d$ 
topological superconductors protected by reflection $R$ with $R^2 = (-1)^F$.
These topological superconductors can be thought of as 
a CPT dual of class BDI topological superconductors. 
The topological classification is given by the $\pinm$ cobordism group \cite{Kirby}
\begin{align}
\Omega_2^{\pinm}(pt) = \Z_8, 
\end{align}
and the generating manifold is the real projective plane $\R P^2$.

\subsubsection{The Klein bottle partition function and the $\Z_4$ invariant}

In the following, 
we first show that 
the action of symmetry (=reflection) on the ground states
in the presence of twisted boundary conditions 
is not sufficient to capture the $\Z_8$ classification. 
In fact, from the space-time path-integral,
this corresponds to the 
partition function defined on the spacetime Klein bottle,
which is not the generating manifold.
The Klein bottle generates a  modulo 4 subgroup of the $\Z_8$ group.

We consider the same model as (\ref{Eq:1D_D_Comp}) on a closed ring. 
This model is invariant under the following reflection symmetry 
\begin{align}
R f^{\dag}_j R^{-1} = i f^{\dag}_{N-j+1}, \quad 
R f_j R^{-1} = -i f_{N-j+1}, 
\end{align}
or equivalently, 
\begin{align}
R c^L_j R^{-1} = c^R_{N-j+1}, \quad 
R c^R_j R^{-1} = -c^L_{N-j+1} 
\end{align}
in terms of real fermions introduced in (\ref{eq:1d_sc_def_real_fermions}). 
In terms of the real fermions $c^L_j, c^R_j$, 
the reflection transformation $R$ is 
explicitly written by 
\begin{align}
R = : \exp \Bigr[ - \frac{\pi}{4} \sum_{j=1}^N c^L_j c^R_{N-j+1} \Bigr]: , 
\end{align}
where we introduced the normal ordering $:\cdots :$ 
with respect to the Fock vacuum $\ket{0_f}$ of $f_j$ 
fermions to fix the overall phase of $R$,
$R \ket{0_f} = \ket{0_f}$. 
Observe that $R^2 = (-1)^F$, 
and hence $R$ here generates a $\Z_4$ symmetry.
From the concrete expressions (\ref{Eq:1D_D_GS}) of the ground states 
$\ket{GS_{r/ns}}$, we obtain 
\begin{align}
R \ket{GS_{r/ns}} 
&= \sum_{n : {\rm odd/even}} \sum_{1 \leq p_1 < p_2 < \cdots < p_n \leq N} i^n f^{\dag}_{N-p_1+1} f^{\dag}_{N-p_2+1} \cdots f^{\dag}_{N-p_n+1} \ket{0_f}  
\nonumber \\
&= \left\{\begin{array}{ll}
i \ket{GS_{r}} \\
\ket{GS_{ns}}
\end{array}\right.
\end{align}
In other words,
the partition function on the Klein bottle ($KB$) with the periodic boundary 
condition for the space direction provides the $\Z_4$ sub group invariant: 
\begin{align}
Z(KB,(ns,r)) = {\rm Tr}_r \big[ R \big] = \braket{GS_r | R | GS_r} = i. 
\end{align}
In other words, 
the full reflection cannot capture the $\Z_8$ invariant. 

\subsubsection{Partial reflection and the $\Z_8$ invariant}
\label{Sec:1D_D+R_Z8}
We have shown that the partial reflection faithfully defines the many-body $\Z_8$ invariant 
in Ref.\ \onlinecite{Shapourian-Shiozaki-Ryu}. 
Here, we give a quick derivation of
the $\Z_8$ invariant by using the ``cut and glue'' construction 
of the reduced density matrix for topological phases.
\cite{Qi2011b}
Let $\ket{GS}$ be a ground state belonging to 
the $1 \in \Z_8$ topological phase. 
We wish to compute the expectation value of partial reflection $\braket{GS | R_I | GS}$, 
where $I = [1,M]$ is an interval in a whole closed chain $S^1 = [0,N]$. 
If $M$ and $N-M$ are sufficiently larger than the correlation length of the bulk, 
the reduced density matrix on the interval $I$, 
$\rho_I = {\rm tr}_{\bar{I} } ( \ket{GS} \bra{GS} )$, 
is approximated by 
the exponential of the entanglement Hamiltonian,
consisting only of edge Majorana fermions: 
\begin{align}
\tilde H = \frac{i}{2} c^R_0 c^L_1 + \frac{i}{2} c^R_M c^L_{M+1}. 
\end{align}
Defining ``in'' and ``out'' complex fermions $f_{in}$ and $f_{out}$ by 
\begin{align}
f^{\dag}_{in} = \frac{c^R_M + i c^L_1}{2},
\quad 
f^{\dag}_{out} = \frac{c^R_0 + i c^L_{M+1}}{2}, 
\end{align}
the ground state of $\tilde H$ is 
\begin{align}
\ket{GS} = \frac{1}{\sqrt{2}} ( f^{\dag}_{in} + f^{\dag}_{out} )\ket{0}. 
\end{align}
The partial reflection acts only on the $f^{\dag}_{in}$ fermion as 
\begin{align}
R_I f^{\dag}_{in} R_I^{-1} = i f^{\dag}_{in}, 
\quad
R_I f^{\dag}_{out} R_I^{-1} = f^{\dag}_{out}, 
\quad 
R_I \ket{0} = \ket{0}.
\end{align}
Hence, we obtain 
\begin{align}
\braket{GS | R_I | GS} 
= {\rm Tr}_{in} (R_I \rho_I)
= \frac{1+i}{2} = \frac{1}{\sqrt{2}} e^{\frac{\pi i}{4}}. 
\label{eq:Brown_z8_inv_kitaev}
\end{align}
The $U(1)$ phase 
$e^{ \frac{\pi i}{4} }$
correctly captures the 
$\mathbb{Z}_8$ classification. 
In fact, 
the phase of
$\braket{GS | R_I | GS}$
coincides with
the Brown invariant of the 
$\pinm$ structure on $\mathbb{R}P^2$.~\cite{Kirby}
(For the description of the Brown invariant,
see Appendix 
\ref{Quadratic form and Z8 Brown invariant}.)
Furthermore,
with the partial reflection, 
the amplitude of 
the wave function overlap
(the partition function)
is reduced from 1 to
$1/\sqrt{2}$.
This ``loss of the amplitude''
is also of topological origin,
which is the quantum dimension of the edge Majorana fermion. 

In the remainder of this section, we present two non-trivial applications of the
partial reflection:
First, we use the partial reflection to map out the phase diagram of the
disordered Kitaev Majorana chain.
Second, we apply it to the exact ground state of an interacting Majorana chain and show that it yields the expected $\Z_8$ phase.

\subsubsection{Case study 1: Robustness of partial reflection against random disorder}
\label{Sec:1D_D+R_dis}

\begin{figure}
\includegraphics[scale=0.5]{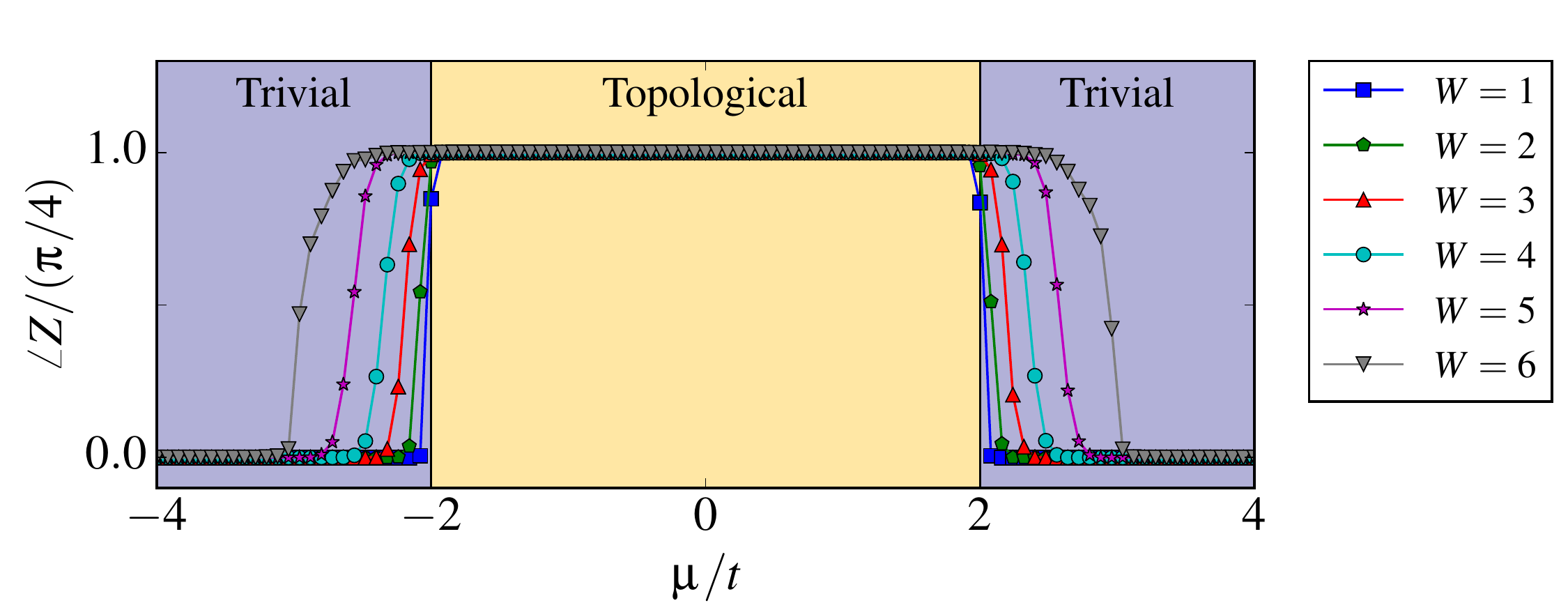}
\caption{\label{fig:Kitaev_dis} (Color online)
  Complex phase of the partial reflection $Z=\braket{GS | R_I | GS}$ for the disordered Kitaev Majorana chain. Each curve represents an ensemble average over $1000$ samples. Solid lines are guides for the eye.  Here, we set $\Delta=t$, $N=200$ and $N_\text{part}=100$.}
\end{figure}

\begin{figure}
\includegraphics[scale=0.5]{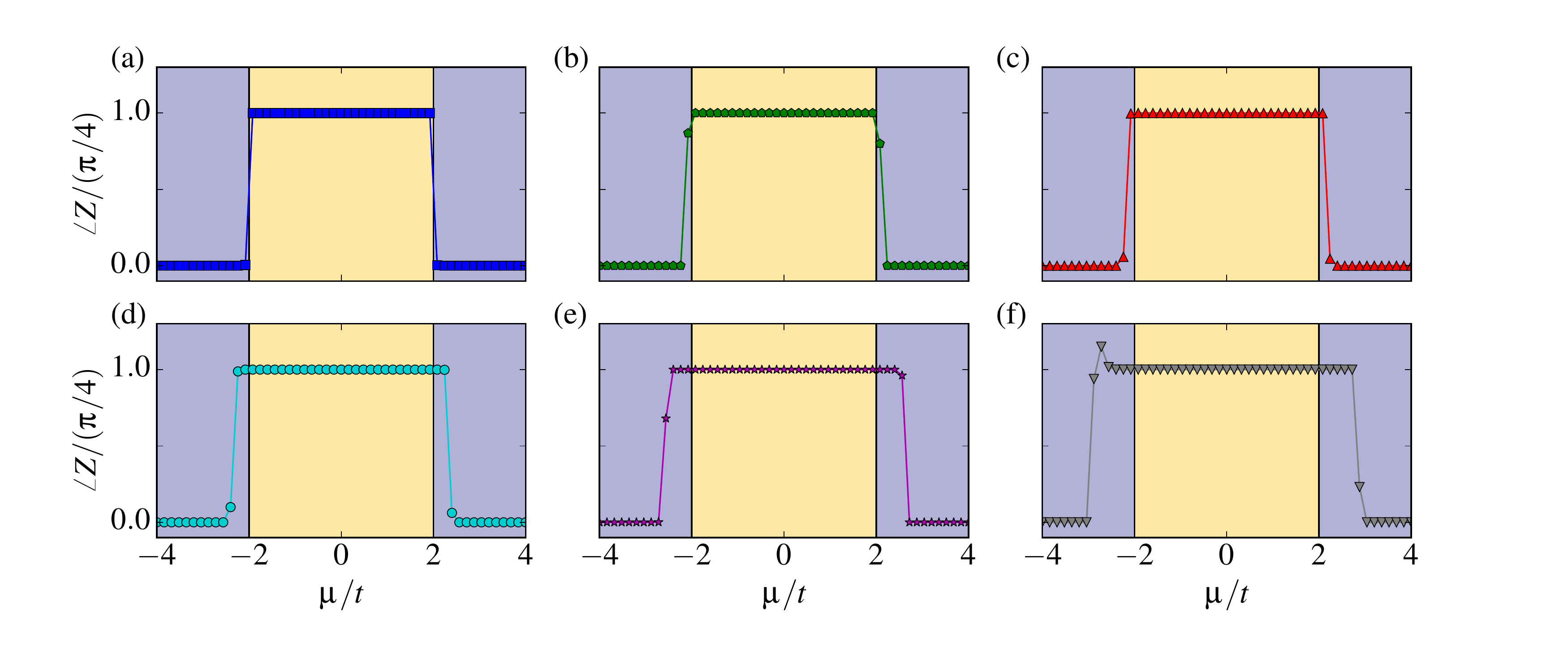}
\caption{\label{fig:Kitaev_one_dis} (Color online)
  Complex phase of the partial reflection $Z=\braket{GS | R_I | GS}$ for one realization of the disorder potential. Panels (a)-(f) represent different disorder strength from $W=1$ to $W=6$ (same as the legend in Fig.~\ref{fig:Kitaev_dis}). Solid lines are guides for the eye. Here, we set $\Delta=t$, $N=200$ and $N_\text{part}=100$.}
\end{figure}

In what follows,
we compute the $\mathbb{Z}_8$ invariant for a microscopic realization of
class D (with reflection symmetry) topological superconductors
in the presence of a random chemical potential disorder.
In particular, we 
study the robustness of the partial reflection against random disorder,
and also show that we can use
the many-body $\mathbb{Z}_8$ topological invariant 
to map out the phase diagram of the
disordered Majorana chain, by reproducing the known results.
\cite{Ian2014,Prodan2014}
The lattice Hamiltonian consists of two parts $H=H_\text{clean}+H_\text{dis}$ where $H_\text{clean}$ is the Hamiltonian of the Kitaev Majorana chain,
\begin{align} \label{eq:BdG1d}
{H}= 
-\sum_{j} \Big[t f_{j+1}^\dagger f^{\ }_{j}-\Delta f_{j+1}^\dagger f^\dagger_{j} +\text{H.c.}\Big] -\mu \sum_{j} f_j^\dagger f^{\ }_j,
\end{align}
and the disorder term is
\begin{align}
H_\text{dis}= \sum_j v_j f_j^\dag f_j,
\end{align}
where $v_j$ is a random number uniformly distributed over the range $[-W/2,W/2]$
and $W$ is the disorder strength.
Figure~\ref{fig:Kitaev_dis} shows the complex phase of the averaged partial
reflection $Z=\braket{GS | R_I | GS}$ for various values of $W$
over a wide range of chemical potential including the trivial and topological
phases.
It is interesting to note that the topological region (characterized by $\angle
Z=\pi/4$) expands a little bit as the disorder strength is increased.
This is similar to the disorder induced topological phase due to quadratic
corrections on the lattice models,
which has been discussed in the context of the 2D and 3D topological insulators.
\cite{Groth_SCBA,Ryu_Nomura_3DTI,Tomi_dis_TI}
In addition, we provide the complex phase of the partial reflection for one
realization of disorder in Fig.~\ref{fig:Kitaev_one_dis}.
It is evident from this figure that the phase is quite robust to moderate disorder.

\begin{figure}
\includegraphics[scale=0.5]{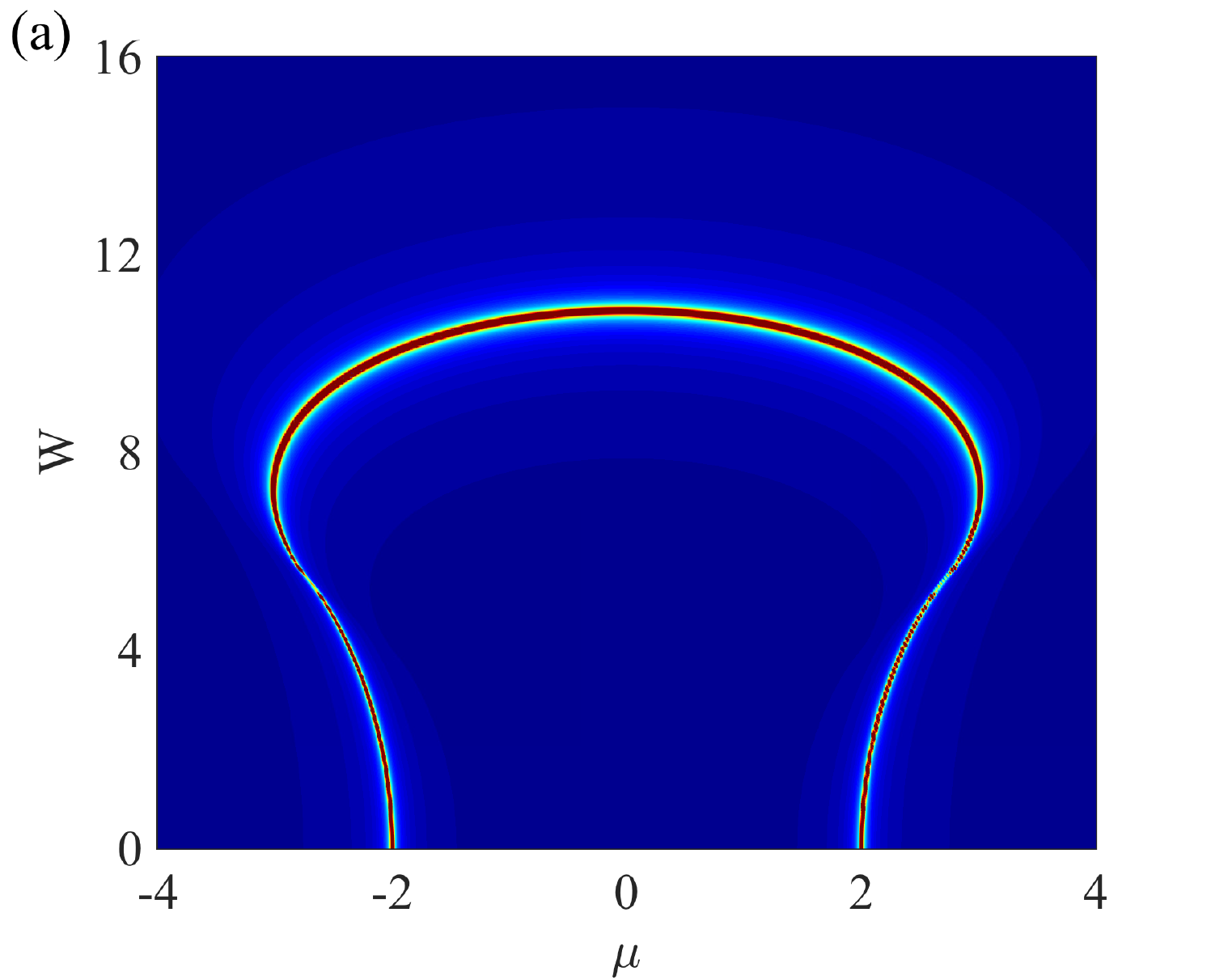}
\includegraphics[scale=0.12]{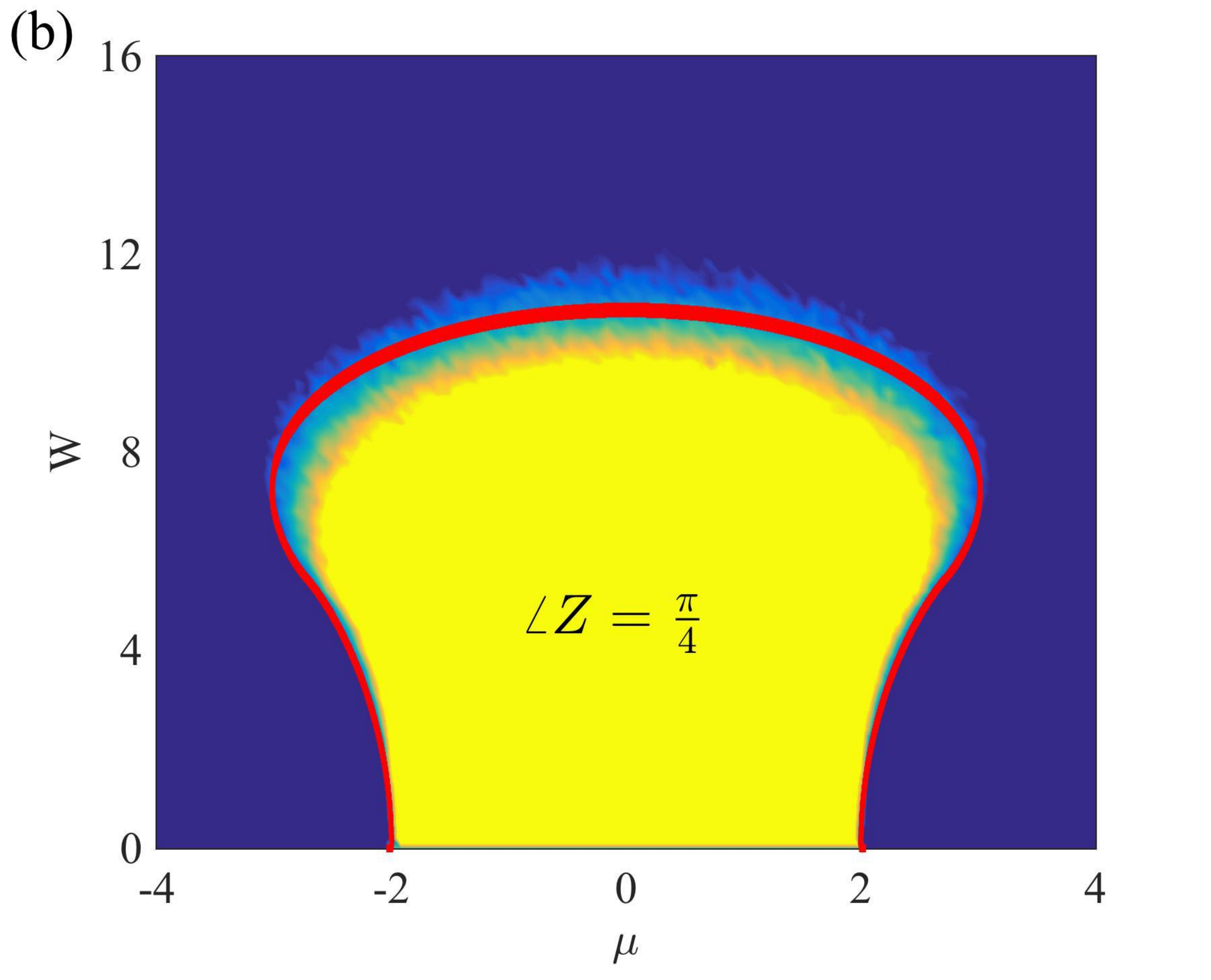}
\caption{\label{fig:Kitaev_dis_phase} (Color online)
  Phase diagram of the disordered Kitaev Majorana chain. (a) Color code is the Lyapunov exponent calculated using the transfer matrix approach (\ref{eq:Tmat}), and (b) color code is the complex phase of the partial reflection $\braket{GS | R_I | GS}$. In panel (b), the red curve shows the phase boundary which is analytically determined by the transfer matrix as shown in (a). Here, we set $\Delta=t$, $N=200$ and $N_\text{part}=100$.}
\end{figure}

We use the complex phase associated with the partial reflection to map out the
phase diagram of the disordered Majorana chain
as shown in Fig.~\ref{fig:Kitaev_dis_phase}(b).
For reference, we compute the phase boundary between the trivial and topological phases as a function of disorder strength $W$, using the transfer matrix approach~\cite{Ian2014,Prodan2014} (see Fig.~\ref{fig:Kitaev_dis_phase}(a)). In the case of our model, the Majorana chain with nearest neighbor hopping, the Lyapunov exponent can be found analytically,
\begin{align} \label{eq:Tmat}
\Lambda^{-1}&= {\Big|} \lim_{n\to\infty} \frac{1}{n} \sum_{j=1}^n \ln| \frac{\mu + v_i}{2t} |  {\Big| } \nonumber \\
&= \left\lvert \int_{-1/2}^{1/2} dx \ln | \frac{\mu + xW}{2t}| \right\lvert \nonumber \\
&= \left\lvert \ln\left[\frac{|2\mu+W|^{\frac{\mu}{W}+\frac{1}{2}}}{|2\mu-W|^{\frac{\mu}{W}-\frac{1}{2}}} \right] - (1+\ln 2t)\right\lvert.
\end{align}
The phase boundary can then be identified as a line of critical points at which
$\Lambda\to \infty$ is diverging.
The diverging $\Lambda$ is indicative of delocalized states at zero chemical potential which are in turn responsible for the topological phase transition. 
We show in Fig.~\ref{fig:Kitaev_dis_phase}(b) that the transfer matrix and our results are in remarkable agreement.  
 
\subsubsection{Case study 2: Partial reflection in the interacting Majorana chain}
\label{Sec:1D_D+R_int}

In this part, 
we consider the many-body $\mathbb{Z}_8$ invariant 
in the presence of interactions. 
We use 
the exact many-body ground state constructed 
for the interacting Majorana chain 
in Ref.\ \onlinecite{katsura2015exact}:
\begin{align}
{H}= -\sum_{j} [t f_{j+1}^\dagger f_{j}-\Delta f_{j+1}^\dagger f^\dagger_{j} +H.c.] -\mu \sum_{j} f_j^\dagger f_j + 
4U \sum_j ( n_j-1/2)( n_{j+1}-1/2).
\end{align}
For parameter values satisfying the condition $\mu=4\sqrt{U^2+tU+(t^2-\Delta^2)/4}$ can be written in a closed form as  
\begin{align} \label{eq:intwf}
\ket{\Psi_{r(ns)}} &= \ket{\Psi^{\alpha}_+} \mp \ket{\Psi^{\alpha}_-},
\end{align}
where
\begin{align}
\ket{\Psi^{\alpha}_\pm} &= \frac{1}{(1+\alpha^2)^{N/2}}\prod_{j=1}^{N} (1\pm\alpha f_j^\dag) \ket{0} \nonumber \\
&= \frac{1}{(1+\alpha^2)^{N/2}} e^{\pm \alpha f_1^\dag} e^{\pm \alpha f_2^\dag} \dots e^{\pm \alpha f_N^\dag} \ket{0},
\end{align}
corresponding to odd (even) fermion parity sectors associated with (anti-)periodic boundary condition where $\alpha=\sqrt{\cot(\theta/2)}$ and $\theta=\text{arctan}(2\Delta/\mu)$.
In the following, we  analytically derive that the amplitudes and complex phases of partial reflection for a generic value of $\alpha$ converge to the anticipated values $1/\sqrt{2}$ and $\pi/4$ in (\ref{eq:Brown_z8_inv_kitaev}), as we approach the long-chain limit (compared to the correlation length).

We prove our result for the case with anti-periodic boundary condition. A similar derivation can be carried out for the case of periodic boundary condition. Consider a long chain with $N$ sites in total and $M$ sites in the subsystem such that $N=2M$ (we take $M$ to be even which means reflection with respect to the central link). The wave function is given by
\begin{align}
\ket{\Psi_{ns}} &= \frac{1}{\sqrt{{\cal A}_+(N)}}  \prod_{j=1}^{N} (1+\alpha f_j^\dag){\Big|}_{n \in \text{even}} \ket{0} \end{align}
where the normalization factor is
\begin{align} \label{eq:amp_prefl}
{\cal A}_\pm (n)= \frac{1}{2} [(1+\alpha^2)^{n} \pm  (1-\alpha^2)^{n}] .
\end{align}
For simplicity we choose the sites $1$ to $M$ to be in the subsystem. The wave function can be rewritten as
\begin{align}
\ket{\Psi_{ns}} =&  \frac{1}{\sqrt{{\cal A}(N)}} (1+\alpha f_1^\dag)F_{in}^\dagger(1+\alpha f_{M}^\dag) F_{out}^\dagger {\Big|}_{n\in \text{even}}  \ket{0} \nonumber \\
=& \frac{1}{\sqrt{{\cal A}(N)}} [(1+ \alpha^2 f_1^\dag f_{M}^\dag) F_{in}^{(e) \dagger} F_{out}^{(e)\dagger}+ (1- \alpha^2 f_1^\dag f_{M}^\dag) F_{in}^{(o) \dagger} F_{out}^{(o)\dagger}] \ket{0} \nonumber \\
&+\frac{\alpha}{\sqrt{{\cal A}(N)}} [(f_1^\dag +f_{M}^\dag) F_{in}^{(e) \dagger} F_{out}^{(o)\dagger} +(f_1^\dag -f_{M}^\dag) F_{in}^{(o) \dagger} F_{out}^{(e)\dagger} ]\ket{0} .
\end{align}
 We define the new operators
\begin{align}
F_{in}^{(o/e)\dag} &= \prod_{j=2}^{M-1} (1+\alpha f_j^\dag) {\Big|}_{n\in \text{odd}/\text{even}} ,\nonumber \\
F_{out}^{(o/e)\dag} &= \prod_{j=M+1}^{N} (1+ \alpha f_j^\dag) {\Big|}_{n\in \text{odd}/\text{even}}.
\end{align}
The partial reflection is defined by
\begin{align}
R_If_j^\dagger R_{I}^{-1}=i f^\dag_{M-(j-1)}, \quad \text{for}\quad 1\leq j\leq M .
\end{align}
Note that
\begin{align}
\bra{0} F_{in}^{(o)} R_IF_{in}^{(o)\dag}\ket{0}&= i {\cal A}_- (M-2), \nonumber \\
\bra{0} F_{in}^{(o)} R_IF_{in}^{(o)\dag}\ket{0}&= {\cal A}_+ (M-2),
\end{align}
where ${\cal A}_\pm (n)$ is defined in (\ref{eq:amp_prefl}).
So, we have
\begin{align} \label{eq:anl}
Z_{N}=\bra{\Psi_{ns}}  {\cal R}_I \ket{\Psi_{ns}} =& \frac{1}{{\cal A}(N)} \left[ (1+\alpha^4){\cal A}_+ (M-2) {\cal A}_+ (N-M) + i (1+\alpha^4){\cal A}_- (M-2) {\cal A}_- (N-M) \right] \nonumber \\
&+ \frac{2\alpha^2}{{\cal A}(N)} \left[i {\cal A}_+ (M-2) {\cal A}_- (N-M) + {\cal A}_- (M-2) {\cal A}_+ (N-M) \right] 
\end{align}
In the thermodynamic limit $N\to \infty$, this becomes
\begin{align}
\lim_{N\to \infty} Z_N &= \frac{(1+\alpha^2)^{N-2}}{2(1+\alpha^2)^N} [(1+\alpha^4)(1+i) +2\alpha^2(i+1) ] \nonumber \\
&= \frac{1+i}{2},
\end{align}
which is identical to (\ref{eq:Brown_z8_inv_kitaev}). Here, we use the fact that
\begin{align}
\lim_{n\to \infty} {\cal A}_\pm (n)= \frac{1}{2} (1+\alpha^2)^n .
\end{align}

\subsection{$(1+1)d$ insulators protected by reflection ($\Omega^{{\rm Pin}^c}(pt)=\mathbb{Z}_4$)
}
\label{sec:(1+1)A+R}
Next, we move on to the system 
described by complex fermions 
in the presence of reflection symmetry. 
Specifically, we consider a reflection transformation 
acting on a multiplet of complex fermions operators
$\psi_i(x,t)$
together with a $U(1)$ transformation as
\begin{align}
U_{\alpha} R \psi^{\dag}_i(x,t) [U_{\alpha} R^{-1}] 
= \psi^{\dag}_j (-x,t) e^{-i \alpha} {\cal R}_{ji}.
\label{eq:1d_A+R_def_ref}
\end{align}
This operation, when used along with twisted boundary conditions,
enables us to put the theory on 
an unoriented manifold with the $\pinc$ structure. 
In $(1+1)$-dimensions,
the corresponding cobordism 
group is
~\cite{Gilkey}
\begin{align}
\Omega^{{\rm Pin}^c}_2(pt) = \Z_4, 
\end{align}
which implies there are $4$ topologically distinct 
phases. 
The generating manifold is $\R P^2$. 
In this section, we will see that 
the expectation value of full reflection captures the $\Z_2$ subgroup 
of the $\Z_4$ topological classification, 
whereas 
the expectation value of partial reflection with an appropriate $U(1)$ phase 
as introduced in Eq.\ \eqref{eq:1d_A+R_def_ref}
provides the $\Z_4$ invariant. 

\subsubsection{The $\Z_2$ equivariant line bundle}
The existence of the many-body $\Z_2$ topological invariant, 
which is a subgroup of the $\Z_4$ classification, 
can be understood by the 
$\Z_2$-equivariant structure of the line bundle over the flux space 
in the same way as Eqs.\ (\ref{eq:1d_A+C_C_action}) and (\ref{Eq:1D_Z2_C}). 
On the closed space circle $S^1$, we have a $\Z_2$-equivariant complex line bundle 
\begin{align}
R \Ket{GS(\theta)} = e^{i \phi(\theta)} \Ket{GS(-\theta)}, 
\end{align}
where $\Ket{GS(\theta)}$ is the ground state under the twisted boundary condition 
$\psi(x+L) = e^{i \theta} \psi(x)$. 
Then, we have the $\Z_2$ quantization of the Berry phase 
\begin{align}
\gamma 
= \exp \Big[ \oint_0^{2 \pi} d \theta \braket{GS(\theta) | \partial_{\theta} GS(\theta)} \Big]
= \frac{\braket{GS(\pi) | R | GS(\pi)}}{\braket{GS(0) | R |GS(0)}}
\in {\pm 1}.
\end{align}

A $\Z_2$ nontrivial model is the same as (\ref{eq:1d_A_C}) with the reflection symmetry 
defined by 
\begin{align}
R a^{\dag}_j R^{-1} = b^{\dag}_{N-j+1}, 
\quad 
R b^{\dag}_j R^{-1} = a^{\dag}_{N-j+1}. 
\end{align}
One can show that $R \ket{GS(\theta)} \sim e^{i \theta} \ket{\Psi(-\theta)}$, 
which implies the nontrivial $\Z_2$ Berry phase $\gamma = -1$.

\subsubsection{Partial reflection and the $\Z_4$ invariant}
In a way similar to Sec.\ \ref{Sec:1D_D+R_Z8}, 
we show that partial reflection provides the $\Z_4$ invariant. 
There are two issues specific to $\pinc$ structures: 
the choice of $U(1)$ phase associated with partial reflection, 
and the choice of the center of reflection, 
i.e., a lattice site or bond center. 

Here, we calculate the partial reflection for the fixed point model (\ref{eq:1d_A_C}). 
The ground state is given by 
\begin{align}
\ket{GS} 
= \frac{b^{\dag}_N - a^{\dag}_{1}}{\sqrt{2}} \frac{b^{\dag}_1 - a^{\dag}_2}{\sqrt{2}} \cdots \frac{b^{\dag}_M - a^{\dag}_{M+1}}{\sqrt{2}} 
\cdots \frac{b^{\dag}_{N-1}- a^{\dag}_{N}}{\sqrt{2}} \ket{0}. 
\end{align}
The partial reflection $[U_{\alpha}R]_I$ acts on interval $I = \{1, \dots, M\}$ as 
\begin{align}
[U_{\alpha}R]_I a^{\dag}_j [U_{\alpha}R]_I^{-1} = e^{-i \alpha} b^{\dag}_{M-j+1}, && 
[U_{\alpha}R]_I b^{\dag}_j [U_{\alpha}R]_I^{-1} = e^{-i \alpha} a^{\dag}_{M-j+1}, && 
[U_{\alpha}R]_I \ket{0} = \ket{0}, && 
(j=1, \dots, M). 
\end{align}
Even $M$ and odd $M$ correspond to the case of
reflection centered at a lattice site
or at a bond, respectively. 
We have 
\begin{align}
\braket{GS | [U_{\alpha}R]_I | GS} 
= \frac{1}{2} (-1)^{\frac{1}{2}M(M-1)} \sin \alpha e^{-i M \alpha}. 
\end{align}
To obtain the quantum dimension of the edge complex fermions, 
$\alpha$ can be chosen as $\alpha = \pm \frac{\pi}{2}$, 
which implies that only bond center partial reflection (odd $M$) 
provides $\Z_4$ phases $\pm i$.

\section{Partial rotations}
\label{Sec:4}

\begin{figure}[!]
 \begin{center}
  \includegraphics[width=0.8\linewidth, trim=0cm 0cm 0cm 0cm]{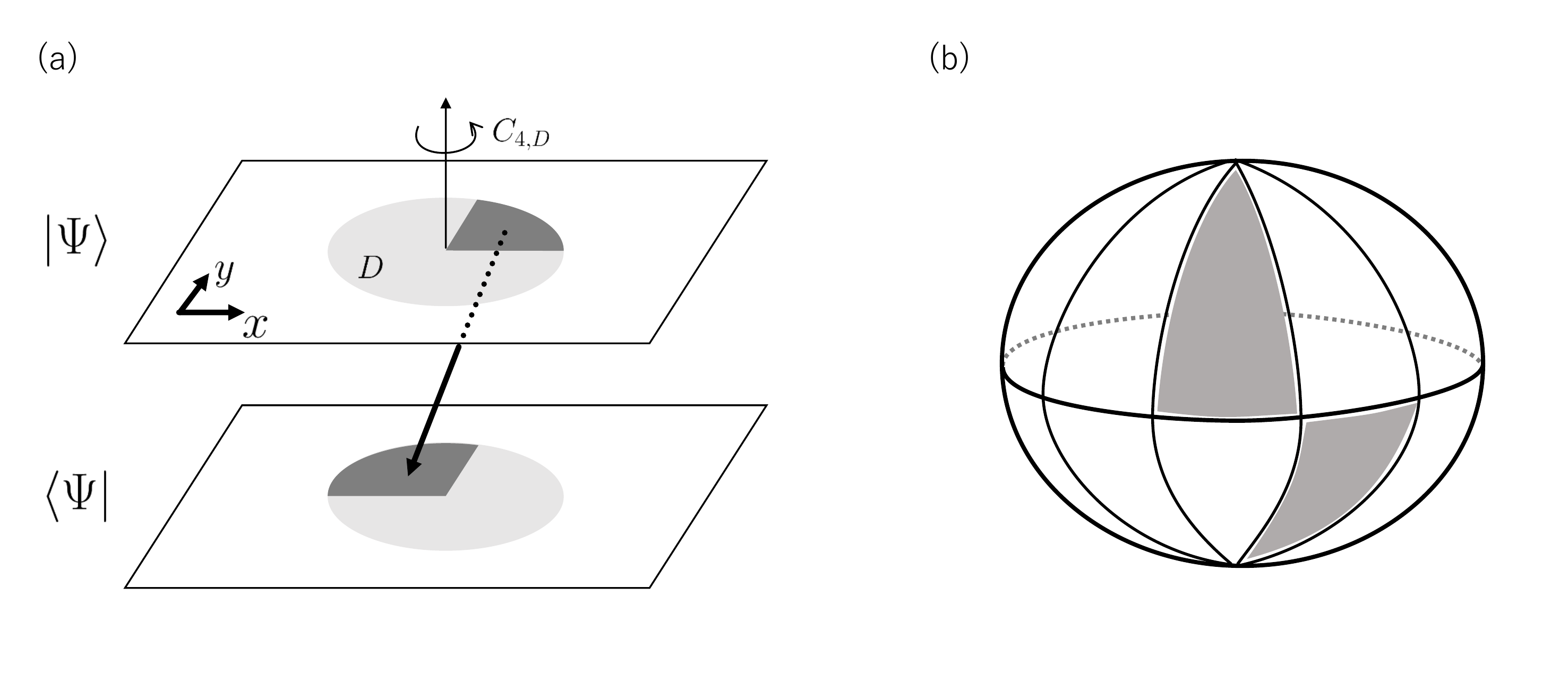}
 \end{center}
 \caption{
 (a) Partial rotation on the ground state $\ket{\Psi}$. The figure shows the partial $C_4$ rotation. 
 (b) A construction of lens space $L(n,1)$. 
The figure shows the boundary ($\cong S^2$) of a 3-ball. 
The boundary of upper hemisphere is rotated by $2\pi/n$ angle, and glued into 
the boundary of lower hemisphere. 
The shadow regions are identified. 
 }
 \label{Fig:Lens_space}
\end{figure}

In this section, we introduce a kind of nonlocal operators, 
{\it partial $n$-fold rotations} 
$C_{n,D}$, acting on a subdisk region $D$ in the 2-dimensional space manifold $M$. 
We find that the expectation value of partial rotations on the ground state 
$\Braket{GS | C_{n,D} | GS }$ (see Fig.\ \ref{Fig:Lens_space} (a)) 
provides 
scale-independent quantities which characterize topological properties of the bulk, 
if the length of the boundary of the subdisk $D$ is sufficiently larger than 
the correlation length of bulk. 
Intuitively,
the action of partial $C_n$ rotation can be thought of as a generalization of 
the procedure introducing a cross-cap, 
which gives rise to lens spaces $L(n,1)$ if we act the partial rotation on $S^2$ 
(see Fig.\ \ref{Fig:Lens_space} (b)). 	
In particular, 
the case of $n=2$ corresponds to $\R P^3 = L(2,1)$. 

In the following, we first 
temporarily depart from SPT phases and consider 
partial rotations in chiral topological phases, 
and present the calculation of their ground state expectation value
using their edge CFT.
In this context, the  ground state expectation value of partial rotation 
can be used to extract a combination of modular $S$ and $T$ matrices: 
It turns out that the low energy quasi particle description is consistent with 
lens space partition functions of $(2+1)d$ TQFTs.~\cite{Witten1989, Jeffrey1992}
Our derivation of the formula for the partial rotation is similar to Ref.\ \onlinecite{Tu2013}, 
where they introduced the partial lattice translation on the cylinder. 
The surgery construction of lens spaces from the solid torus is summarized 
in Appendix \ref{App:1}. 


\subsection{$(2+1)d$ chiral topological phase and partial rotation and lens space}
Let us consider a $(2+1)d$ chiral topological phase. 
For simplicity, we assume that the ground state is unique. 
We further assume the bulk is symmetric under $C_n$ rotation symmetry. 
Here we consider a right moving CFT with dispersion $\varepsilon(k) \sim v k$ 
for the boundary theories,
where $v > 0$ is the velocity of excitations.
The cases with left moving chiral CFT will be briefly discussed later. 

We would like to estimate the expectation value of the partial rotation 
$\Braket{GS | C_{n,D} | GS}$ on a subdisk $D$ for the ground state over 
a space manifold $M$. 
The edge chiral CFT is described by Hamiltonian and momentum operators 
\begin{align}
H = \frac{2 \pi v}{L} (L_0 - \frac{c}{24}), 
\qquad 
P = \frac{H}{v}, 
\end{align}
where $c$ is the central charge, 
and $L = |\partial D|$ is the length of the boundary. 
By using the cut and glue construction,~\cite{Qi2011b} the reduced density matrix 
$\rho_D$ for the subdisk $D$ is approximated by the edge CFT with temperature determined 
by the correlation length $\xi$ of bulk, 
\begin{align}
\rho_D = \frac{e^{- \frac{\xi}{v} H} }{ {\rm Tr} \big[ e^{- \frac{\xi}{v} H} \big]}. 
\end{align}
Then, partial $C_n$ rotation is nothing but 
translation by ${L}/{n}$ 
on the edge CFT, 
\begin{align}
\braket{GS| C_{n,D} |GS }
&= \frac{{\rm Tr} \big[ e^{- i \widetilde{P} \frac{L}{n}} e^{- \frac{\xi}{v} H} \big]}{ {\rm Tr} \big[ e^{- \frac{\xi}{v} H} \big]} 
= \frac{e^{\frac{2\pi i}{n} ( \braket{L_0} - \frac{c}{24})} \sum_{a\in {\rm irreps.}} \chi_a(\frac{i \xi}{L}-\frac{1}{n})}{\sum_{a\in {\rm irreps.}} \chi_a(\frac{i \xi}{L})}, 
\label{Eq:PR_Cn}
\end{align}
where 
\begin{align}
\chi_a(\tau) = {\rm tr}_a[e^{2 \pi i \tau (L_0-\frac{c}{24})}]
\end{align}
is the Virasoro character of the $a$-th irreducible representation,
and the summation $\sum_{a\in {\rm irreps.}}$ runs over Virasoro representations 
of the CFT realized in the boundary $\partial D$. 
The generator of the translation, $\widetilde P$, was normalized as 
\begin{align}
\widetilde P
:= \frac{1}{v} (H - E_0) 
= \frac{2 \pi}{L} \Big[ L_0 - \frac{c}{24} - \braket{L_0 - \frac{c}{24}} \Big] 
\end{align}
so that $\widetilde P \ket{\rm vac} = 0$ on the vacuum state $\ket{\rm vac}$
of the CFT. 
Here, $E_0 = \braket{{\rm vac} | H | {\rm vac}}$ is the vacuum energy. 

We want to estimate (\ref{Eq:PR_Cn}) for 
a sufficiently large subdisc, $\xi/L \ll 1$. 
Using modular transformations, 
\begin{align}
\chi_a\left(i \ell - \frac{1}{n}\right) 
= S_{ab} \chi_b\left(- \frac{1}{i \ell - \frac{1}{n}}\right)
= (ST^n)_{ab} \chi_b\left(- \frac{-n i \ell}{i \ell + \frac{1}{n}} \right)
= (ST^nS)_{ab} \chi_b\left( \frac{i}{n^2 \ell} + \frac{1}{n}\right), 
\label{Eq:PR_Modular}
\end{align}
(\ref{Eq:PR_Cn}) can be expressed as a low-temperature partition function as 
\begin{align}
\Braket{GS| C_{n,D} |GS }
= \frac{e^{\frac{2\pi i}{n} ( \braket{L_0} - \frac{c}{24})} \sum_{a\in {\rm irreps.}} \sum_b (S T^n S)_{ab} \chi_b(\frac{i L}{n^2 \xi}+\frac{1}{n})}{\sum_{a\in {\rm irreps.}} \sum_b S_{ab} \chi_b(\frac{i L}{\xi})} 
\label{Eq:PR_Cn_2}
\end{align}
where $S$ and $T$ are modular matrices. 
Notice that $\sum_b$ runs over all irreps.\ of the theory, 
which should be contrasted with $\sum_{a\in {\rm irreps.}}$. 
This formula enables us to estimate the 
ground state expectation value of the partial rotation 
through the highest weight state $\ket{h_b}$, 
\begin{align}
\chi_b\left(\frac{i L}{n^2 \xi}+\frac{1}{n}\right) 
&= \sum_{m=0}^{\infty} p(m) e^{2 \pi i (\frac{i L}{n^2 \xi}+\frac{1}{n}) (h_b + m - \frac{c}{24})}
\nonumber \\
&= e^{\frac{2 \pi i}{n} (h_b - \frac{c}{24})} e^{- \frac{2 \pi L}{n^2 \xi} (h_b - \frac{c}{24})} \Big( 1 + \sum_{m \geq 1}^{\infty} p(m) e^{2 \pi i (\frac{i L}{n^2 \xi}+\frac{1}{n}) m} \Big)
\nonumber \\
& \sim e^{\frac{2 \pi i}{n} (h_b - \frac{c}{24})} e^{- \frac{2 \pi L}{n^2 \xi} (h_b - \frac{c}{24})}, 
\end{align}
where $h_b$ is the conformal weight and $p(m)$ is the number of states 
with energy $h_b + m$. 
Subleading contributions are suppressed by the factor $e^{- \frac{2 \pi L}{n^2 \xi}}$. 
In the same way, 
\begin{align}
\chi_b\left(\frac{i L}{\xi}\right) 
= e^{- \frac{2 \pi L}{\xi} (h_b - \frac{c}{24})} \Big( 1 + \sum_{m \geq 1}^{\infty} p(m) e^{- \frac{2 \pi L}{n^2 \xi} m} \Big) 
\sim e^{- \frac{2 \pi L}{\xi} (h_b - \frac{c}{24})}. 
\end{align}

The mapping torus $T^2 \times_{ST^n S} S^1$ built from the modular transformation 
$S T^n S$ is knows as the lens space $L(n,1)$.~\cite{Jeffrey1992}
Thus, the partial $C_n$ rotation is related to the partition function on the lens space.  



\subsubsection{Left mover chiral CFT}
For a left-mover chiral edge excitation with dispersion $\varepsilon(k) \sim - v k$ ($v>0$), 
the momentum operator is changed to $P = - H/v$, which leads to 
the formula of the partial rotation for the left-mover chiral topological phases 
\begin{align}
\braket{GS| C_{n,D} |GS }
&= \frac{{\rm Tr} \big[ e^{i \widetilde{P} \frac{L}{n}} e^{- \frac{\xi}{v} H} \big]}{ {\rm Tr} \big[ e^{- \frac{\xi}{v} H} \big]} 
\nonumber \\
&= \frac{e^{\frac{2\pi i}{n} ( \braket{L_0} - \frac{c}{24})} \sum_{a\in {\rm irreps.}} \chi_a(\frac{i \xi}{L}+\frac{1}{n})}{\sum_{a\in {\rm irreps.}} \chi_a(\frac{i \xi}{L})} 
\nonumber \\
&= \frac{e^{-\frac{2\pi i}{n} ( \braket{L_0} - \frac{c}{24})} \sum_{a\in {\rm irreps.}} \sum_b (S T^{-n} S)_{ab} \chi_b(\frac{i L}{n^2 \xi}-\frac{1}{n})}{\sum_{a\in {\rm irreps.}} \sum_b S_{ab} \chi_b(\frac{i L}{\xi})}, 
\end{align}
This is the complex conjugate of Eq.\ (\ref{Eq:PR_Cn}). 
Thus, the non-chiral CFT without on-site symmetry 
cannot provide nontrivial $U(1)$ phases in partial rotations. 
However, if there is an on-site symmetry, we can associate the partial on-site 
transformation with the partial rotation, which offers a 
nontrivial $U(1)$ phase as shown in Sec. \ref{Sec:4-3}.


\subsection{$(2+1)d$ $(p_x - i p_y)$ chiral superconductor with rotation symmetry}
\label{$(2+1)d$ $(p_x - i p_y)$ chiral superconductor with rotation symmetry}
For an application of partial rotations, 
let us consider a $(p_x-i p_y)$ superconductor: 
\begin{align} \label{Eq:pip_sc}
H = \sum_{\bk} \Big[ \psi^{\dag}(\bk) (\frac{k^2}{2m}-\mu) \psi(\bk) + \frac{\Delta}{2} \psi^{\dag}(\bk) (k_x-i k_y) \psi^{\dag}(-\bk) + \frac{\Delta}{2} \psi(-\bk) (k_x+ i k_y) \psi(\bk) \Big] 
\end{align}
where $\Delta(\bk) = \Delta (k_x - i k_y)\ (\Delta>0)$ is the gap function and 
$\bk$ is momentum. 
We introduce a polar coordinate $(x,y) = (r \cos \phi, r \sin \phi)$. 
This model has the following continuum rotation symmetry $C_{\theta}$, 
\begin{align}
C_{\theta} \psi^{\dag}(r,\phi) C_{\theta}^{-1} =  e^{-\frac{i \theta}{2}} \psi^{\dag}(r,\phi+\theta), 
\quad 
C_{2 \pi} = (-1)^F. 
\label{eq:2d_px-ipy_rot_sym}
\end{align}
Note that the $2 \pi$ rotation is the fermion parity. 

On the disk geometry, 
the system supports a right-moving  gapless chiral real fermion mode $\gamma(\ell)$ localized at the boundary.
The fermion mode $\gamma(\ell)$ can be constructed explicitly as
\begin{align}
\gamma(\frac{L \phi}{2 \pi}) \sim \Big[ e^{ \frac{i \phi}{2} + \frac{\pi i}{4}} \psi(r,\phi) + e^{ -\frac{i \phi}{2} - \frac{\pi i}{4}} \psi^{\dag}(r,\phi) \Big] e^{\int^{r} \frac{\mu(r')}{\Delta} d r'}, 
\label{Eq:2D_p-ip_Edge_State}
\end{align}
up to a normalization constant, where the chemical potential $\mu(r)$ is 
chosen such that a finite disk geometry is realized,  
$\mu(r) > 0$ for $r < \frac{L}{2 \pi}$ and $\mu(r) <0$ for $r> \frac{L}{2 \pi}$, 
and $L$ is the circumference. 
(See Appendix~\ref{Edge theory on the boundary of disc} for
the derivation of Eq.\ \eqref{Eq:2D_p-ip_Edge_State}.)  
$\gamma(\ell)$ obeys the real condition $\gamma^{\dag}(\ell) = \gamma(\ell)$ and 
the antiperiodic boundary condition $\gamma(\ell + L) = - \gamma(\ell)$, 
which reflects the absence of exact zero energy states 
since there is no $\pi$-flux inside the disk $D$.~\cite{Stone2004}
The Hamiltonian of the edge theory is given by
\begin{align}
&H_{ns} = \frac{2 \pi \Delta}{L} 
\Big( \sum_{\substack{m>0\\ m \in \Z+\frac{1}{2}}} m \gamma_{-m} \gamma_m - \frac{1}{48} \Big), 
\quad  
P = \frac{H}{\Delta}. 
\end{align}
The free real chiral fermion CFT 
is characterized by the data
\cite{ginsparg1988applied}
\begin{align}
\label{CFT data}
&c = \frac{1}{2}, \qquad 
(h_1, h_{\psi}, h_{\sigma}) = (0, \frac{1}{2}, \frac{1}{16}), \\
&S= \frac{1}{2} \begin{pmatrix}
1 & 1 & \sqrt{2} \\
1 & 1 & -\sqrt{2} \\
\sqrt{2} & \sqrt{2} & 0
\end{pmatrix}, 
\qquad 
T = e^{-\frac{\pi i}{24}} \begin{pmatrix}
1 & & \\
& -1 & \\
& & e^{\frac{\pi i}{8}}
\end{pmatrix}, 
\end{align}
where $c$ is the chiral central charge,
$(h_1, h_{\psi}, h_{\sigma})$ 
is the set of dimensions (topological spin) in the CFT 
(for the vacuum, fermion, and Ising spin sectors, respectively), 
and the modular $S$ and $T$ matrices are 
given in the basis $(1,\psi,\sigma)$.
The Virasoro representations that appear in 
the NS sector is  $[1] \oplus [\psi]$.

\subsubsection{Partial $C_n$ rotation}
Let us consider partial $C_n := C_{\theta = \frac{2 \pi}{n}}$ rotation. 
First, we need to specify 
the action of the $C_{\theta}$ rotation on the gapless edge excitation 
$\gamma(\ell)$, which can be read off from the concrete expression 
(\ref{Eq:2D_p-ip_Edge_State}) as 
\begin{align}
C_{\theta} \gamma(\ell) C_{\theta}^{-1} = \gamma(\ell + \frac{\theta L}{2 \pi}). 
\end{align}
This is consistent with $C_{2 \pi} = (-1)^F$ because of the anti-periodic boundary condition. 
From the formula (\ref{Eq:PR_Cn_2}), 
the expectation value of partial $C_n$ rotation on the $(p_x - i p_y)$ superconductor is given by 
\begin{align}
\label{partial CnD}
\braket{GS| C_{n,D} |GS }
= \frac{e^{- \frac{2\pi i}{n} \frac{1}{48}} 
\sum_{a = 1, \psi} \sum_b (S T^n S)_{ab} \chi_b(\frac{i L}{n^2 \xi}+\frac{1}{n})}{\sum_{a = 1, \psi} \sum_b S_{ab} \chi_b(\frac{i L}{\xi})}. 
\end{align}
From \eqref{CFT data}, the matrix elements 
$(ST^nS)_{ab}$
are given by 
\begin{align}
&
S T^{n} S 
= e^{-\frac{n}{24} \pi i} \cdot \frac{1}{4} \begin{pmatrix}
1+(-1)^n + 2 e^{\frac{n}{8} \pi i} & 1+(-1)^n + 2 e^{\frac{n}{8} \pi i} & \sqrt{2}-(-1)^n \sqrt{2} \\
1+(-1)^n - 2 e^{\frac{n}{8} \pi i} & 1+(-1)^n + 2 e^{\frac{n}{8} \pi i} & \sqrt{2}-(-1)^n \sqrt{2} \\
\sqrt{2} + (-1)^n \sqrt{2} & \sqrt{2} + (-1)^n \sqrt{2} & 2 - 2 (-1)^n \\
\end{pmatrix},
\end{align}
and hence we have 
\begin{align}
&e^{\frac{2 \pi i}{n} (h_{1}-\frac{1}{24})} \sum_{a=1,\psi} (ST^{n}S)_{a1} 
= \left\{ \begin{array}{ll}
e^{-\frac{(n^2+2) \pi i}{24 n}} & (n : {\rm\ even}) \\
0 & (n : {\rm\ odd}) \\
\end{array} \right. \\
&e^{\frac{2 \pi i}{n} (h_{\psi}-\frac{1}{24})} \sum_{a=1,\psi} (ST^{n}S)_{a \psi} 
= \left\{ \begin{array}{ll}
2 \cos (\frac{n \pi}{16}) e^{\frac{(n^2+44) \pi i}{48 n}} & (n : {\rm\ even}) \\
e^{\frac{(n^2+11) \pi i}{12 n}} & (n : {\rm\ odd}) \\
\end{array} \right.  \\
&e^{\frac{2 \pi i}{n} (h_{\sigma}-\frac{1}{24})} \sum_{a=1,\psi} (ST^{n}S)_{a \sigma} 
= \left\{ \begin{array}{ll}
0 & (n : {\rm\ even}) \\
\frac{1}{\sqrt{2}} e^{-\frac{(n^2-1) \pi i}{24 n}} & (n : {\rm\ odd}) \\
\end{array} \right.   
\end{align}
To evaluate Eq.\ \eqref{partial CnD} 
to the leading order, 
we note 
\begin{align}
&
e^{- \frac{2\pi i}{n} \frac{1}{48}} \chi_b\left(\frac{i L}{n^2 \xi}+\frac{1}{n}\right)
\sim e^{\frac{2 \pi i}{n}(h_b - \frac{1}{24})} e^{-\frac{2 \pi L}{n^2 \xi}(h_b - \frac{1}{48})}. 
\end{align}
There is an even odd effect:
The leading contributions come from the vacuum sector $b=1$ for even $n$, 
and the $b=\sigma$ quasiparticle for odd $n$, 
whereas the $b = \psi$ quasipaticle contributes to the next leading contribution. 
Summarizing, we have 
\begin{align}
\braket{GS| C_{n,D} |GS }
\sim 
\left\{ \begin{array}{ll}
\exp \Big[ -\frac{(n^2+2) \pi i}{24 n} - (1-\frac{1}{n^2}) \frac{1}{48} \frac{2 \pi L}{\xi} \Big]  & (n : {\rm\ even}) \\
\exp \Big[ -\frac{(n^2-1) \pi i}{24 n} - \ln \sqrt{2} -(1+\frac{2}{n^2}) \frac{1}{48} \frac{2 \pi L}{\xi} \Big]  & (n : {\rm\ odd}) \\
\end{array} \right.   
\label{Eq:pip_CnD}
\end{align}
For odd $n$, 
in addition to the topological $U(1)$ phase, 
there is a topological amplitude $e^{-\ln \sqrt{2}}$, 
which is an analog of the topological entanglement entropy.~\cite{Kitaev2006a, Levin2006}
Here we show some examples of partial $C_{n,D}$ rotations: 
\begin{align}
\braket{GS| C_{2,D} |GS }
&\sim \exp \Big[ -\frac{\pi i}{8} - \frac{3}{4} \frac{2 \pi L}{\xi} \frac{1}{48} \Big] , 
\label{eq:2d_c2_pr}\\
\braket{GS| C_{3,D} |GS }
&\sim \exp \Big[ -\frac{\pi i}{9} - \ln \sqrt{2} -\frac{11}{9} \frac{2 \pi L}{\xi} \frac{1}{48} \Big] , \\
\braket{GS| C_{4,D} |GS }
&\sim \exp \Big[ -\frac{3 \pi i}{16} - \frac{15}{16} \frac{2 \pi L}{\xi} \frac{1}{48} \Big] , \\
\braket{GS| C_{5,D} |GS }
&\sim \exp \Big[ -\frac{\pi i}{5} - \ln \sqrt{2} -\frac{27}{25} \frac{2 \pi L}{\xi} \frac{1}{48} \Big] , \\
\braket{GS| C_{6,D} |GS }
&\sim \exp \Big[ -\frac{19}{72} \pi i - \frac{35}{36} \frac{2 \pi L}{\xi} \frac{1}{48} \Big] , \\
\braket{GS| C_{7,D} |GS }
&\sim \exp \Big[ -\frac{2 \pi i}{7} - \ln \sqrt{2} -\frac{51}{49} \frac{2 \pi L}{\xi} \frac{1}{48} \Big] , \\
\braket{GS| C_{8,D} |GS }
&\sim \exp \Big[ - \frac{11}{32} \pi i - \frac{63}{64} \frac{2 \pi L}{\xi} \frac{1}{48} \Big] , \\
\braket{GS| C_{9,D} |GS }
&\sim \exp \Big[ -\frac{10}{27}\pi i - \ln \sqrt{2} -\frac{83}{81} \frac{2 \pi L}{\xi} \frac{1}{48} \Big]. 
\end{align}

By including higher energy states in $\chi_b(\frac{i L}{n^2 \xi} + \frac{1}{n})$, 
we can estimate the expectation value of partial rotation 
to higher orders in ${\xi}/{L} \ll 1$. 
For example, the expectation value of 
partial $C_n$ rotations
at next-to-leading order is given by 
\begin{align}
  &\braket{GS| C_{n,D} |GS }
    \nonumber \\
&    \sim
\left\{ \begin{array}{ll}
e^{-\frac{(n^2+2) \pi i}{24 n}} e^{-(1-\frac{1}{n^2}) \frac{1}{48} \frac{2 \pi L}{\xi}} 
+ 2 \cos (\frac{n \pi}{16}) e^{\frac{(n^2+44) \pi i}{48 n}} e^{-\left[ (1-\frac{1}{n^2}) \frac{1}{48} + \frac{1}{2 n^2} \right] \frac{2 \pi L}{\xi}},
& (n : {\rm\ even}, \ n \neq 8 ({\rm mod\ }16)) \\
e^{-\frac{(n^2+2) \pi i}{24 n}} e^{-(1-\frac{1}{n^2}) \frac{1}{48} \frac{2 \pi L}{\xi}} (1-e^{-\frac{1}{2} \frac{2 \pi L}{\xi}} ),
& (n : {\rm\ even}, \ n = 8 ({\rm mod\ }16)) \\
\frac{1}{\sqrt{2}} e^{-\frac{(n^2-1) \pi i}{24 n}} e^{-(1+\frac{2}{n^2}) \frac{1}{48} \frac{2 \pi L}{\xi}} 
+ e^{ \frac{(n^2+11) \pi i}{12 n}} e^{-\left[ (1-\frac{1}{n^2}) \frac{1}{48} + \frac{1}{2 n^2} \right] \frac{2 \pi L}{\xi}},
& (n : {\rm\ odd}) \\
\end{array} \right.
\end{align}
For example, 
\begin{align}
\braket{GS| C_{4,D} |GS }
\sim e^{-\frac{3 \pi i}{16}} e^{- \frac{15}{16} \frac{2 \pi L}{\xi} \frac{1}{48}} (1+ \sqrt{2} e^{\frac{\pi i}{2}} e^{-\frac{1}{32} \frac{2 \pi L}{\xi}}). 
\end{align}

\subsubsection{Partial fermion parity}
\label{Sec:2D_PFP}
In fermionic topological phases, 
the fermion parity symmetry is always preserved. 
Here, we consider the {\it partial fermion parity flip} $(-1)^F_D$ on the 
ground state of the $(p_x - i p_y)$ superconductor. 
The partial fermion parity flip was discussed in detection of quantum phases 
in some literature.~\cite{Shindou2006, RyuHatsugai2006} 
The bulk fermion parity transformation induces the edge fermion parity transformation 
through (\ref{Eq:2D_p-ip_Edge_State}) as 
\begin{align}
(-1)^F \gamma(\ell) (-1)^F = - \gamma(\ell). 
\end{align}
Then, the partial fermion parity flip on the disk $D$ is given by 
\begin{align}
\braket{GS| (-1)^F_D |GS }
&= \frac{{\rm Tr} \big[ (-1)^F e^{- \frac{\xi}{v} H} \big]}{ {\rm Tr} \big[ e^{- \frac{\xi}{v} H} \big]} 
= \frac{\chi_1(\frac{i \xi}{L}) - \chi_{\psi}(\frac{i \xi}{L})}{\chi_1(\frac{i \xi}{L}) + \chi_{\psi}(\frac{i \xi}{L})} 
= \frac{\sum_b (S_{1b} - S_{\psi b}) \chi_b(\frac{i L}{\xi})}{\sum_b (S_{1b} + S_{\psi b}) \chi_b(\frac{i L}{\xi})} 
\nonumber \\
&= \frac{\sqrt{2} \chi_{\sigma}(\frac{i L}{\xi})}{\chi_1(\frac{i L}{\xi}) + \chi_{\psi}(\frac{i L}{\xi})} 
\sim \exp \Big[ \ln \sqrt{2} - \frac{1}{16} \frac{2 \pi \xi}{L} \Big] . 
\label{Eq:top_parity}
\end{align}
We observe that there emerges a scale-independent,
topological contribution to the amplitude,
$e^{\ln \sqrt{2}}$. 
In topologically trivial phases there is no such topological amplitude associated with 
the partial fermion parity flip since the entanglement Hamiltonian is trivial.
Hence, 
the existence of a finite topological amplitude in the 
partial fermion parity flip is a hallmark of topologically nontrivial phases.


\begin{figure}
\centering
  \includegraphics[scale=.5]{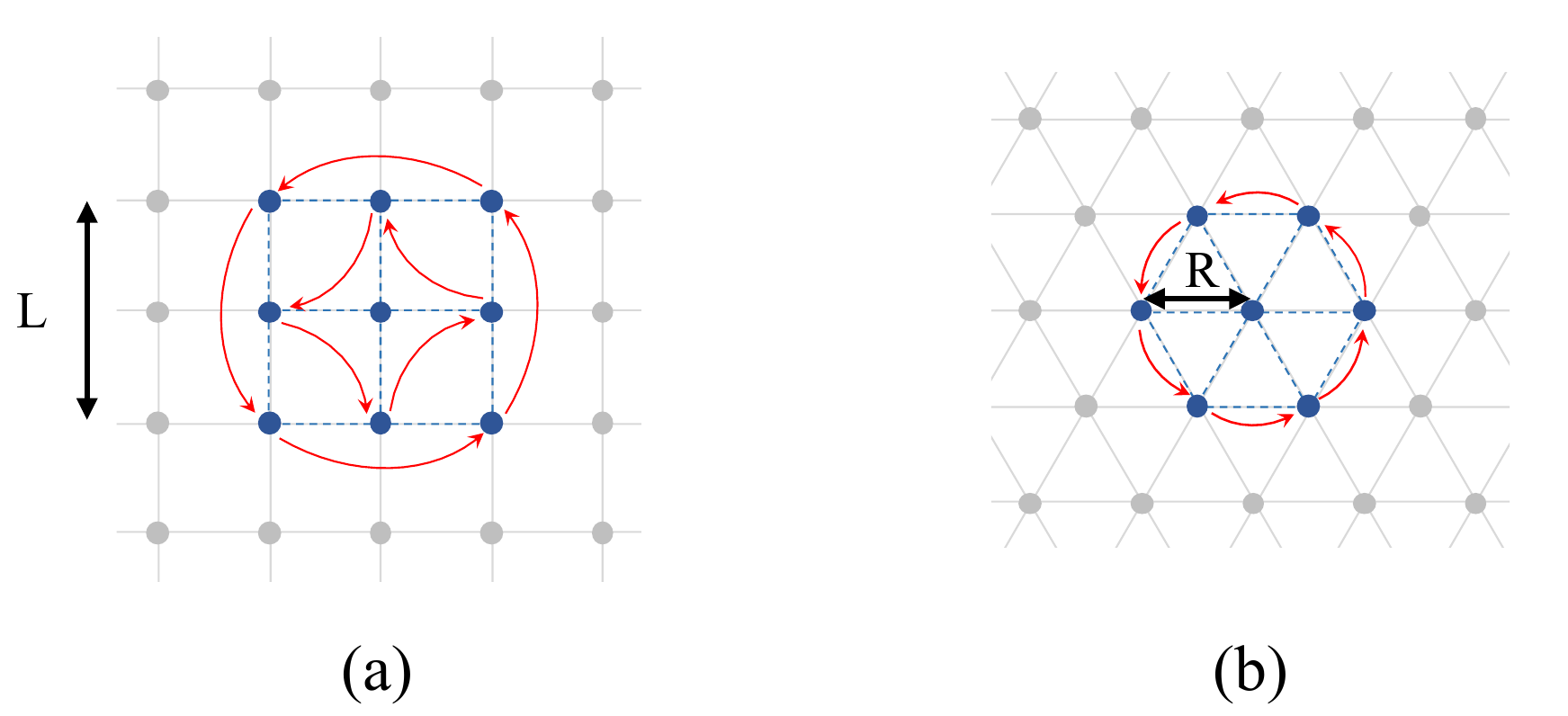}
 \caption{ \label{Fig:lattice} (Color online) (a) Partial $C_4$ rotation on the square lattice, and (b) Partial $C_6$ rotation on the hexagonal lattice. }
\end{figure}

\begin{figure}
\centering
  \includegraphics[scale=.5]{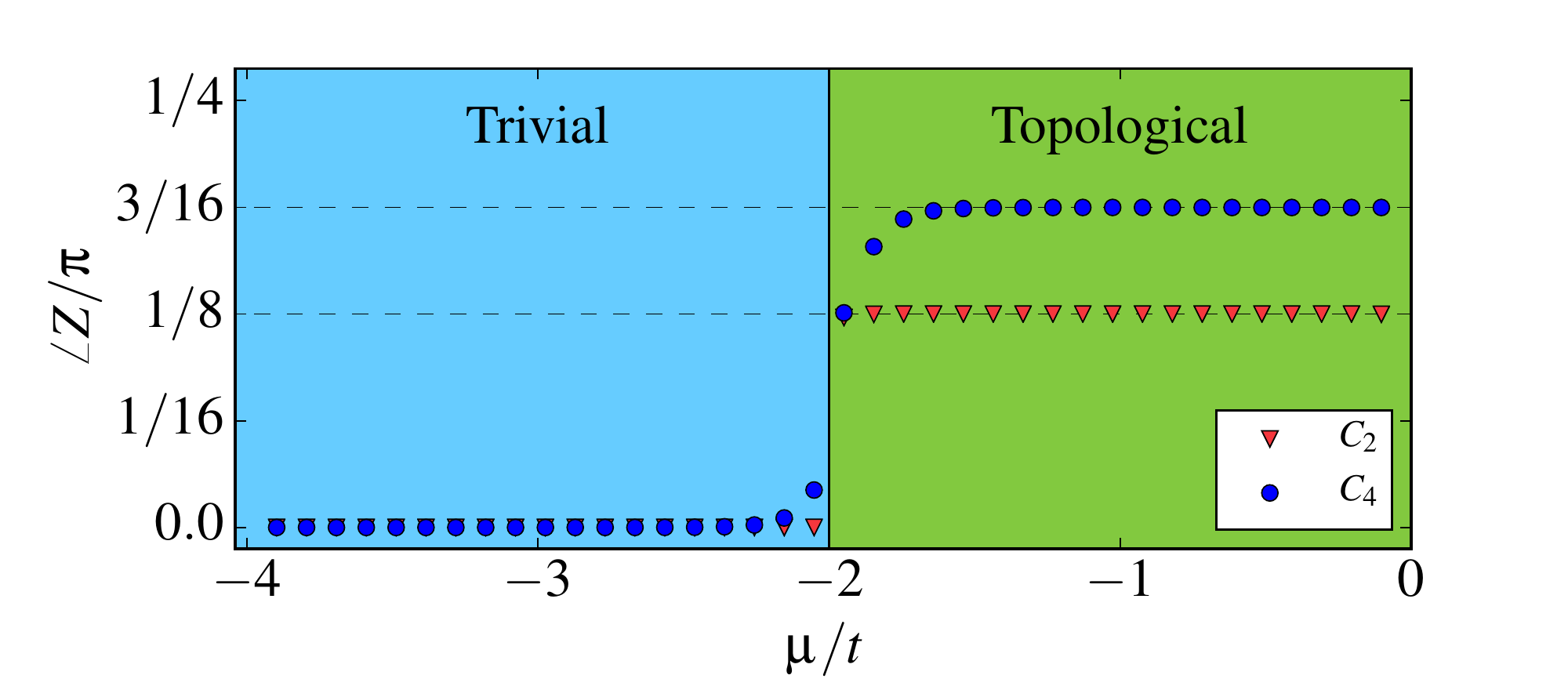}
  \includegraphics[scale=.5]{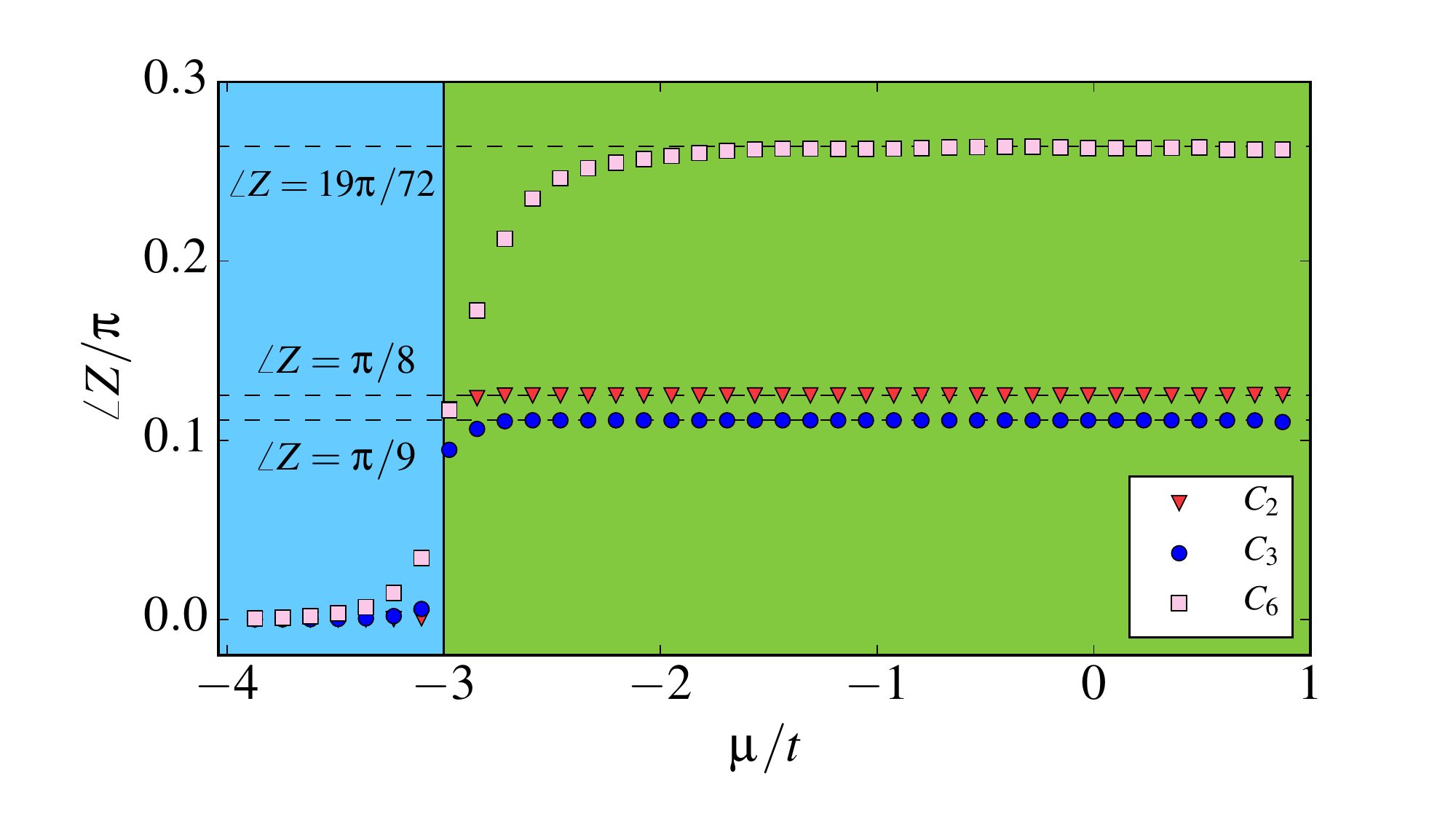}
  \caption{ \label{Fig:phase}
    (Color online)
 Complex phase $\angle Z = {\rm Im} \ln \braket{GS | C_{n,D} | GS}$ of the partial rotation over (a) square lattice, and (b) hexagonal lattice.
 We set $t=\Delta$. The total number of sites is $N=32^2$ and the size of the transformed subsystem is $N_\text{part}=16^2$ for (a) and $N_\text{part}=217$ where $R=8$ for (b).  }
\end{figure}

\subsubsection{Numerical results for lattice models}

In this section, we provide numerical results for partial rotation using the lattice realizations of topological superconductors in class D. We should note that point group symmetries in two dimensional Bravais lattices are limited to four possible cases which are $C_2, C_3, C_4,$ and $C_6$.
Here, we study two models: $(p_x-ip_y)$ superconductors (\ref{Eq:pip_sc}) on 
the square lattice and on the hexagonal lattice. 
The former can be furnished with $C_2$ or $C_4$ symmetry groups and the latter can have $C_2$, $C_3$, and $C_6$ symmetries. 
In the following, we verify that the partial rotation in these systems obeys the generic form of Eq.~(\ref{Eq:PPGT}). In particular, we show that the complex phase $\theta$ and 
the area law coefficient $\alpha$
match those predicted in Eq.~(\ref{Eq:pip_CnD}). 
We find the expectation value of the partial rotation by rearranging  the position of lattice sites  inside the subsystem as shown in Fig.~\ref{Fig:lattice} and then compute the inner product of the two wave functions $Z =\bra{GS} C_{n,D} \ket{GS}$. One important fact here is that the subsystem must be invariant under $C_n$ rotation, otherwise we cannot perform this procedure, since the full lattice after performing the partial  $C_n$ rotation will not be the same as the original one, if the subsystem is not $C_n$ symmetric.

As the first model, we consider the tight-binding Hamiltonian
\begin{align} \label{eq:BdG_sq_latt}
H= -\frac{t}{2} \sum_\xv 
[\psi_{\xv+\iv}^\dagger \psi^{\ }_{\xv}
+\psi_{\xv+\jv}^\dagger \psi^{\ }_{\xv}+{h.c.}] +\frac{1}{2} \sum_\xv [i\Delta \psi_{\xv+\iv}^\dagger \psi^\dagger_{\xv}+\Delta \psi_{\xv+\jv}^\dagger \psi^\dagger_{\xv}+{h.c.}] -\mu \sum_{\xv} \psi_\xv^\dagger \psi^{\ }_\xv
\end{align}
on a square lattice with the basis vectors $\iv$ and $\jv$. For $-2t<\mu<0$, this model describes a $(p_x+ip_y)$ superconductor where the long-wavelength theory around $\kv=0$ is given by Eq.~(\ref{Eq:pip_sc}). Figure~\ref{Fig:phase}(a) shows the complex phase of the partial $C_2$ and $C_4$ rotations as a function of $\mu$. It is evident that 
$\angle Z = {\rm Im} \ln \braket{GS | C_{n,D} | GS}$ 
is zero in the trivial phase, while it is quantized in the topological phase and the numerical values are in perfect agreement with the predicted values from Eq.~(\ref{Eq:pip_CnD}).

In order to realize $C_6$ point group symmetry and its subgroups $C_3$ and $C_2$, we study the hexagonal lattice version of the $(p_x-ip_y)$ superconductor, given by the Hamiltonian
\begin{align} \label{eq:BdG_hex_latt}
H= -\frac{t}{2} 
\sum_{\xv,m}
[\psi_{\xv+\ev_m}^\dagger \psi^{\ }_{\xv}+{h.c.}]
+\frac{1}{3} 
\sum_\xv [i\Delta e^{i\theta_m} \psi_{\xv+\ev_m}^\dagger \psi^\dagger_{\xv}+{h.c.}]
-\mu \sum_{\xv} \psi_\xv^\dagger \psi^{\ }_\xv
\end{align}
where $m=1,2,3$ denotes the nearest-neighbor lattice vectors $\ev_1=\iv$, $\ev_2=\frac{1}{2}\iv+\frac{\sqrt{3}}{2}\jv$, and $\ev_3=-\frac{1}{2}\iv+\frac{\sqrt{3}}{2}\jv$. The phase $\theta_m$ is the angle between nearest-neighbor link and the horizontal axis, i.e. $\cos\theta_m=\ev_m\cdot \iv$. Notice that $\ev_1-\ev_2+\ev_3=0$ which means only two vectors out of the three are independent and can be used to construct a basis for the lattice. Here, the chiral $(p_x-ip_y)$ superconductor is realized in the limit $-3t<\mu<t$ where the long wavelength theory is determined by expanding near $\kv=0$. The results are shown in Fig.~\ref{Fig:phase}(b) which conform with Eq.~(\ref{Eq:pip_CnD}).

\begin{figure}
\centering
  \includegraphics[scale=.5]{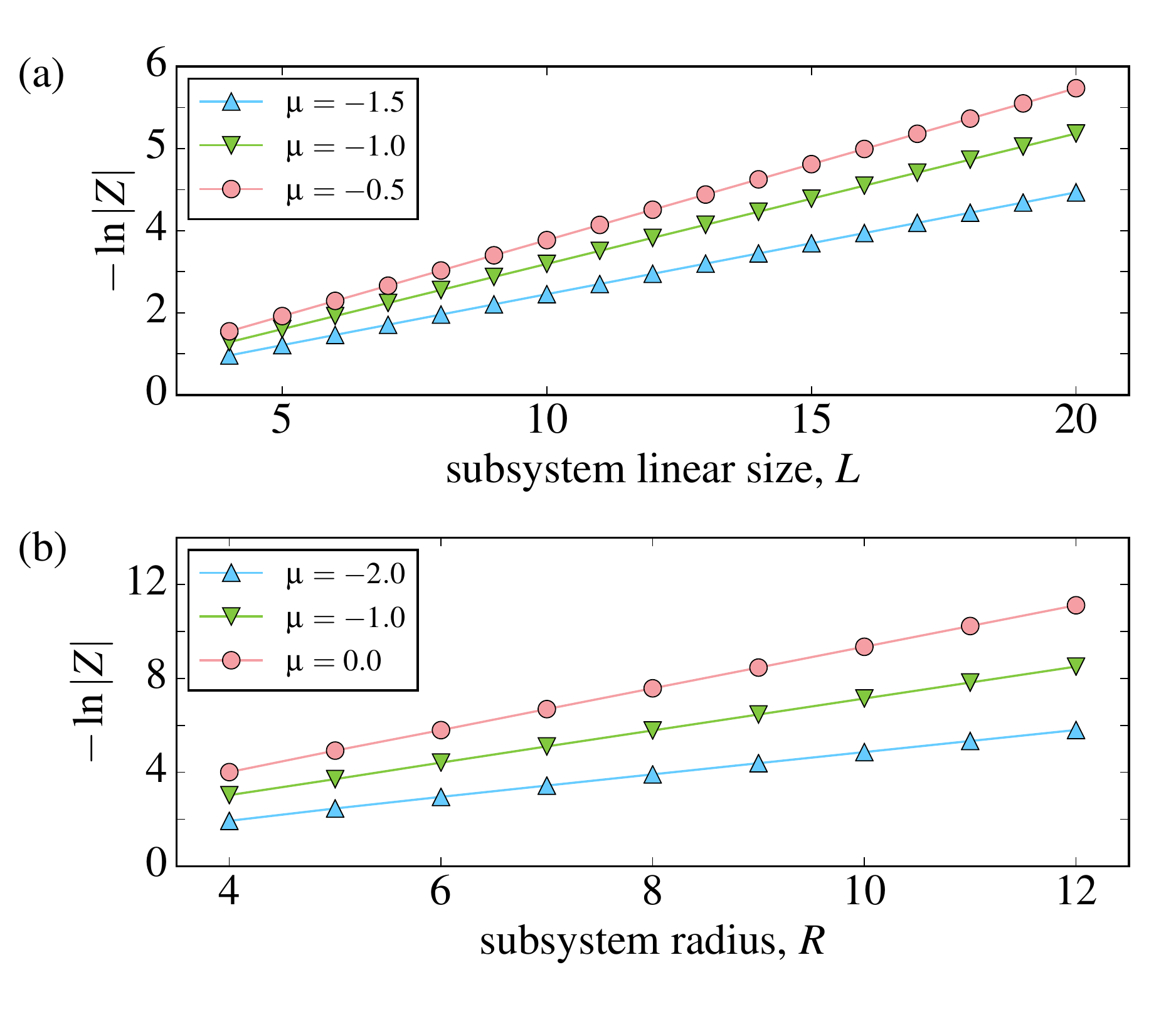}
 \caption{ \label{Fig:alaw}
 (Color online) 
 Amplitude $|Z| = |\braket{GS | C_{n,D} | GS}|$ of  (a) partial $C_4$ rotation on the square lattice (\ref{eq:BdG_sq_latt}), and (b) partial $C_6$ rotation on the hexagonal lattice (\ref{eq:BdG_hex_latt}) for various values of the chemical potential $\mu$.
Here, we set $t=\Delta$. The total number of sites is $N=32^2$ and the dimension of the subsystem is given in terms of $L$ and $R$ (see Fig.~\ref{Fig:lattice}) in each case. Solid lines are linear fits to data points.}
\end{figure}

We also check the exponentiated area law behavior by looking at the amplitude of the partial rotation as a function of subsystem size. We show some typical results for the partial $C_4$ and $C_6$ rotations in Figs.~\ref{Fig:alaw}(a) and (b), respectively. The linear behavior in terms of subsystem perimeter is evident. In addition, we look at the ratio of the area law coefficients given by Eq.~(\ref{Eq:pip_CnD}) where the microscopic quantity $\xi$ is cancelled and as a result we should get a model-independent (topological) value. Table~\ref{tab:arealaw} shows that the numerically calculated ratios are quite close to the values predicted by Eq.~(\ref{Eq:pip_CnD}).

\begin{table}
	\begin{tabular}{cccc}
	\hline
	 Ratio    & Eq.~(\ref{Eq:pip_CnD})  & Numerics & Difference (\%) \\ \hline 
	$C_2/C_4$ & 4/5 &  0.810 & 1.2  \\	
	$C_2/C_6$ & 27/35 & 0.754 & 2.3 \\
	$C_3/C_6$ & 44/35 & 1.262 & -0.4 \\\hline
	\end{tabular}
	\caption{\label{tab:arealaw} 
	Ratios of area-law coefficients in Eq. (\ref{Eq:pip_CnD}) for Hamiltonian (\ref{eq:BdG_sq_latt}).
	$t=\Delta$ and $\mu=-1$.}
\end{table}

Let us now make our final remark in this section regarding the topological contribution $\gamma$ in Eq.~(\ref{Eq:PPGT}). A direct way to compute this quantity is by looking at the $y$-intercept of the area law plots ($\ln Z$ versus $L$). However, we should note that the expression (\ref{Eq:PPGT}) is derived for subsystems with smooth edges and any partitioning of a lattice inevitably results in sharp corners. Therefore, the $y$-intercept method is likely to fail on lattice models due to extra contributions from the corners which add to $\gamma$ and make the evaluation of $\gamma$ imprecise. Fortunately, there is a scheme~\cite{Kitaev2006a,YZhang2011} to overcome this issue and remove the corner terms by writing the $\gamma$ in the following way
\begin{figure}
\centering
  \includegraphics[scale=0.42]{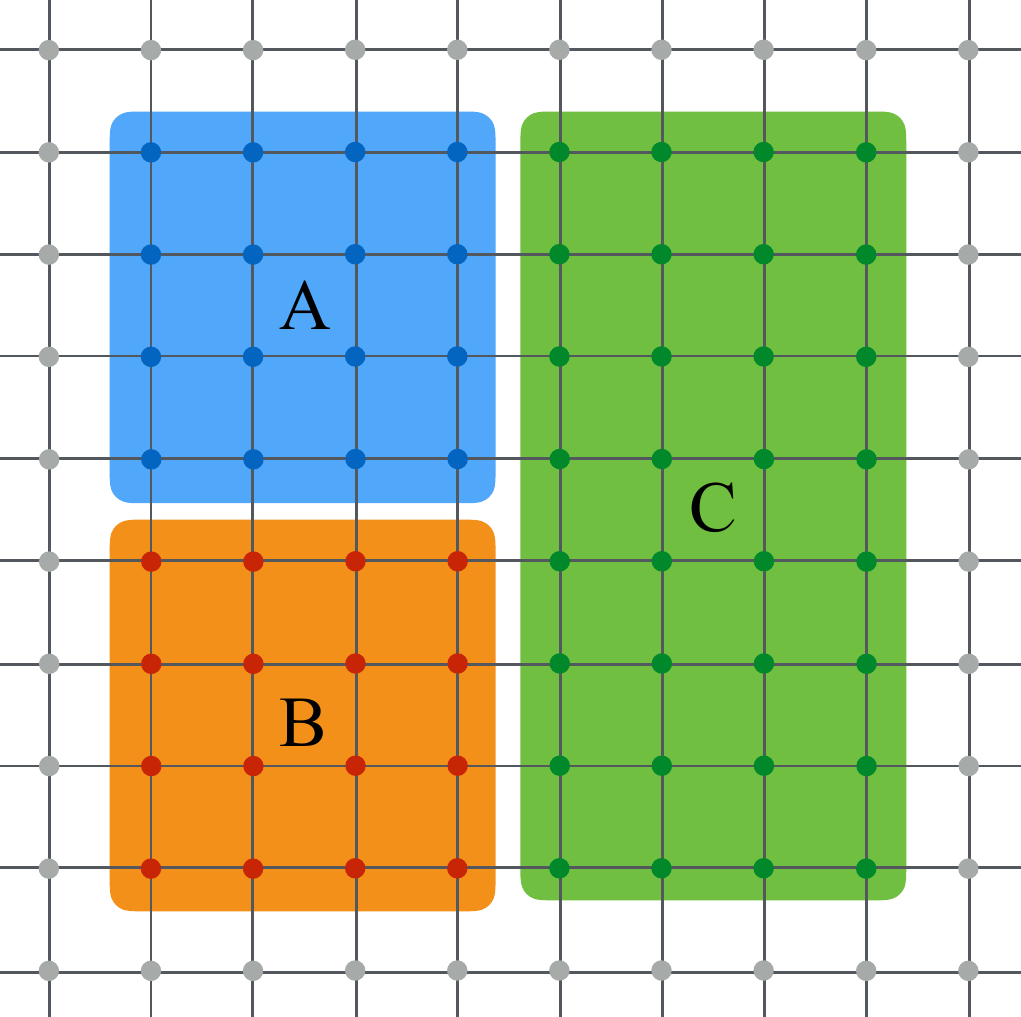}
  \includegraphics[scale=0.5]{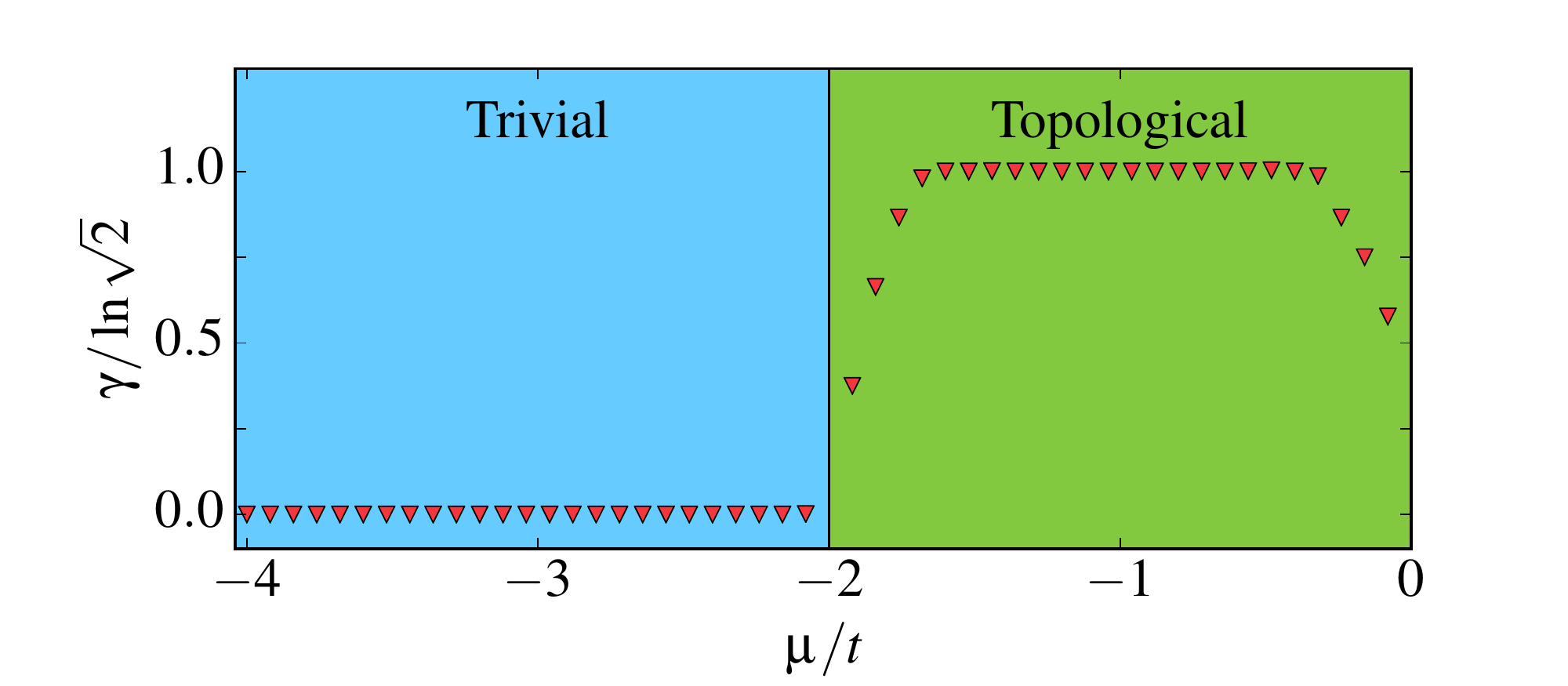}
  \caption{ \label{Fig:top_parity} (Color online)
    (a) The scheme to extract topological contribution $\gamma$ in the partial
    fermion parity transformation (\ref{eq:top_cont_Cn}).
    (b) Numerical results for the $(p_x-ip_y)$ superconductor (Eq.~(\ref{eq:BdG_sq_latt})) as a function of the chemical potential $\mu$.
 We set $t=\Delta$.  Here, the total system size is $N=40^2$ and the size of subsystem $A$ is $N_A=16^2$. }
\end{figure}
\begin{align} \label{eq:top_cont_Cn}
\gamma=\ln |Z_A| + \ln |Z_B| + \ln |Z_C| - \ln |Z_{AB}|- \ln |Z_{AC}|- \ln |Z_{BC}| + \ln |Z_{ABC}|   
\end{align}
where $A, B$, and $C$ are three corner sharing subsystems as shown in Fig.~\ref{Fig:top_parity}(a).
We use this scheme to extract the topological contribution $\ln \sqrt{2}$ in the partial fermion parity flip as given in Eq.~(\ref{Eq:top_parity}). We compute $\gamma$ over a wide range of $\mu$ and the results are shown in Fig.~\ref{Fig:top_parity}. Away from the critical point in the topological phase, $\gamma$ is quantized to $\ln \sqrt{2}$. The same procedure can be done for the hexagonal lattice which yields the same results (we do not show them here). 
It is worth noting that the above scheme cannot be used for partial $C_n$ rotation since, as mentioned earlier, in order to compute the partial rotation all subsystems must be $C_n$-invariant and this is not the case for some segments such as $C$ and $BC$.

\subsection{$(2+1)d$ nonchiral superconductors with $C_2$ rotation symmetry ($\Omega^{\spin}_3(B \Z_2)$)}
\label{Sec:4-3}

Fermionic SPT phases (which are non-chiral) with on-site $\Z_2$ symmetry 
are classified by the spin cobordism group 
$\Omega^{\rm Spin}_3(B \Z_2) = \Z_8$ where $B \Z_2$ is the classifying space 
of $\Z_2$ group.~\cite{ryu2012interacting, gu2014effect, Kapustin2015a}
The generating manifold of
the $\Omega^{\rm Spin}_3(B \Z_2)$ group is $3d$ real projective space $\R P^3$ 
with an appropriate background $\Z_2$ gauge field. 
A generating model Hamiltonian is given by the 
$(p_x + i p_y)_{\uparrow} \oplus (p_x - i p_y)_{\downarrow}$ superconductor 
\begin{align}
&H = H_{\ua} + H_{\da}, \label{eq:2d_D_nonchiral} 
\nonumber \\
&H_{\ua/\da} 
= \sum_{\bk} \psi_{\ua/\da}^{\dag}(\bk) (\frac{k^2}{2m}-\mu) \psi_{\ua/\da}(\bk) 
+ \sum_{\bk} \big[ \frac{\Delta}{2} \psi_{\ua/\da}^{\dag}(\bk) (k_x\pm i k_y) \psi_{\ua/\da}^{\dag}(-\bk) + h.c. \big], 
\end{align}
with the on-site $\Z_2$ flavor symmetry 
\begin{align}
U \psi^{\dag}_{\ua/\da}(\bx) U^{-1} = \mp \psi^{\dag}_{\ua/\da}(\bx). 
\end{align}
That is to say, 
the fermion parity symmetry 
for each spin up and down fermions
is separately preserved. 

To introduce a cross-cap to have $\R P^3$ as the spacetime, 
we need to consider partial $C_2$ rotation
in the canonical formalism 
(i.e., in terms of a given ground state wave function). 
However, there is no $C_2$ rotation symmetry a priori in this model.
Thus, instead of imposing the non-spatial $\Z_2$ symmetry,  
we consider $C_2$ symmetry in advance. 
The breakdown of the $\Z$ classification of $(2+1)d$ class D non-chiral superconductors 
with $C_2$ rotation symmetry~\cite{Shiozaki2014} is also given by $\Z_8$~\cite{
you2014symmetry}, 
which is natural from the point of view of TQFTs 
since $C_2$ 
rotation symmetry is an orientation preserving symmetry. 
The translation from the non-spatial $\Z_2$ symmetry 
to the $C_2$ rotation symmetry 
in the same topological class is as follows: 
We introduce a combined rotation symmetry $\tilde C_2 = U C_2$ 
from the on-site $\Z_2$ symmetry $U$ and the {\it inherent} continuum 
rotation symmetry for chiral $(p_x \pm i p_y)$ superconductors 
$C_{\theta} \psi_{\uparrow/\downarrow}^{\dag}(\bx) C_{\theta}^{-1} 
= e^{\pm i \theta/2} \psi_{\uparrow/\downarrow}^{\dag}(C_{\theta} \bx) $
introduced in (\ref{eq:2d_px-ipy_rot_sym}), 
where $C_2 = C_{\pi}$. 
Then, the $\tilde C_2$ rotation is defined as 
\begin{align}
\tilde C_2 \psi_{\uparrow/\downarrow}^{\dag}(\bx) \tilde C_2^{-1} 
= -i \psi_{\uparrow/\downarrow}^{\dag}(- \bx). 
\end{align}
Under this $\tilde C_2$ rotation symmetry, 
the model Hamiltonian (\ref{eq:2d_D_nonchiral}) 
is the generating model of the $\Z_8$ group. 
Finally, we can forget the on-site $\Z_2$ symmetry $U$ and the continuum rotation 
symmetry $C_{\theta}$: the $\Z_8$ classification is ensured only by the $\tilde C_2$ 
rotation symmetry. 

Let us evaluate 
the ground state expectation value of partial $\widetilde C_2$ rotation. 
The contribution from 
the $(p_x-i p_y)_{\downarrow}$ sector
is the same as (\ref{eq:2d_c2_pr}). 
On the other hand, 
for the $(p_x + i p_y)_{\ua}$ sector,
since the gapless edge excitation has left-moving chirality, 
the expectation value of
partial $\widetilde C_2$ rotation is given by 
\begin{align}
\braket{GS_{\uparrow}| \widetilde C_{2,D} |GS_{\uparrow}}
&= \frac{{\rm Tr} \big[ (-1)^F e^{i \widetilde{P} \frac{L}{2}} e^{- \frac{\xi}{v} H} \big]}{ {\rm Tr} \big[ e^{- \frac{\xi}{v} H} \big]}  
\nonumber \\
&= \frac{e^{ \pi i \frac{1}{48}} \sum_b \big\{ (S T^{-2} S)_{1b} -(S T^{-2} S)_{\psi b} \big\} \chi_b(\frac{i L}{4 \xi}-\frac{1}{2}) }
{\sum_{a = 1, \psi} \sum_b S_{ab} \chi_b(\frac{i L}{\xi})}
\nonumber \\
&\sim e^{- \frac{\pi i}{8}} e^{- \frac{3}{4} \frac{2 \pi L}{\xi} \frac{1}{48}}. 
\end{align}
Therefore, 
the total expectation value is 
\begin{align}
\braket{GS| \widetilde C_{2,D} |GS}
= \braket{GS_{\uparrow}| \widetilde C_{2,D} |GS_{\uparrow}} \braket{GS_{\downarrow}| \widetilde C_{2,D} |GS_{\downarrow}}
\sim e^{- \frac{\pi i}{4}} e^{- \frac{3}{2} \frac{2 \pi L}{\xi} \frac{1}{48}}. 
\end{align}
Thus, the $U(1)$ phase in the expectation value of partial $\widetilde C_2$ 
rotation does capture the $\Z_8$ 
classification.

\subsection{$(2+1)d$ topological insulators with rotation symmetry}

In this section, we discuss
partial rotations in 
$(2+1)d$ Chern insulators. 
In Chern insulators (particle number conserving systems),
partial rotation can be combined with
continuous $U(1)$ phase rotation,
which should be contrasted 
with 
the case of superconductors 
where 
a partial rotation can be combined with
the discrete $\Z_2$ 
fermion parity transformation.
To be specific,
let us consider 
the following simple model
realizing a chiral Chern insulator 
\begin{align}
H 
&= \sum_{\bk} \psi^{\dag}(\bk) \Big[ (\frac{k^2}{2m} - \mu) \sigma_z + v k_x \sigma_x + v k_y \sigma_y \Big] \psi(\bk)
\nonumber\\
&= \sum_{\bk} \Big[ \psi^{\dag}(\bk) (\frac{k^2}{2m} - \mu) \sigma_z \psi(\bk) + v (k_x-i k_y) \psi_1^{\dag}(\bk) \psi_2(\bk) + h.c. \Big], 
\quad 
\psi(\bk) = (\psi_1(\bk), \psi_2(\bk))^T. 
\label{eq:2d_chiral_chern_insulator}
\end{align}
This model is invariant 
under 
the continuous spatial rotation 
and the $U(1)$ charge rotation
\begin{align}
&C_{\theta} \psi^{\dag}(\bx) C_{\theta} = \psi^{\dag}(C_{\theta} \bx) 
\begin{pmatrix}
e^{-i \theta/2} & 0 \\
0 & e^{i \theta/2} \\
\end{pmatrix}, 
\qquad 
U_{b} \psi^{\dag}(\bx) U_b^{-1} = e^{-2 \pi i b }\psi^{\dag}(\bx) , 
\quad 
b \in \R/\Z. 
\end{align}

On the disk geometry, 
the chiral Chern insulator
supports
a chiral (right-moving)
gapless excitations
localized on the boundary,
which can be created by 
the following complex fermion 
operator $\gamma(\ell)$  
\begin{align}
\gamma^{\dag}(\frac{L \phi}{2 \pi}) \sim (e^{-i \phi/2} \psi^{\dag}_1(r,\phi) +i e^{i \phi/2} \psi^{\dag}_2(r,\phi)) e^{-\int^r d r' m(r')}, 
\label{eq:2d_chern_edge}
\end{align}
where 
$\ell =L\phi/2\pi$ is the spatial coordinate 
along the boundary,
and $L$ is the circumference of the boundary. 
We have chosen the gauge of $\gamma(\ell)$
such that $\gamma(\ell)$ 
satisfies the anti-periodic boundary condition 
$\gamma(\ell + L) = - \gamma(\ell)$. 
With this boundary condition, 
the Hamiltonian and momentum operator 
for the edge mode
can be written as
\begin{align}
H = \frac{2 \pi v}{L} \sum_{m \in \Z + 1/2} m : \gamma^{\dag}_m \gamma_m : 
- \frac{1}{24},  
\qquad 
P = \frac{H}{v},
\label{Eq:2D_Edge_A}
\end{align}
where $\gamma_m$ is the $m$-th
Fourier mode
of $\gamma(\ell)$.
Here, 
the Hamiltonian and momentum 
operator are normal ordered 
with respect to the Fermi sea 
which is filled with
$\gamma^{\dag}_m (m < 0)$ fermions. 
A derivation of (\ref{eq:2d_chern_edge}) and (\ref{Eq:2D_Edge_A}) 
are summarized in Appendix \ref{Derivation of surface theories}.

\subsubsection{Partial rotation with $U(1)$ charge transformation}

We now calculate the expectation value of the partial $C_{\theta}$ rotation 
together with the partial $U(1)$ charge transformation. 
The partial rotation $C_{\theta,D}$ 
on the disk $D$  
combined 
with the partial 
$U(1)$ 
transformation 
$U_{b,D}$ 
acts on the boundary 
fermion operators as 
\begin{align}
&U_b C_{\theta} \gamma^{\dag}(\ell) (U_b C_{\theta})^{-1} = e^{-2 \pi i b} \gamma^{\dag}(\ell + \frac{\theta L}{2 \pi}). 
\end{align}
These transformations
are generated by the momentum and 
the $U(1)$ charge defined by 
\begin{align}
\widetilde P = \frac{2 \pi}{L} \sum_{m \in \Z+1/2} m : \gamma_n^{\dag} \gamma_m:, 
\qquad  
\widetilde Q = \sum_{m \in \Z+1/2} : \gamma_m^{\dag} \gamma_m:. 
\end{align}
Here we regularized $\widetilde P$ and $\widetilde Q$ so that 
$\widetilde P \ket{FS} = \widetilde Q \ket{FS} = 0$ 
where $\ket{FS}$ is the Fermi sea
of the edge theory.  
The expectation value 
of 
$U_{b,D} C_{\theta,D}$
is given by
\begin{align}
\braket{GS| U_{b,D} C_{\theta,D} |GS } 
\sim \frac{{\rm Tr}_{a=\frac{1}{2}}\Big[ e^{-2 \pi i \widetilde Q b} e^{-i \widetilde P \frac{\theta L}{2 \pi}} e^{- \frac{\xi}{v} H} \Big]}{{\rm Tr}_{a=\frac{1}{2}}\Big[ e^{- \frac{\xi}{L} H} \Big]}
= \frac{e^{- \frac{i \theta}{24}} Z_{\frac{1}{2},b+\frac{1}{2}}(\frac{i \xi}{L} - \frac{\theta}{2 \pi})}{ Z_{\frac{1}{2},\frac{1}{2}}(\frac{i \xi}{L})}. 
\end{align}
Here, $Z_{a,b}(\tau)$ is the partition function of
the right-mover complex fermion theory
defined on the spacetime torus with
twisted boundary conditions
\begin{align}
Z_{a,b}(\tau) = {\rm Tr}_a \Big[ e^{-2 \pi i (\widetilde Q + a - \frac{1}{2}) (b-\frac{1}{2})} e^{2 \pi i \tau (L_0 - \frac{1}{24})} \Big] 
= \frac{\theta_{a-\frac{1}{2},\frac{1}{2}-b}(0|\tau)}{\eta(\tau)}, 
\end{align}
where $\theta_{a,b}(z|\tau)$ and $\eta(\tau)$ is the generalized theta function and the Dedekind eta function, respectively.~\cite{DiFrancesco1997}

We now specialize to
partial $n$-fold rotation $C_n := C_{\frac{2 \pi}{n}}$ 
combined with
with the $n$-fold $U(1)$ transformation $U_{b = \frac{p}{n}}$ ($p=0, \dots, n-1$). 
By noting the $a,b$ dependence and 
the modular transformation of $Z_{a,b}(\tau)$, 
\begin{align}
&Z_{a+1,b}(\tau) = Z_{a,b}(\tau), && 
Z_{a,b+1}(\tau) = e^{- 2 \pi i(a-\frac{1}{2})} Z_{a,b}(\tau), 
\nonumber \\
&Z_{a,b}(\tau+1) = e^{- \pi i a^2 + \frac{\pi i}{6}} Z_{a,b-a}(\tau), && 
Z_{a,b}(-\frac{1}{\tau}) = e^{- 2 \pi i (a-\frac{1}{2}) (b-\frac{1}{2})} Z_{1-b,a}(\tau), 
\label{eq:2d_chern_n_fold}
\end{align}
one can show  
\begin{align}
\braket{GS| U_{\frac{p}{n},D} C_{n,D} |GS } 
&= \frac{e^{- \frac{i \pi}{12 n}} Z_{\frac{1}{2},\frac{p}{n}+\frac{1}{2}}(\frac{i \xi}{L} - \frac{1}{n})}{ Z_{\frac{1}{2},\frac{1}{2}}(\frac{i \xi}{L})} 
= \frac{e^{- \frac{i \pi}{12 n} - \frac{n i \pi}{12} + \frac{p^2 \pi i}{n}} Z_{\frac{1}{2}+\frac{n}{2},\frac{1}{2}-\frac{p}{n}}(\frac{i L}{n^2 \xi} + \frac{1}{n})}{ Z_{\frac{1}{2},\frac{1}{2}}(\frac{i L}{\xi})} 
\label{eq:2d_chern_chiral_cn}
\end{align}
by the same modular transformation as (\ref{Eq:PR_Modular}). 
When the circumference of the disk is sufficiently larger than 
the bulk correlation length, $L\gg \xi$, 
we can approximate (\ref{eq:2d_chern_chiral_cn}) by taking 
lowest energy states.
The denominator is approximated as 
\begin{align}
Z_{\frac{1}{2},\frac{1}{2}}\left(\frac{i L}{\xi}\right) \sim e^{\frac{2 \pi L}{\xi} \frac{1}{24}}. 
\end{align}
As for the numerator,
when $n=\mbox{even}$, the unique vacuum state $\ket{0}_{\frac{1}{2}}$ gives the leading contribution 
\begin{align}
Z_{\frac{1}{2} + \frac{n}{2}, \frac{1}{2}-\frac{p}{n}} \left( \frac{i L}{n^2 \xi} + \frac{1}{n} \right)
=Z_{\frac{1}{2}, \frac{1}{2}-\frac{p}{n}} \left( \frac{i L}{n^2 \xi} + \frac{1}{n} \right)
\sim e^{- \frac{\pi i}{12 n}} e^{\frac{2 \pi L}{n^2 \xi} \frac{1}{24}} 
\qquad 
(n : {\rm even}). 
\label{eq:partial_tr_4.56}
\end{align}
On the other hand, 
when $n$ is odd, the ground state 
associated with the torus partition function 
has double degeneracy originated from the zero mode. 
We have
\begin{align}
Z_{\frac{1}{2}+\frac{n}{2},\frac{1}{2}-\frac{p}{n}}
\left(\frac{i L}{n^2 \xi} + \frac{1}{n}\right) 
&= Z_{0,\frac{1}{2}-\frac{p}{n}}
\left(\frac{i L}{n^2 \xi} + \frac{1}{n}\right)  
\nonumber \\
&\sim ( e^{\pi i p/n} + e^{-\pi i p/n}) e^{2 \pi i (\frac{i L}{n^2 \xi} + \frac{1}{n})(\frac{1}{8} - \frac{1}{24})} 
= 2 \cos \frac{\pi p}{n} e^{\frac{\pi i}{6n}} e^{- \frac{\pi L}{6 n^2 \xi}} && (n : {\rm odd}). 
\label{eq:partial_rot_chern_odd_n_boundary_cond}
\end{align}
Combining these contributions, 
\begin{align}
\braket{GS| U_{\frac{p}{n},D} C_{n,D} |GS } 
\sim \left\{\begin{array}{ll}
e^{ \frac{12 p^2 - n^2 - 2}{12 n} \pi i} e^{- (1-\frac{1}{n^2}) \frac{2 \pi L}{\xi} \frac{1}{24}} & (n : {\rm even}) \\
2 \cos \frac{\pi p}{n} e^{\frac{12 p^2 - n^2 + 1}{12 n} \pi i} e^{- (1+\frac{2}{n^2}) \frac{2 \pi L}{\xi} \frac{1}{24}} & (n : {\rm odd}) \\
\end{array}\right.
\label{eq:2d_chern_n_fold_result}
\end{align}
The $U(1)$ phases in (\ref{eq:2d_chern_n_fold_result}) for $n = 2, \dots, 9$
and 
$p=0,\dots, 8$
are summarized in Table \ref{tab phases}. 
\begin{table}
\renewcommand{\arraystretch}{1.5}
\begin{tabular}{c|cccccccccccc}
 & $p=0$ & $p=1$ & $p=2$ & $p=3$ & $p=4$ & $p=5$ & $p=6$ & $p=7$ & $p=8$ \\
\hline
$n=2$ & $-\frac{\pi}{4}$ & $\frac{\pi}{4}$ & & & & & & & \\
$n=3$ & $-\frac{2\pi}{9}$ & $\frac{\pi}{9}$ & $\frac{\pi}{9}$ & & & & & & \\
$n=4$ & $-\frac{3\pi}{8}$ & $-\frac{\pi}{8}$ & $\frac{5\pi}{8}$ & $-\frac{\pi}{8}$ & & & & & \\
$n=5$ & $-\frac{2\pi}{5}$ & $-\frac{\pi}{5}$ & $\frac{2 \pi}{5}$ & $\frac{2 \pi}{5}$ & $-\frac{\pi}{5}$ & & & &  \\
$n=6$ & $-\frac{19 \pi}{36}$ & $-\frac{13\pi}{36}$ & $\frac{5\pi}{36}$ & $\frac{35\pi}{36}$ & $\frac{5\pi}{36}$ & $-\frac{13\pi}{36}$ & & &  \\
$n=7$ & $-\frac{4 \pi}{7}$ & $-\frac{3 \pi}{7}$ & 0 & $\frac{5 \pi}{7}$ & $\frac{5 \pi}{7}$ & 0 & $-\frac{3 \pi}{7}$ & & \\
$n=8$ & $-\frac{11\pi}{16}$ & $-\frac{9\pi}{16}$& $-\frac{3\pi}{16}$ & $\frac{7\pi}{16}$ & $\frac{21\pi}{16}$ & $\frac{7\pi}{16}$ & $-\frac{3\pi}{16}$ & $-\frac{9\pi}{16}$ & \\
$n=9$ & $-\frac{20 \pi}{27}$ & $-\frac{17 \pi}{27}$ & $-\frac{8 \pi}{27}$ & $\frac{7 \pi}{27}$ & $\frac{28 \pi}{27}$ & $\frac{28 \pi}{27}$ & $\frac{7 \pi}{27}$ & $-\frac{8 \pi}{27}$ & $-\frac{17 \pi}{27}$ \\
\end{tabular}
\renewcommand{\arraystretch}{1}
\caption{
The $U(1)$ phases of the partial $U_{\frac{p}{n}} C_n$ rotation
(\ref{eq:2d_chern_n_fold_result}) of
the topological insulator with rotational symmetry defined in (\ref{eq:2d_chiral_chern_insulator}) for $n = 2, \dots, 9$ and $p=0,\dots, 8$. 
\label{tab phases}
}
\end{table}
We have confirmed that all these results in the above table match with numerical calculations 
in two dimensional lattice models for $n=2, 3, 4,$ and $6$.

The expectation value of the partial $U_{\frac{p}{n}} C_n$ rotation is related to the appropriate TQFT partition function on $L(n,1)$. 
In the same way, the partial $(U_{\frac{p}{n}} C_n)^m$ rotation with $n$ and $m$ being coprime
is related to the the lens space $L(n,m)$. 
The expectation value of the partial $(U_{\frac{p}{n}} C_n)^m$ can be evaluated 
by using an appropriate modular transformation,
which is determined
by the fraction expansion of $\frac{n}{m}$ (see Appendix~\ref{App:1}.)
For instance, for odd $n=2k+1$, the partial $(U_{\frac{p}{2k+1}} C_{2k+1})^2$
rotation is computed by using
the modular transformation $ST^{k}ST^{-2}S$ as 
\begin{equation}\begin{split}
\braket{GS| (U_{\frac{p}{2k+1},D} C_{2k+1,D})^2 |GS } 
&= \frac{e^{- \frac{i \pi}{6 (2k+1)}} Z_{\frac{1}{2},\frac{2p}{2k+1}+\frac{1}{2}}(\frac{i \xi}{L} - \frac{2}{2k+1})}{ Z_{\frac{1}{2},\frac{1}{2}}(\frac{i \xi}{L})} \\
&= \frac{e^{-\frac{i \pi  k \left(12 k^2+k (8-48 p)+48 p^2-24 p-3\right)}{24 k+12}} Z_{\frac{1}{2},\frac{1}{2}+\frac{k}{2}+\frac{p}{2k+1}}(\frac{i L}{(2k+1)^2 \xi} - \frac{k}{2k+1})}{ Z_{\frac{1}{2},\frac{1}{2}}(\frac{i L}{\xi})} \\
&\sim e^{-\frac{i \pi  k \left(3 k^2+k (2-12 p)+12 p^2-6 p-1\right)}{6 k+3}} e^{- (1-\frac{1}{(2k+1)^2}) \frac{2 \pi L}{\xi} \frac{1}{24}}.
\label{eq:2d_chern_n_fold_result_double_odd}
\end{split}\end{equation}
The $U(1)$ phases in (\ref{eq:2d_chern_n_fold_result_double_odd}) for $n = 3,5,7,9$ and $p=0,\dots, 8$ are summarized in Table \ref{tab phases double}. 
Notice that the boundary condition of the space direction is still anti-periodic after the modular transformation in (\ref{eq:2d_chern_n_fold_result_double_odd}), which implies that the partition function is approximated by the single vacuum state as in (\ref{eq:partial_tr_4.56}). 
For even $n = 2k$, the partial $(U_{\frac{p}{2k}}C_{2k})^2$ transformation is the same as the partial $U_{\frac{p}{k}} C_k$ transformation computed in (\ref{eq:partial_rot_chern_odd_n_boundary_cond}). 
\begin{table}
\renewcommand{\arraystretch}{1.5}
\begin{tabular}{c|cccccccccccc}
& $p=0$ & $p=1$ & $p=2$ & $p=3$ & $p=4$ & $p=5$ & $p=6$ & $p=7$ & $p=8$ \\
\hline
$n=3$ & $\frac{14\pi}{9}$ & $\frac{2\pi}{9}$ & $\frac{2\pi}{9}$ & & & & & & \\
$n=5$ & $0$ & $\frac{2\pi}{5}$ & $\frac{8 \pi}{5}$ & $\frac{8 \pi}{5}$ & $\frac{2\pi}{5}$ & & & &  \\
$n=7$ & $\frac{10 \pi}{7}$ & $\frac{12 \pi}{7}$ & $\frac{4 \pi}{7}$ & $0$ & $0$ & $\frac{4 \pi}{7}$ & $\frac{12 \pi}{7}$ & & \\
$n=9$ & $\frac{50 \pi}{27}$ & $\frac{2 \pi}{27}$ & $\frac{20 \pi}{27}$ & $\frac{50 \pi}{27}$ & $\frac{38 \pi}{27}$ & $\frac{38 \pi}{27}$ & $\frac{50 \pi}{27}$ & $\frac{20 \pi}{27}$ & $\frac{2 \pi}{27}$ \\
\end{tabular}
\renewcommand{\arraystretch}{1}
\caption{
The $U(1)$ phases of the partial $(U_{\frac{p}{n}} C_n)^2$ rotation (\ref{eq:2d_chern_n_fold_result_double_odd}) of topological insulator with rotational symmetry defined in (\ref{eq:2d_chiral_chern_insulator}) for $n = 3,5,7,9$ and $p=0,\dots, 8$. 
\label{tab phases double}
}
\end{table}

\subsubsection{$(2+1)d$ non-chiral topological insulators with rotation symmetry
($\Omega^{\spinc}_3(B\mathbb{Z}_n)$)}
\label{Sec:4-4-2}

The calculations similar to the above results can be deployed to discuss a classification of SPT phases with $C_n$ rotation 
symmetry in non-chiral topological insulators: 
We first construct a model Hamiltonian and the $C_n$ symmetry 
in a similar way to Sec.\ \ref{Sec:4-3}; 
we then introduce the combined $\widetilde C_n$ symmetry from 
the model Hamiltonian with $\Z_n$ on-site symmetry and continuum rotation symmetry. 
The topological classification of non-chiral Chern insulators with 
$\Z_n$ symmetry is given by the $\spinc$ bordsim group $\Omega_3^{\spinc}(B \Z_n)$. 
The cobordism groups $\Omega_3^{\spinc}(B \Z_n)$ and their generating manifolds are 
derived by Bahri and Gilkey~\cite{Bahri1987, Gilkey} by use of the $\eta$-invariant 
of Dirac operators with a $\spinc$ structure. 
In Appendix \ref{App:2},
we briefly summarize their results and a relation to the 
equivariant $K$-theory classification at the free fermion level. 

The third column in Table \ref{Tab:Spin^c_Cobordism} shows the 
cobordism group $\Omega_3^{\spinc}(B \Z_n)$ for $n=2, \dots, 9$. 
$(1-t)$ and $(1-t^2)$ represent generating model Hamiltonians as follows: 
We consider a 4 flavor non-chiral Chern insulator 
\begin{align}
H_{1-t^p} 
&= \sum_{\bk} \psi_{p}^{\dag}(\bk) \Big[ (\frac{k^2}{2m} - \mu) \sigma_z + v k_x \sigma_x + v k_y \sigma_y \Big] \psi_{p}(\bk) 
\nonumber \\
&\quad
+ \sum_{\bk} \psi_{0}^{\dag}(\bk) \Big[ (\frac{k^2}{2m} - \mu) \sigma_z + v k_x \sigma_x - v k_y \sigma_y \Big] \psi_{0}(\bk), 
\label{eq:2d_nonchiral_chern} \\
\psi_{s}(\bk) &= (\psi_{s,1}(\bk),\psi_{s,2}(\bk))^T, 
\quad 
s = p, 0, 
\quad
p \geq 1. 
\end{align}
Here, subscripts of $\psi_{s}(\bx), s \in \{0, \dots n-1\}$ mean 1-dimensional 
representations of the $\Z_n$ symmetry as 
$U \psi_{s}(\bx) U^{-1} = e^{- \frac{2 \pi i s}{n}} \psi_s(\bx)$. 
As shown in Table \ref{Tab:Spin^c_Cobordism}, 
the generating models of $\Omega_3^{\spinc}(B \Z_n)$ for $n \geq 3$ are given by 
$H_{1-t}$ and $H_{1-t^2}$ which generate independent cyclic groups. 
In addition to the $\Z_n$ symmetry, 
the Hamiltonian (\ref{eq:2d_nonchiral_chern}) has the continuum rotation symmetry 
\begin{align}
C_{\theta} \psi^{\dag}_p(\bx) C_{\theta}^{-1} = \psi^{\dag}_p(C_{\theta} \bx) e^{- i \theta \sigma_z/2}, 
\quad  
C_{\theta} \psi^{\dag}_0(\bx) C_{\theta}^{-1} = \psi^{\dag}_0(C_{\theta} \bx) e^{i \theta \sigma_z/2}. 
\label{eq:def_rotation_complex fermion_2d}
\end{align}
We introduce the combined $n$-fold rotation symmetry by 
$\widetilde C_n := U C_{n}$ where $C_n = C_{\theta=\frac{2 \pi}{n}}$. 

Let us evaluate
the expectation value of the partial $\widetilde C_n$ rotation
with respect to the ground state of (\ref{eq:2d_nonchiral_chern}).
The expectation value is expected to simulate 
the path integral (the partition function)
on lens spaces with background $\spinc$ structures with a $\Z_n$ gauge field. 
The contribution from the $\psi_{p}$ fermion,  
$\langle GS_p| \widetilde C_{n,D} |GS_p\rangle$,  
is the same as the previous section. 
On the other hand, 
for the $\psi_{0}$ fermion sector, 
since the gapless edge excitation has left-moving chirality, 
the expectation value of 
the partial $C_n$ rotation is given by 
the complex conjugate of (\ref{eq:2d_chern_chiral_cn}), 
\begin{align}
\langle GS_0| \widetilde C_{n,D} |GS_0 \rangle
= \frac{{\rm Tr}_{a=\frac{1}{2}}\Big[ e^{i \widetilde P \frac{L}{n}} e^{- \frac{\xi}{v} H} \Big]}{{\rm Tr}_{a=\frac{1}{2}}\Big[ e^{- \frac{\xi}{v} H} \Big]}
= \braket{GS| C_{n,D} |GS}^*.
\end{align}
In the same way, the expectation value of the partial $(\wt C_n)^2$ rotation with respect to the ground state (\ref{eq:2d_nonchiral_chern}) is computed. 
The total $U(1)$ phases 
of the expectation value 
$\langle GS| \widetilde C_{n,D} |GS \rangle$
and 
$\langle GS| (\widetilde C_{n,D})^2 |GS \rangle$
are summarized in Table \ref{Tab:Spin^c_Cobordism}. 
To compare with SPT phases with on-site $\Z_n$ symmetry, we list the $\spin^c$
cobordisms
in the rightmost column of Table~\ref{Tab:Spin^c_Cobordism}.

It is interesting to note that, from Table~\ref{Tab:Spin^c_Cobordism},
the partial $(\wt C_n)^2$ rotation provides less information than the partial $\wt C_n$ rotation. 
Except for $n=2,6$, the $U(1)$ phase of the partial $(\wt C_n)^2$ rotation is twice that of the partial $\wt C_n$ rotation. 
As an example,
let us focus on the $C_3$ rotation symmetry.
  The $\spin^c$ cobordism with $\Z_3$ on-site symmetry,
  $\Omega^{\spin^c}(B \mathbb{Z}_3)= \Z_3 \oplus \Z_3$, 
  suggests the existence of
two inequivalent/independent SPT phases
which are generated by Hamiltonians
(\ref{eq:2d_nonchiral_chern}) with $p=1$ and $2$, respectively. 
In general,
in order to confirm that the classification of SPT phases is given by two or more Abelian groups (e.g., $\Z_p \oplus \Z_q$),
multiple many-body invariants are needed.
(I.e., multiple many-body invariants are required to distinguish all possible SPT phases.) 
However, the partial $\wt C_3$ and $(\wt C_3)^2$ rotations give rise to the same $U(1)$
phases for ground states of both Hamiltonians $H_{1-t}$ and $H_{1-t^2}$, which
implies that the partial $\wt C_3$ and $(\wt C_3)^2$ rotations cannot
distinguish these two series of SPT phases.
The ground state expectation value of the {\it full} $\wt C_3$
rotation~\cite{PhysRevB.86.115112} is a candidate
for the manybody topological invariant to differentiate these two groups of SPT phases.

  More generically, 
for even $n$, the expectation value of the partial $\widetilde C_n$ rotations gives rise to the $U(1)$ phases which are consistent with the cobordism classification of fermionic SPT phases protected by the on-site $\Z_n$ symmetry. 
On the other hand, for odd $n$, there is a mismatch between partial $\widetilde C_n$ rotations and cobordism groups: 
The partial $\wt C_n$ rotation gives at least one $\Z_{2n}$ SPT phase
whereas $\Omega^{\spin^c}_{3}(B \Z_n)=\Z_{n} \oplus \Z_{n}$.
This mismatch suggests that either one of the following statements can be true: 
(i)
The SPT phase protected by $C_n$ rotation symmetry for odd $n$ includes a
$\Z_{2n}$ phase, which differs
from the SPT phase protected by {\it on-site} $\Z_n$ symmetry. 
I.e., the statement/conjecture made in the end of Sec.\ \ref{Partial point group operation and spacetime path-integral}
is not correct in this case. 
(ii)
The $\Z_{2n}$ $U(1)$ phases associated with the partial $\wt C_n$ rotation for
odd $n$ is not stable; that is, it reduces to the $\Z_{n}$ $U(1)$ phase under perturbations and/or disorder. 
We wish to clarify this point in a future work.

\begin{table}
\renewcommand{\arraystretch}{1.5}
\begin{tabular}{c|cc|cc|c}
& \multicolumn{2}{c|}{Partial $\wt C_n$ transformation}  & \multicolumn{2}{c|}{The partial $(\wt C_n)^2$ transformation}  & Cobordism \\
 & $H_{1-t}$ & $H_{1-t^2}$ & $H_{1-t}$ & $H_{1-t^2}$ & $\Omega_3^{{\rm Spin}^c}(B \Z_n)$ \\
\hline
$n=2$ & $\frac{\pi}{2}$ & 0 & $0$ & $0$ & $\Z_4[1-t]$ \\
$n=3$ & $\frac{\pi}{3}$ & $\frac{\pi}{3}$ & $\frac{2 \pi}{3}$ & $\frac{2 \pi}{3}$ & $\Z_3[1-t] \oplus \Z_3[1-t^2]$ \\
$n=4$ & $\frac{\pi}{4}$ & $\pi$ & $\frac{\pi}{2}$ & 0 & $\Z_8[1-t] \oplus \Z_2[1-t^2]$ \\
$n=5$ & $\frac{\pi}{5}$ & $\frac{4\pi}{5}$ & $\frac{2 \pi}{5}$ & $\frac{8 \pi}{5}$ & $\Z_5[1-t] \oplus \Z_5[1-t^2]$ \\
$n=6$ & $\frac{\pi}{6}$ & $\frac{2\pi}{3}$ & $\frac{\pi}{3}$ & $\frac{\pi}{3}$ & $\Z_{12}[1-t] \oplus \Z_3[1-t^2]$ \\
$n=7$ & $\frac{\pi}{7}$ & $\frac{4\pi}{7}$ & $\frac{2 \pi}{7}$ & $\frac{8 \pi}{7}$ & $\Z_{7}[1-t] \oplus \Z_7[1-t^2]$ \\
$n=8$ & $\frac{\pi}{8}$ & $\frac{\pi}{2}$ & $\frac{\pi}{4}$ & $\pi$ & $\Z_{16}[1-t] \oplus \Z_4[1-t^2]$ \\
$n=9$ & $\frac{\pi}{9}$ & $\frac{4\pi}{9}$ & $\frac{2 \pi}{9}$ & $\frac{8 \pi}{9}$ & $\Z_{9}[1-t] \oplus \Z_9[1-t^2]$ \\
\end{tabular}
\renewcommand{\arraystretch}{1}
	\caption{
	\label{Tab:Spin^c_Cobordism}
	The 2nd and 3rd (4th and 5th) columns show the $U(1)$ complex phases of the expectation 
	value of the partial $\widetilde C_n$ ($(\wt C_n)^2$) rotation $\braket{GS | \widetilde C_n | GS}$ ($\braket{GS | (\widetilde C_n)^2 | GS}$) on the 
	Hamiltonian (\ref{eq:2d_nonchiral_chern}) with $p=1$ and $2$, respectively. 
	The rightmost column show the $\spinc$ cobordsim group classification of SPT phases with on-site $\Z_n$ symmetry. 
	The notation $\Z_q[1-t^p]$ means that the $\spin^c$ cobordism group 
	$\Omega_3^{{\rm Spin}^c}(B \Z_n)$ 
	consists of $\Z_q$ groups generated by the Hamiltonian $H_{1-t^p}$ 
	defined in (\ref{eq:2d_nonchiral_chern}). 
	}
\end{table}

\subsection{$(2+1)d$ nonchiral superconductors with $C_n$ rotation symmetry ($n \geq 3$)}
\label{sec:(2+1)D+Cn_nonchiral}

Here, we briefly comment on
how to construct $n$-fold rotation symmetry in 
non-chiral superconductors for $n \geq 3$. 
To this end, let us first consider 
how to realize an on-site $\Z_n$ symmetry 
in real fermion systems. 
Since a single-component real fermion field
does not have the $U(1)$ phase degree of freedom, 
the $\Z_n$ symmetry on the one-component real fermion cannot be introduced
except for $n=2$. 
To define $\Z_n$ symmetry on real fermions, 
it is necessary to 
introduce a complex fermion 
operator $\phi(\bx)$ consisting of two real fermion
operators 
$\chi(\bx), \eta(\bx)$ as 
\begin{align}
\phi^{\dag}(\bx) = \chi(\bx) + i \eta(\bx), 
\quad 
\phi(\bx) = \chi(\bx) - i \eta(\bx). 
\end{align}
The on-site $\Z_n$ symmetry is now defined 
such that the complex fermion operator 
transforms as a 1-dimensional representation of 
$\Z_n$, 
$U \phi^{\dag}(\bx) U^{-1} = e^{-2 \pi i p/n} \phi^{\dag}(\bx)$, $p=0, \dots, n-1$, 
which is equivalent to 
in terms of the real fermions,
\begin{align}
U \chi(\bx) U^{-1} = \cos \frac{2 \pi p}{n} \chi(\bx) + \sin \frac{2 \pi p}{n} \eta(\bx), \quad
U \eta(\bx) U^{-1} = -\sin \frac{2 \pi p}{n} \chi(\bx) + \cos \frac{2 \pi p}{n} \eta(\bx). 
\end{align}
As the combined transformation
$\tilde{C}_2$
in Sec.\ \ref{Sec:4-4-2}, 
the on-site $\mathbb{Z}_n$ phase rotation can 
be combined with $C_n$ rotation to define 
$\widetilde C_n$ rotation. 
The computation of the ground state expectation 
value of the partial $\widetilde C_n$ rotation is recast into 
the calculation presented in Sec.\ \ref{Sec:4-4-2}.

\section{Partial inversions}
\label{Sec:5}

\begin{figure}[!]
 \begin{center}
  \includegraphics[width=0.4\linewidth, trim=0cm 0cm 0cm 0cm]{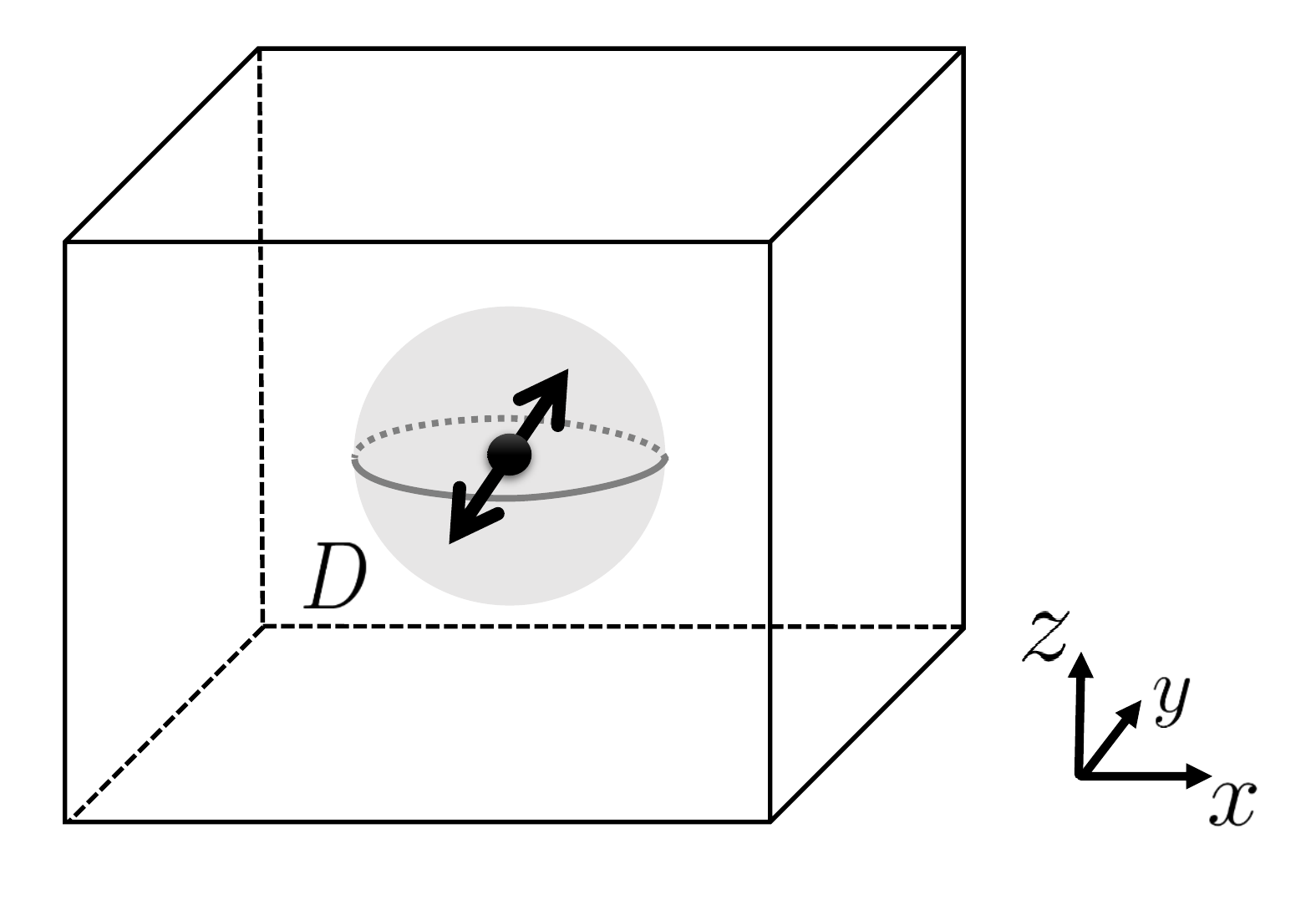}
 \end{center}
 \caption{Partial inversion on a ground state on 3d space torus $T^3$. 
The partial inversion transformation is performed only on inside of the 3-ball $D$
(the shadow region).}
 \label{Fig:RP4}
\end{figure}

There are SPT phases in $(3+1)d$, 
which are protected by orientation-reversing symmetry,  
and 
the generating manifold of the relevant cobordism group 
is the $4d$ real projective space, $\R P^4$. 
For example, 
$(3+1)d$ topological superconductors with inversion/reflection symmetry, 
which are  
the CPT dual of class DIII time-reversal symmetric 
topological superconductors, 
are 
classified by 
the $\pinp$ cobordism group,
$\Omega^{{\rm Pin}^+}_4(pt) = \Z_{16}$.
The abelian group 
$\Z_{16}$ is generated by $\R P^4$.~\cite{Kirby, Kapustin2015a}
In this section,
given a ground state wave function and symmetry actions, 
we aim at directly computing 
the many-body topological invariant associated to $\R P^4$. 
$\R P^4$ is not a mapping torus, 
and hence we need to employ a partial symmetry operation 
similar to partial reflection introduced in Sec.\ \ref{Sec:3}. 
Topologically, $\R P^4$ is realized by inserting a cross-cap in $S^4$. 
The path-integral on $\R P^4$ 
is expected to be simulated 
by considering an expectation value of the 
partial inversion operator $I_D$
defined for a subregion $D$,
which is a three-ball of the total system
(Fig.~\ref{Fig:RP4}).

Taking $(3+1)d$ topological superconductors
and insulators as an example, 
we will demonstrate below 
that the expectation value of
the partial inversion
correctly reproduces the 
known topological classification. 
We will evaluate the expectation value of
partial inversion both numerically and analytically. 
For our analytical calculations, 
we will again make use 
of the cut and glue construction:~\cite{Qi2011b}
We assume that 
the reduced density matrix 
for the 3-ball $D$,
obtained by taking the 
partial trace $\mathrm{Tr}_{\bar{D}}$
for the complimentary region $\bar{D}=T^3\backslash D$
of the pure state $|GS\rangle\langle GS|$, 
is given 
approximately 
by the canonical thermal density matrix of a gapless theory 
(CFT) 
realized on the boundary $S^2 = \p D$.
The fictitious temperature is determined by 
the correlation length $\xi$ of the bulk.
Namely, 
\begin{align}
\rho_{D} 
= {\rm Tr}_{T^3 \backslash D} \big[ \ket{GS} \bra{GS} \big] 
\sim \frac{ e^{- \frac{\xi}{v} H_{S^2}} }{{\rm Tr}\big[ e^{- \frac{\xi}{v} H_{S^2}} \big]}, 
\end{align}
where $v$ is a velocity of gapless theory on $S^2$. 
The gapless theory is defined on the spacetime manifold $S^2(R) \times S^1(\xi/v)$  
where $R$ is the radius of the 3-ball $D$. 
We assume,
for simplicity, 
that the Hamiltonian $H_{S^2}$ is rotation symmetric,
and exclude the possibility of surface topological order.
\cite{vishwanath2013physics} 
The expectation value of the partial inversion is given 
in terms of the gapless surface 
theory 
as the expectation value of an antipodal map $I_{S^2}$ on $S^2$: 
\begin{align}
\braket{GS | I_D | GS}
\sim \frac{ {\rm Tr} \big[ I_{S^2} e^{- \frac{\xi}{v} H_{S^2}} \big] }{{\rm Tr}\big[ e^{- \frac{\xi}{v} H_{S^2}} \big]}, 
\qquad  
I_{S^2} : (\theta,\phi) \mapsto (\pi - \theta, \phi + \pi), 
\label{eq:3d_pi_def}
\end{align}
where $(\theta, \phi)$ is the polar coordinates of $S^2$. 
We will be interested in the behavior of
(\ref{eq:3d_pi_def}) 
for sufficiently large $R$,   $R \gg \xi$.

In this section, we deal with two examples of free theories: 
class D superconductors with inversion symmetry 
and class A insulators with inversion symmetry. 
It will turn out that the surface CFT calculations of partial inversions indeed provide  
$\Z_{16}$ and $\Z_{8}$ topological invariants.~\cite{
Wang2014, you2014symmetry, Kapustin2015a, Morimoto2015, Metlitski2015, Freed2016} 
We also show numerical calculations for lattice models, in which the results are  
consistent with the calculation of the surface CFTs. 
A generalization to higher spacetime dimensions will be discussed at last.

\subsection{$(3+1)d$ superconductors with inversion symmetry 
($\Omega_4^{\pinp}(pt)=\mathbb{Z}_{16}$)
}
\label{Sec:5-1}
Let us consider $(3+1)d$ topological superconductors
protected by inversion symmetry $I$ with $I^2 = (-1)^F$. 
The topological classification is given by the $\pinp$ cobordism group, 
$\Omega^{\pinp}_4(pt) = \Z_{16}$. 
Notice that the $\pi$ rotation $C_{\pi}$ of the real fermions are associated with 
$\pm i$ phase as shown in (\ref{eq:2d_px-ipy_rot_sym}), which implies 
that the inversion transformation $I=C_{\pi}R$ with $I^2 = (-1)^F$ in $3$-space dimensions 
is equivalent to the reflection transformation with $R^2 = 1$, that is, the $\pinp$ structure.
The generating manifold is $\R P^4$.~\cite{you2014symmetry, Kapustin2015a, Kirby}
A convenient model Hamiltonian,
which describes 
the ${}^3$He-B phase,
is given by 
\begin{align}
H = \sum_{\bk} \Psi^{\dag}(\bk) \Big[ \big( \frac{k^2}{2m} - \mu \big) \tau_z + \Delta \tau_x \bm{\sigma} \cdot \bk \Big] \Psi(\bk), 
\quad  
\Psi(\bk) = (\psi_{\uparrow}(\bk), \psi_{\downarrow}(\bk), \psi^{\dag}_{\downarrow}(-\bk), -\psi^{\dag}_{\uparrow}(-\bk))^T. 
\label{Eq:3D_SC_Model}
\end{align}
The model is invariant under inversion 
defined by 
\begin{align}
I \psi_{\sigma}^{\dag}(\bx) I^{-1} = i \psi_{\sigma}^{\dag}(-\bx), 
\quad  
(\sigma = \uparrow, \downarrow). 
\label{Eq:5-1-a}
\end{align}


To compute the expectation value of partial inversion 
using the cut and glue construction, 
we first look for the effective surface theory 
on the boundary of the $3$-ball.~\cite{imura2012spherical}
We consider the Hamiltonian (\ref{Eq:3D_SC_Model}) on the 
open $3$-ball with radius $R$. 
We introduce a polar coordinate 
$(x,y,z) = (r \sin \theta \cos \phi, r \sin \theta \sin \phi, r \cos \theta)$. 
Instead of specifying a boundary condition, 
we consider the following Jackiw-Rebbi type domain wall one-particle Hamiltonian 
\begin{align}
{\cal H} = -i \Delta \tau_x (\sigma_x \partial_x + \sigma_y \partial_y + \sigma_z \partial_z) + \mu(r) \tau_z. 
\end{align}
with $\mu(r)<0$ for $r<R$ and $\mu(r)>0$ for $r>R$. 
From a straightforward calculation 
(Appendix \ref{App:Der_Surf_(3+1)d_D}), 
we obtain
the explicit form of the complex fermion operators $\gamma^{\dag}(\theta, \phi)$ 
creating gapless surface excitations
\begin{align}
\label{Eq:5-1-b1}
\gamma^{\dag}(\theta,\phi) 
& \sim \Big[ 
-e^{-i \frac{\phi}{2}} \sin \frac{\theta}{2} \big\{ i \psi^{\dag}_{\uparrow}(r,\theta,\phi) + \psi_{\downarrow}(r,\theta,\phi) \big\} 
+ e^{i \frac{\phi}{2}} \cos \frac{\theta}{2} \big\{ i \psi^{\dag}_{\downarrow}(r,\theta,\phi) - \psi_{\uparrow}(r,\theta,\phi) \big\}
\Big] e^{- \int^r  \frac{\mu(r')}{\Delta} dr'}, 
\nonumber \\
\gamma(\theta,\phi) 
& \sim \Big[ 
-e^{-i \frac{\phi}{2}} \cos \frac{\theta}{2} \big\{ \psi^{\dag}_{\uparrow}(r,\theta,\phi) + i \psi_{\downarrow}(r,\theta,\phi) \big\} 
-e^{i \frac{\phi}{2}} \sin \frac{\theta}{2} \big\{ \psi^{\dag}_{\downarrow}(r,\theta,\phi) - i \psi_{\uparrow}(r,\theta,\phi) \big\}
\Big] e^{- \int^r \frac{\mu(r')}{\Delta} dr'}. 
\end{align}
Notice that the anti-periodic boundary condition 
in the $\phi$ direction, 
$\gamma^{\dag}(\theta,\phi + 2 \pi ) = -\gamma^{\dag}(\theta,\phi)$, 
is satisfied.
In terms of these fermion operators, 
the effective surface BdG Hamiltonian is given by 
\begin{align}
\label{Eq:5-1-b2}
&H_{S^2}
=\int \sin \theta d \theta d \phi 
\big( \gamma^{\dag}(\theta,\phi), -\gamma(\theta,\phi) \big) 
{\cal H}
\begin{pmatrix}
\gamma(\theta,\phi) \\
-\gamma^{\dag}(\theta,\phi) \\
\end{pmatrix}, 
\nonumber \\ 	
&{\cal H} = \frac{\Delta}{R} 
\begin{pmatrix}
0 & -i \partial_{\theta} - \frac{1}{\sin \theta} \partial_{\phi} - \frac{i \cot \theta}{2} \\
-i \partial_{\theta} + \frac{1}{\sin \theta} \partial_{\phi} - \frac{i \cot \theta}{2} & 0 \\
\end{pmatrix}. 
\end{align}

The one-particle Hamiltonian ${\cal H}$ is solved by using the monopole harmonics $Y_{l,m}^g(\theta,\phi)$ with $g=\pm \frac{1}{2}$ 
(see, for example, Ref. \onlinecite{Hasebe2015}):
\begin{align}
&
{\cal H} \Psi_{\pm n,m}(\theta,\phi) = \pm \frac{n \Delta}{R} \Psi_{\pm n,m}(\theta,\phi), 
\qquad 
\Psi_{\pm n,m}(\theta,\phi) = \frac{1}{\sqrt{2}} \begin{pmatrix}
Y^{-\frac{1}{2}}_{n-\frac{1}{2},m}(\theta,\phi) \\
\mp i Y^{\frac{1}{2}}_{n-\frac{1}{2},m}(\theta,\phi) \\
\end{pmatrix}, 
\nonumber\\
& n=1,2,\dots, 
\qquad 
m = 
-(n-\frac{1}{2}),  \dots, 
n-\frac{3}{2}, 
n-\frac{1}{2}.
\end{align}
There is no zero mode, which is consistent with the absence of a monopole inside of the 
3-ball $D$. 
The degeneracy of 
the states with the eigenvalue 
$\pm \frac{n \Delta}{R}$ is $2 |n|$. 
The explicit form of the monopole harmonics is \cite{Hasebe2015}
\begin{align}
Y^{g}_{l,m}(\theta,\phi)
&= 2^m \sqrt{\frac{(2l+1)(l-m)!(l+m)!}{4 \pi (l-g)! (l+g)!}} (\sin \frac{\theta}{2})^{-(m+g)} (\cos \frac{\theta}{2})^{-(m-g)} P^{(-m-g, -m+g)}_{l+m}(\cos \theta) e^{i m \phi} 
\nonumber\\
&= (-1)^{l+m} \sqrt{\frac{(2l+1)(l-m)!(l+m)!}{4 \pi (l-g)! (l+g)!}} e^{i m \phi} \sum_n (-1)^n \begin{pmatrix}
l-g \\
n \\
\end{pmatrix}
\begin{pmatrix}
l+g \\
g-m+n \\
\end{pmatrix}
(\sin \frac{\theta}{2})^{2l-2n-g+m} (\cos \frac{\theta}{2})^{2n+g-m}
\end{align}
where $P_n^{\alpha,\beta}(x)$ is the Jacobi polynominals and sum $\sum_n$ runs over all 
possible integers. 
The periodicities of eigenstates agree with anti-periodic boundary condition of 
$\gamma(\theta,\phi)$, 
$\Psi_{\pm n,m}(\theta,\phi+2 \pi) = - \Psi_{\pm n,m}(\theta,\phi)$. 
In terms of these eigen functions, we introduce the Bogoliubov operators $\chi^{\dag}_{n,m}$ for the positive energy states 
by 
\begin{align}
\chi^{\dag}_{n,m} := \int \sin \theta d \theta d \phi \Big[ 
\frac{1}{\sqrt{2}} Y^{- \frac{1}{2}}_{n-\frac{1}{2}, m}(\theta,\phi) \gamma^{\dag}(\theta,\phi) 
+ i \frac{1}{\sqrt{2}} Y^{\frac{1}{2}}_{n-\frac{1}{2}, m}(\theta,\phi) \gamma(\theta,\phi) \Big], 
\quad (n>0). 
\label{Eq:5-1-c}
\end{align}
In terms of 
the Bogoliubov operators, 
the Hamiltonian $H_{S^2}$ can be written as 
\begin{align}
H_{S^2} = \frac{\Delta}{R} 
\sum_{n\in \Z,n>0} 
\sum_{ 
m= 
-(n-\frac{1}{2}), 
\cdots,
n-\frac{3}{2}, 
n-\frac{1}{2}
} 
n \chi^{\dag}_{n,m} \chi_{n,m}.  
\label{Eq:Surf_Ham_SC}
\end{align}

\subsubsection{Partial inversion}
\label{Partial inversion}
To compute the partial inversion on the surface theory (\ref{Eq:Surf_Ham_SC}), 
we first derive the antipodal transformation $I_{S^2}$ on the Bogoliubov operators $\chi_{n,m}$. 
It is induced by the inversion transformation in 
the bulk (\ref{Eq:5-1-a}) through 
(\ref{Eq:5-1-b1}),  (\ref{Eq:5-1-b2}), and (\ref{Eq:5-1-c}) as 
\begin{align}
I_{S^2} \chi^{\dag}_{n,m} I^{-1}_{S^2} = i (-1)^n \chi^{\dag}_{n,m}, \quad 
I_{S^2} \chi_{n,m} I_{S^2}^{-1} = -i (-1)^n \chi_{n,m}, 
\quad (n>0). 
\end{align}
Here, we have used 
\begin{align}
&I \gamma^{\dag}(\theta,\phi) I^{-1} = -i \gamma(\pi-\theta,\phi+\pi), && 
I \gamma(\theta,\phi) I^{-1} = i \gamma^{\dag}(\pi-\theta,\phi+\pi), 
\nonumber \\
&Y^{\frac{1}{2}}_{n-\frac{1}{2}, m}(\pi-\theta,\phi+\pi)
= (-1)^{n} i Y^{-\frac{1}{2}}_{n-\frac{1}{2}, m}(\theta,\phi) , &&
Y^{-\frac{1}{2}}_{n-\frac{1}{2}, m}(\pi-\theta,\phi+\pi)
= (-1)^{n} i Y^{\frac{1}{2}}_{n-\frac{1}{2}, m}(\theta,\phi). 
\label{Eq:Inversion_Monopole_Hram}
\end{align}

As explained around (\ref{eq:3d_pi_def}), 
the expectation value of the partial inversion is given by that of the antipodal map 
within the surface theory: 
\begin{align}
\braket{GS | I_D | GS}
\sim 
\frac{{\rm Tr} \big[ I_{S^2} e^{- \frac{\xi}{\Delta} H_{S^2}} \big] }{{\rm Tr} \big[ e^{- \frac{\xi}{\Delta} H_{S^2}} \big] }
= \frac{\prod_{n=1}^{\infty} (1+i (-q)^n)^{2n}}{\prod_{n=1}^{\infty} (1+q^n)^{2n}}, 
\quad 
q=e^{- \frac{\xi}{R}}. 
\label{Eq:5_PI_SC}
\end{align}
Here we normalized the antipodal transformation $I_{S^2}$ so that 
$I_{S^2} \ket{0_{\chi}} = \ket{0_{\chi}}$ where 
$\ket{0_{\chi}}$ is the Fock vacuum of $\chi_{n,m}$ fermions. 

Equation (\ref{Eq:5_PI_SC}) can be evaluated in the same way as in 
Ref.\ \onlinecite{Cardy1991}. 
Here we briefly sketch the method. 
What we want to compute are 
\begin{align}
I_1(q) = \sum_{n=1}^{\infty} n \ln (1+q^n), 
\quad 
I_2(q) = \sum_{n=1}^{\infty} n \ln (1+i (-q)^n)
\end{align}
for $\delta \ll 1$ with $q=e^{-\delta}$. 
In terms of $I_{1,2}(q)$, the partial inversion is given by 
\begin{align}
\braket{GS | I_D | GS} \sim \exp \Big[ 2 I_2(e^{- \frac{\xi}{R}}) - 2 I_1(e^{- \frac{\xi}{R}}) \Big]. 
\end{align}
By using the Cahen-Mellin integral 
\begin{align}
e^{-y} = \frac{1}{2 \pi i} \int_{c-i \infty}^{c+i \infty} y^{-s} \Gamma(s) d s
\end{align}
for $c>0$, ${\rm Re}(y) > 0$ and $y^{-s}$ on the principal branch ($\Gamma(s)$ is the Gamma function), 
$I_1(q)$ is written as 
\begin{align}
I_1(q = e^{- \delta})
&= - \sum_n n \sum_{r=1}^{\infty} r^{-1} (-1)^r e^{- \delta n r} 
\nonumber \\
&= - \sum_n n \sum_{r=1}^{\infty} r^{-1} (-1)^r \frac{1}{2 \pi i} \int_{c- i \infty}^{c+i \infty} ds (\delta n r)^{-s} \Gamma(s) 
\quad (c \gg 0)
\nonumber \\
&= \frac{1}{2 \pi i} \int_{c- i \infty}^{c+i \infty} ds \delta^{-s} \Big[ \Gamma(s) \zeta(s-1) (1-2^{-s}) \zeta(s+1) \Big], 
\end{align}
where $c$ is sufficiently far to the right. 
$\zeta(s) = \sum_{n=1}^{\infty} n^{-s}$ is the Riemann zeta function. 
From the contour integral, one finds 
the contributions from the poles of the integrand 
$F(s) = 
\Gamma(s) \zeta(s-1) (1-2^{-s}) \zeta(s+1)$ 
as 
\begin{align}
I_1(q = e^{- \delta}) = \sum_{s \in {\rm poles}} \frac{{\rm Res}(F,s)}{\delta^s}.
\end{align}
(There is no multiple pole in this case.)
To estimate this for small $\delta >0$, 
it is sufficient to include only poles with ${\rm Re}(s) \geq 0$. 
Furthermore, the pole at $s=0$ is scale independent, 
i.e., it is a topological contribution. 
There are two simple poles at $s=0$ and $s=2$. 
(Recall that $\zeta(s)$ has a single pole at $s=1$ with residue $1$, $\Gamma(s)$ has 
single poles at integers $n \leq 0$ with residue $\frac{(-1)^n}{n!}$, 
and $\zeta(s)$ has zeros at negative even integers. 
One of the poles at $s=0$ is canceled by zero of $(1-2^{-s})$.) 
One can show 
\begin{align}
I_1(q=e^{- \delta}) 
= \frac{3}{4} \zeta(3) \delta^{-2} -\frac{1}{12} \ln(2) + \frac{\delta^2}{960} +  \cdots.
\end{align}
Similarly, we obtain (see Appendix \ref{App:I(q,z)})
\begin{align}
I_2(q=e^{- \delta}) 
&= \frac{1}{2 \pi i} \int_{c- i \infty}^{c+i \infty} ds \delta^{-s} \Big[ 
-2^{1-s} \Gamma(s) \zeta(s-1) {\rm Li}_{s+1}(-i)
- (1-2^{1-s}) \Gamma(s) \zeta(s-1) {\rm Li}_{s+1}(i)
\Big] 
\nonumber \\
&= \frac{3}{32} \zeta(3) \delta^{-2} - \frac{\pi i}{16} - \frac{1}{24} \ln(2) + \frac{\delta^2}{480} + \cdots. 
\end{align}
Here, ${\rm Li}_s(z) := \sum_{n=1}^{\infty} \frac{z^n}{n^s}$ is the polylogarithm function. 
Let us consider the imaginary part of $I_2(q)$, 
\begin{align}
{\rm Im} \big[ I_2(q=e^{- \delta}) \big] 
&= \frac{1}{2 \pi i} \int_{c- i \infty}^{c+i \infty} ds 
\delta^{-s} \Gamma(s) \zeta(s-1) (2^{2-s}-1) \beta(s+1),  
\end{align}
where $\beta(s) = \sum_{n=0}^{\infty} \frac{(-1)^n}{(2n+1)^s}$ is the Dirichlet beta function. 
There is a dramatic cancellation:
Since $\beta(s+1)$ has zeros at even negative integers $s=-2, -4, -6, \dots$, 
all the poles from $\Gamma(s)$ are canceled with zeros! 
Moreover, the zero from $(2^{2-s}-1)$ at $s=2$ is canceled with the pole 
from $\zeta(s-1)$. 
Eventually, there remains only one pole at $s=0$, 
which implies an exactly scale-independent value 
\begin{align}
{\rm Im} \big[ I_2(q=e^{- \delta}) \big] = - \frac{\pi i}{16}. 
\end{align}

Finally, we get the formula of partial inversion 
\begin{align}
&\braket{GS | I_D | GS} = |\braket{GS | I_D | GS}| e^{i \theta_{\rm top}}, 
\quad 
\theta_{\rm top} = - \frac{\pi}{8}, 
 \\ 
&|\braket{GS | I_D | GS}| \sim \exp \Big[ \frac{1}{12} \ln(2) - \frac{21}{16} \zeta(3) \left( \frac{R}{\xi} \right)^2 \Big]. 
\label{Eq:Formula_PI_3D_SC}
\end{align}
The topological $U(1)$ phase $e^{- \frac{\pi i}{8}}$ is 
indeed consistent with 
the cobordism classification $\Omega^{{\rm Pin}^+}_4(pt) = \Z_{16}$. 
Also, observe that in addition to the topological $U(1)$ phase, 
a topological amplitude $e^{\frac{1}{12} \ln(2)}$ appears.

\subsubsection{Numerical results for lattice systems}

In this section, we provide a direct numerical evidence for the partial inversion of the three-dimensional lattice models.
A generating model in class D is given by the BdG Hamiltonian 
\begin{align}
\hat H = \frac{1}{2} \sum_\kv \Psi^{\dag}(\kv) h (\kv) \Psi(\kv), 
\end{align}
on a cubic lattice, where
\begin{align}
h(\kv) 
= \left[-t( \cos k_x + \cos k_y+ \cos k_z)-\mu\right] \tau_z + \Delta \left[ \sin k_x \tau_x \sigma_x + \sin k_y \tau_x \sigma_y+ \sin k_z \tau_x \sigma_z\right].
\end{align}
in which the $\tau$ and $\sigma$ matrices act on particle-hole and spin subspaces, respectively. As mentioned earlier, the above Hamiltonian also describes the ${}^3$He-B phase. 
The inversion symmetry in this model is defined as in Eq.~(\ref{Eq:5-1-a}).
This model exhibits three different topological phases depending on the chemical $\mu$ potential as follows:
\begin{enumerate}
\item $|\mu|<t$: Top. II. This phase supports an even number of 2d gapless Majorana surface states. It is topologically equivalent to a stack of 2d topological superconductors in the same symmetry class.
\item $t<|\mu|<3t$:  Top. I. This phase hosts a 2d gapless Majorana surface states. 
\item $|\mu|>3t$: Trivial. No topological surface states.
\end{enumerate}

\begin{figure}
\centering
  \includegraphics[scale=.5]{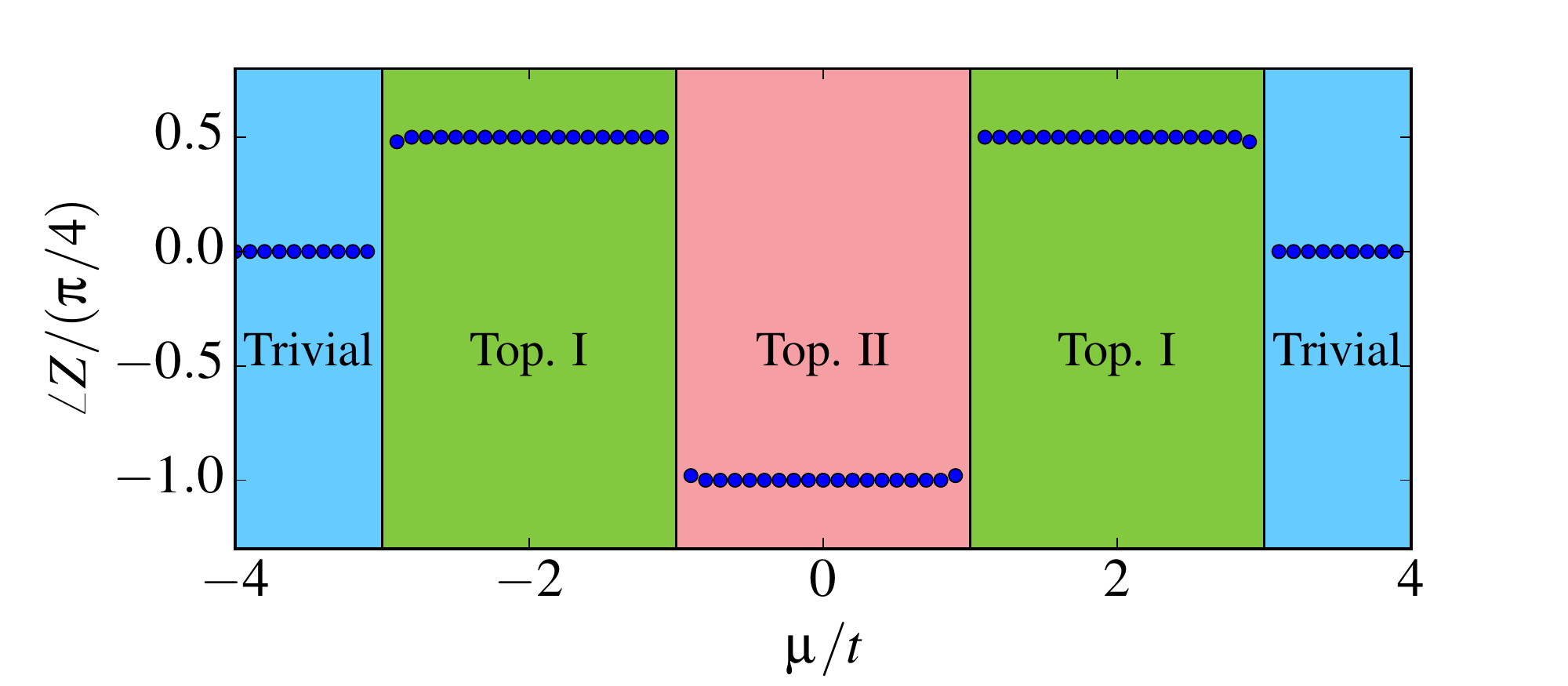}
\caption{\label{fig:3Dsc} (Color online) Complex phase of the partial inversion
  $\angle Z= {\rm Im} \ln \braket{GS | I_D | GS}$ computed
for 3D inversion symmetric topological superconductor (class D). Top. I (II) corresponds to the phase with odd (even) number of gapless Majorana surface states. Here, we set $t=\Delta$. The size of total system and  subsystem are $N=12^3$ and $N_\text{part}=6^3$, respectively.}
\end{figure}

Figure~\ref{fig:3Dsc} shows the calculated complex phases 
$\angle Z = {\rm Im} \ln \braket{GS | I_{D} | GS}$ of the partial inversion for
various values of $\mu$. 
This quantity is computed in a similar fashion to the two-dimensional
case that is to calculate the inner product
$Z=\bra{GS}I_D\ket{GS}$ after rearranging the lattice sites in
the subsystem (to get $I_D\ket{GS}$). Remarkably, the partial inversion gives
the correct $\mathbb{Z}_{16}$ and $\mathbb{Z}_{8}$ phases in the topological
phases characterized by odd and even number of gapless Majorana surface modes,
respectively. We should note that the latter case is topologically equivalent to
stacking two dimensional reflection symmetric class D SPT layers which obey a
$\mathbb{Z}_8$ classification.
 
\subsubsection{Other partial symmetry operations}

In addition to partial inversion,
we now discuss 
three additional partial symmetry operations:
partial fermion parity, 
partial reflection,
and 
partial $\pi$-rotation.
In the partial fermion parity flip and the partial reflection, 
we have a logarithmic term $\log (R/\xi)$ as will be shown in Eq.\ (\ref{result partial fermion parity 4d}). 
The partial $\pi$-rotation gives rise to a scale-independent contribution (\ref{eq:result_partial_pi_rotation_3d_sc}). 

\paragraph{Partial fermion parity}

Following our discussion on 
$(2+1)d$ topological superconductors in Sec.\ \ref{Sec:2D_PFP}, 
let us consider the partial fermion parity flip on $(3+1)d$ topological superconductors. 
Since the partial fermion parity flip induces the fermion parity flip 
on the surface fermions as $(-1)^F \chi_{n,m} (-1)^F = - \chi_{n,m}$, 
we have 
\begin{align}
\braket{GS | (-1)^F_D | GS} 
\sim \frac{{\rm Tr} \big[ (-1)^F e^{- \frac{\xi}{\Delta} H_{S^2}} \big] }{{\rm Tr} \big[ e^{- \frac{\xi}{\Delta} H_{S^2}} \big] }
= \frac{\prod_{n=1}^{\infty} (1-q^n)^{2n}}{\prod_{n=1}^{\infty} (1+q^n)^{2n}} 
= \exp \Big[ 2 I_3(q=e^{- \frac{\xi}{R}}) - 2 I_1(q= e^{-\frac{\xi}{R}}) \Big] 
\label{eq:3dsc_fermion_parity}
\end{align}
with 
\begin{align}
I_3(q = e^{- \delta}) 
&= \sum_{n=1}^{\infty} n \ln (1-q^n) 
= - \frac{1}{2 \pi i} \int_{c- i \infty}^{c+i \infty} ds \delta^{-s} 
\Big[ \Gamma(s) \zeta(s-1) \zeta(s+1) \Big]. 
\end{align}
The integrand has a double pole at $s=0$, 
which leads to an algebraic power law for $\delta = \frac{\xi}{R} \sim 0$, 
\begin{align}
  \label{530}
I_3(q=e^{- \delta}) \sim - \zeta(3) \delta^{-2} - \frac{1}{12} \ln (\delta) - \frac{1}{12} + \ln (A), 
\end{align}
where $A \cong 1.2824\dots$ is the Glaisher-Kinkelin constant. 
Finally, we obtain 
the following formula for the 
expectation value of the partial fermion parity 
in the $(3+1)d$ topological superconductor 
\begin{align}
\braket{GS | (-1)^F_D | GS} \sim 
\exp \Big[ -\frac{1}{6} + \frac{1}{6} \ln(2) + 2 \ln(A) + \frac{1}{6} \ln \left( \frac{R}{\xi} \right) - \frac{7}{4} \zeta(3) \left( \frac{R}{\xi} \right)^2  \Big].
\label{result partial fermion parity 4d}
\end{align}
It should be noted that the logarithmic term appears in addition to the area law term. 
A similar logarithmic contribution to the the entanglement entropy in gapped phases is discussed 
in Ref.~\onlinecite{GroverTurnerVishwanath2011}, 
where the authors pointed out that the non-flatness of the curvature of the boundary 
of a region $D$ is a necessary condition to give a constant part of the entanglement entropy.

\paragraph{Partial reflection}

The topological classification of 
$(3+1)d$ superconductors 
with reflection symmetry
with $R^2 = 1$ 
is the same 
as 
the topological classification of 
inversion symmetric superconductors discussed in 
Sec.~\ref{Sec:5-1}, 
i.e., it is given by $\Omega_4^{\pinp}(pt)=\Z_{16}$, 
and the generating manifold is $\R P^4$. 
However, it seems difficult to make a 4d analog of the cross-cap leading to $\R P^4$ by using the reflection.  
Let us consider a general partial point group transformation on a subregion $D$ in a 3-space manifold. 
In order to make the resulting 4-manifold free from a singularity, the partial transformation should act freely on the boundary $\p D$. 
In the case of the reflection $x \mapsto -x$, 
only the partial reflection on the subregion in the form of $[-L,L] \times M_2$ 
($M_2$ is a 2-space manifold) meets this condition, which leads to the 4-manifold $\R P^2 \times M_2$, not $\R P^4$. 
Here, we discuss the partial reflection on a $3$-ball $D$. 
Let us consider the reflection symmetry $R_z$ for the model (\ref{Eq:3D_SC_Model}), 
\begin{align}
R_z \psi^{\dag}_i(x,y,z) R_z^{-1} 
= \psi^{\dag}_j(x,y,-z) [\sigma_z]_{ji}. 
\end{align}
By using the following relations
\begin{align}
&R_z \gamma^{\dag}(\theta,\phi) R_z^{-1} 
= i \gamma(\pi-\theta,\phi), && 
R_z \gamma(\theta,\phi) R_z^{-1} = -i \gamma^{\dag}(\pi-\theta,\phi), 
\nonumber \\
&Y^{\frac{1}{2}}_{n-\frac{1}{2}, m}(\pi-\theta,\phi)
= (-1)^{n+\frac{1}{2}-m} Y^{-\frac{1}{2}}_{n-\frac{1}{2}, n-\frac{1}{2} - m}(\theta,\phi) , &&
Y^{-\frac{1}{2}}_{n-\frac{1}{2}, m}(\pi-\theta,\phi)
= (-1)^{n+\frac{1}{2}-m} Y^{\frac{1}{2}}_{n-\frac{1}{2}, n-\frac{1}{2} - m}(\theta,\phi), 
\label{Eq:Reflection_Monopole_Hram}
\end{align}
We see that the Bogoliubov operators 
are transformed as 
\begin{align}
R_z \chi^{\dag}_{n,m} R_z^{-1} = (-1)^{n+\frac{1}{2}-m} \chi^{\dag}_{n,m}, 
\qquad 
R_z \chi_{n,m} R_z^{-1} = (-1)^{n+\frac{1}{2}-m} \chi_{n,m}, 
\quad (n>0).
\end{align}
The half of degenerate energy states labeled by $n$ 
have the negative reflection parity $R_z = -1$. 
Then, the partial reflection on the $3$-ball $D$ is given by 
\begin{align}
\braket{GS | R_{z,D} | GS}
\sim 
\frac{{\rm Tr} \big[ R_z e^{- \frac{\xi}{\Delta} H_{S^2}} \big] }{{\rm Tr} \big[ e^{- \frac{\xi}{\Delta} H_{S^2}} \big] }
= \frac{\prod_{n=1}^{\infty} (1-q^n)^{n}}{\prod_{n=1}^{\infty} (1+q^n)^{n}}, 
\quad 
q=e^{- \frac{\xi}{R}}, 
\label{Eq:5_PR_SC}
\end{align}
which is exactly the square root of the partial fermion parity flip (\ref{eq:3dsc_fermion_parity}). 
This is real positive number and has no information for the $\Z_{16}$ classification.

\paragraph{Partial $\pi$-rotation}
Finally, 
we consider the $\pi$-rotation symmetry of the model (\ref{Eq:3D_SC_Model}), 
\begin{align}
C_z \psi^{\dag}_i(x,y,z) C_z^{-1} 
= - i \psi^{\dag}_j(-x,-y,z) [\sigma_z]_{ji}. 
\end{align}
From the point of view of TQFTs, 
the existence of $\pi$-rotation symmetry is equivalent to the on-site $\Z_2$ 
symmetry, thus we expect the topological classification is given by the 
$\spin$ cobordism $\Omega_4^{\spin}(B \Z_2) = \Z$.~\cite{Kapustin2015a} 
Here, the integer cobordism group means that there is a topological action 
parameterized by $a \in U(1)$ through ${\rm Hom}\big[ \Omega_4^{\spin}(B \Z_2), U(1) \big] = U(1)$. 
$a \in U(1)$ is a material parameter determined by a model Hamiltonian 
similar to axion $\theta$ term in $(3+1)d$ insulators.~\cite{Qi_3DTI}
In the context of SPT phases, 
the topological action parameterized by $U(1)$ does not mean 
the existence of a nontrivial SPT phase since 
all phases labeled by $a \in U(1)$ are adiabatically connected 
unless there is a symmetry fixing $a \in U(1)$ to a discrete value.~\cite{Kapustin2014symmetry}

Here we compute the partial $\pi$-rotation on a 3-ball $D$. 
One can show 
\begin{align}
&C_z \gamma^{\dag}(\theta,\phi) C_z^{-1} 
= \gamma^{\dag}(\theta,\phi+\pi), && 
C_z \gamma(\theta,\phi) C_z^{-1} = \gamma(\theta,\phi+\pi), 
\nonumber \\
&Y^{\frac{1}{2}}_{n-\frac{1}{2}, m}(\pi,\phi+\pi)
= -i (-1)^{m+\frac{1}{2}} Y^{\frac{1}{2}}_{n-\frac{1}{2}, n-\frac{1}{2} - m}(\theta,\phi) , &&
Y^{-\frac{1}{2}}_{n-\frac{1}{2}, m}(\theta,\phi+\pi)
= -i (-1)^{m+\frac{1}{2}} Y^{-\frac{1}{2}}_{n-\frac{1}{2}, n-\frac{1}{2} - m}(\theta,\phi), 
\label{Eq:Rotation_Monopole_Hram}
\end{align}
then, we have 
\begin{align}
C_z \chi^{\dag}_{n,m} C_z^{-1} = i (-1)^{m+\frac{1}{2}} \chi^{\dag}_{n,m}, 
\quad  
C_z \chi_{n,m} C_z^{-1} = i (-1)^{m+\frac{1}{2}} \chi_{n,m}, 
\quad n>0). 
\end{align}
The expectation value of the partial $\pi$-rotation on the $3$-ball $D$ is given by 
\begin{align}
\braket{GS | C_{z,D} | GS}
&\sim 
\frac{{\rm Tr} \big[ C_z e^{- \frac{\xi}{\Delta} H_{S^2}} \big] }{{\rm Tr} \big[ e^{- \frac{\xi}{\Delta} H_{S^2}} \big] }
= \frac{\prod_{n=1}^{\infty} (1 + i q^n)^n (1 - i q^n)^n}{\prod_{n=1}^{\infty} (1+q^n)^{2n}} 
= \exp \Big[ I_1(q=e^{- \frac{2 \xi}{R}}) - 2 I_1(q= e^{-\frac{\xi}{R}}) \Big] 
\nonumber \\
&\sim \exp \Big[ \frac{1}{12} \ln (2) - \frac{21}{16} \zeta(3) \left( \frac{R}{\xi} \right)^2 \Big].  
\label{eq:result_partial_pi_rotation_3d_sc}
\end{align}
Note that this coincides with the amplitude part of the partial inversion (\ref{Eq:Formula_PI_3D_SC}).


\subsection{$(3+1)d$ insulators with inversion symmetry 
($\Omega^{\pinc}_4(pt)=\mathbb{Z}_8\oplus \mathbb{Z}_2$)
}
\label{Sec:5-2}

In this section,
we consider $(3+1)d$ topological insulators protected by inversion symmetry. 
The topological classification is given by the $\pinc$ cobordism \cite{Gilkey}
\begin{align}
\Omega^{{\rm Pin}^c}_4(pt) = \Z_8 \oplus \Z_2. 
\end{align}
The latter direct summand $\Z_2$ arises from
bosonic SPT phases corresponding to one of $\Z_2$ of the unoriented cobordism group 
$\Omega^{O}_4(pt) = \Z_2 \oplus \Z_2$.~\cite{Wang2014a}
Our focus here is on the former direct summand $\Z_8$.
This part is generated by $\R P^4$,
and the 
following four-orbital free fermion model 
\begin{align}
H = \sum_{\bk} \psi^{\dag} (\bk) {\cal H}(\bk) \psi(\bk), 
\quad 
\psi(\bk)= \{ \psi_{\tau,\sigma} \}_{\tau,\sigma = 1,2}, 
\nonumber\\
{\cal H}(\bk) = (\frac{k^2}{2m} - \mu) \tau_z + \Delta \tau_x \bk \cdot \bm{\sigma}, 
\quad
(m, \mu, \Delta>0), 
\end{align}
which is equivalent to the two copies of the 
$(3+1)d$ superconductor (\ref{Eq:3D_SC_Model}). 
The inversion symmetry is defined by 
\begin{align} \label{Eq:invTI}
I \psi^{\dag}(\bx) I^{-1} = \psi^{\dag}(-\bx) \tau_z. 
\end{align}
In addition to the inversion symmetry,
there is the $U(1)$ charge conservation symmetry, 
\begin{align} \label{Eq:U1}
U^{\ }_b \psi^{\dag}(\bx) U_b^{-1} = e^{- 2 \pi i b} \psi^{\dag}(\bx). 
\end{align}

As in the case of 
$(3+1)d$ topological superconductors protected by
inversion, 
the expectation value of 
the partial inversion is
a candidate of the $\mathbb{Z}_8$ SPT invariant. 
The $U(1)$ phase in 
the expectation value of 
the partial inversion is
simply twice that of the $(3+1)d$ superconductors.
On the other hand,
in the topological insulator system, 
there is the additional charge $U(1)$ symmetry,
which can be combined with the partial inversion
to introduce 
$
I_D U_{b,D}
$,
where $I_D$ and $U_{b,D}$ is the
partial inversion and partial $U(1)$ transformation
for the 3-ball $D$. 
We will focus on the role of this $U(1)$ twist.

\subsubsection{Partial inversion with $U(1)$ transformation}

We first give an analytical 
evaluation of 
$\langle GS| I_D U_{b,D}|GS\rangle$.
In the same way as in Sec.\ \ref{Sec:5-1}, 
we have a surface entanglement Hamiltonian on $S^2$ 
(see Appendix \ref{App:Der_Surf_(3+1)d_A})
\begin{align}
H_{S^2} 
&= \sum_{n=1}^{\infty} 
\sum_{m= 
-(n-\frac{1}{2})}^{n-\frac{1}{2} } \frac{\Delta n}{R} \big[ \chi^{\dag}_{n,m} \chi_{n,m} + \chi_{-n,m} \chi^{\dag}_{-n,m} \big]. 
\label{Eq:Surf_Ham_Ins}
\end{align}
Here, we have normal-ordered
the fermion operators
with respect to the fermi sea $\Ket{FS}$,
which is defined by
fully occupying states created by $\chi^{\dag}_{-n,m} (n>0)$. 
In general, an entanglement chemical potential $\mu_e$ 
can be added, which is determined by 
the geometries of the ball $D$ and other region. 
Here, for simplicity, we assume $\mu_e = 0$. 
The quasiparticles operators
$\chi^{\dag}_{n,m}$ 
are obtained from
the surface 
Dirac fermion operators 
in the spherical coordinate
$\gamma_{1,2}(\theta,\phi)$ as
\begin{align}
&\chi^{\dag}_{n,m} := \int \sin \theta d \theta d \phi \Big[ 
\frac{1}{\sqrt{2}} Y^{- \frac{1}{2}}_{n-\frac{1}{2}, m}(\theta,\phi) \gamma^{\dag}_1(\theta,\phi) 
- i \frac{1}{\sqrt{2}} Y^{\frac{1}{2}}_{n-\frac{1}{2}, m}(\theta,\phi) \gamma^{\dag}_2(\theta,\phi) \Big], 
\nonumber \\
&\chi^{\dag}_{-n,m} := \int \sin \theta d \theta d \phi \Big[ 
\frac{1}{\sqrt{2}} Y^{- \frac{1}{2}}_{n-\frac{1}{2}, m}(\theta,\phi) \gamma^{\dag}_1(\theta,\phi) 
+ i \frac{1}{\sqrt{2}} Y^{\frac{1}{2}}_{n-\frac{1}{2}, m}(\theta,\phi) \gamma^{\dag}_2(\theta,\phi) \Big], 
\end{align}
where $n>0$.
The surface Dirac fermion operators 
are related to the bulk fermion operators as 
\begin{align}
\gamma^{\dag}_1(\theta,\phi) 
& \sim \Big[ 
-e^{-i \frac{\phi}{2}} \sin \frac{\theta}{2} \big\{ i \psi^{\dag}_{\tau=1,\sigma=1}(r,\theta,\phi) + \psi^{\dag}_{\tau=2,\sigma=1}(r,\theta,\phi) \big\}
\nonumber \\
& \ \ \ \ \ \ + e^{i \frac{\phi}{2}} \cos \frac{\theta}{2} \big\{ i \psi^{\dag}_{\tau=1,\sigma=2}(r,\theta,\phi) + \psi^{\dag}_{\tau=2,\sigma=2}(r,\theta,\phi) \big\}
\Big] e^{- \int^r \frac{m(r')}{v} dr'},
\\
\gamma^{\dag}_2(\theta,\phi) 
& \sim \Big[ 
e^{-i \frac{\phi}{2}} \cos \frac{\theta}{2} \big\{ \psi^{\dag}_{\tau=1,\sigma=1}(r,\theta,\phi) + i \psi^{\dag}_{\tau=2,\sigma=1}(r,\theta,\phi) \big\}
\nonumber \\
& \ \ \ \ \ \ +e^{i \frac{\phi}{2}} \sin \frac{\theta}{2} \big\{ \psi^{\dag}_{\tau=1,\sigma=2}(r,\theta,\phi) + i \psi^{\dag}_{\tau=2,\sigma=2}(r,\theta,\phi) \big\}
\Big] e^{- \int^r \frac{m(r')}{v} dr'}. 
\end{align}

From the transformation law 
\begin{align}
I \gamma_1^{\dag}(\theta,\phi) I^{-1} = \gamma_2^{\dag}(\pi-\theta,\phi+\pi), 
\qquad 
I \gamma_2^{\dag}(\theta,\phi) I^{-1} = -\gamma_1^{\dag}(\pi-\theta,\phi+\pi), 
\end{align}
and Eq. (\ref{Eq:Inversion_Monopole_Hram}), 
we note 
the partial inversion and 
partial $U(1)$ charge transformation 
act on 
the quasiparticle operators $\chi^{\dag}_{n,m}$ as 
\begin{align}
&I \chi^{\dag}_{n,m} I^{-1} = (-1)^n \chi^{\dag}_{n,m}, 
\quad
I \chi^{\dag}_{-n,m} I^{-1} = - (-1)^n \chi^{\dag}_{-n,m}, 
\quad (n>0), 
\nonumber \\
&U_b \chi^{\dag}_{n,m} U_b^{-1} = e^{-2 \pi i b} \chi^{\dag}_{n,m}. 
\end{align}
We fix constant phases associated with the antipodal map $I$ and $U(1)$ transformation 
so that $I \ket{FS} =  mbU_b \ket{FS} = \ket{FS}$. 
Then,
the expectation value of 
the 
combined 
partial inversion with partial $U(1)$ charge transformation 
can be evaluated as 
\begin{align}
\braket{GS | U_b I | GS} 
\sim \frac{{\rm Tr} \Big[ U_b I e^{- \frac{\xi}{\Delta} \widetilde H_{S^2}} \Big] }{{\rm Tr} \Big[ e^{- \frac{\xi}{\Delta} \widetilde H_{S^2}} \Big] } 
= \frac{
\prod_{n=1}^{\infty} (1+e^{- 2 \pi i b} (-q)^n)^{2n} (1-e^{2 \pi i b} (-q)^n)^{2n} }{
\prod_{n=1}^{\infty} (1+q^{n})^{2n} (1+q^{n})^{2n} }. 
\end{align}
From Appendix \ref{App:I(q,z)}, 
the phase and amplitude 
of this expectation value,
$
\braket{GS | U_b I | GS} = |\braket{GS | U_b I | GS} | e^{i \theta_{\rm top}}
$, 
are evaluated as
\begin{align}
\theta_{\rm top} = \left\{\begin{array}{ll}
\frac{\pi}{4} & (0 < b < \frac{1}{2}) \\
-\frac{\pi}{4} & (- \frac{1}{2} < b < 0) \\
\end{array}\right. , 
\label{eq:partial_inversion_A+I_theta}
\end{align}
\begin{align}
|\braket{GS | U_b I | GS} | 
&= \exp \Big[ - \Big( 3 \zeta(3) + \frac{1}{4} \big\{ {\rm Li}_3(e^{4 \pi i b}) + {\rm Li}_3(e^{-4 \pi i b}) \big\} \Big) \Big( \frac{R}{\xi} \Big)^2
- \frac{1}{6} \ln \Big| \frac{\sin (2 \pi b)}{2} \Big| 
\nonumber  \\
&\qquad \qquad 
+ \frac{3+\cos(4 \pi b)}{480 \sin^2 (2 \pi b)} \Big( \frac{\xi}{R} \Big)^2 
+ \frac{125+68 \cos (4 \pi  b)-\cos (8 \pi  b)}{96768 \sin^4(2 \pi b)} \Big( \frac{\xi}{R} \Big)^4
+ \cdots \Big], 
\label{eq:partial_inversion_A+I_amp}
\end{align}
for $e^{-2 \pi i b} \neq \pm 1$. 
Notice that the result of the $U(1)$ phase part $e^{i \theta_{\rm top}}$ is exact, which is 
independent of the scale $\xi/R$. 
The quantized scale-independent $U(1)$ phase
(\ref{eq:partial_inversion_A+I_theta}) is somewhat unexpected from the viewpoint
of $\pinc$ structure. 
In the $\pinc$ structure on $\R P^4$, the holonomy associated to the
$\mathbb{Z}_2$ nontrivial loop threading the cross-cap 
is quantized to $\pm i$. 
However, (\ref{eq:partial_inversion_A+I_theta}) means that even if the holonomy
is not properly chosen to be 
$e^{-2 \pi i b} = \pm i$, 
the $U(1)$ phase of the partial inversion is well quantized. 
This agrees with the numerical calculation (see Fig.~\ref{Fig:TI_vs_b}), 
where the plateau structure of the $U(1)$ phase becomes sharper 
as one increases the sizes of the subsystem. 
As $b$ approaches the ``phase transition'' points $e^{-2 \pi i b} = \pm 1$, 
the higher-order terms proportional to $(\xi/\sin(2 \pi b) R)^{2\ell}$ in (\ref{eq:partial_inversion_A+I_amp}) contribute to the amplitude.
At the points $e^{-2 \pi i b} = \pm 1$, 
there appears a double pole at $s=0$ and it gives an algebraic correction to the amplitude, 
in addition to the area law decay, as 
\begin{align}
&
|\braket{GS | U_{b=0,1/2} I | GS} | \nonumber \\
&= \exp \Big[ - \frac{7 \zeta(3)}{2} \Big( \frac{R}{\xi} \Big)^2
+ \frac{1}{6} \ln \Big( \frac{R}{\xi} \Big) 
+ \frac{12 \ln (A)-1+\ln (2)}{6}
- \frac{1}{720} \Big( \frac{\xi}{R} \Big)^2 
- \frac{1}{25920} \Big( \frac{\xi}{R} \Big)^4 
- \cdots \Big], 
\label{eq:partial_inversion_A+I_amp_transition_pt}
\end{align}
c.f., \eqref{F6} and \eqref{530}.

%


\subsubsection{Numerical results for lattice systems}
\begin{figure}
\centering
  \includegraphics[scale=.5]{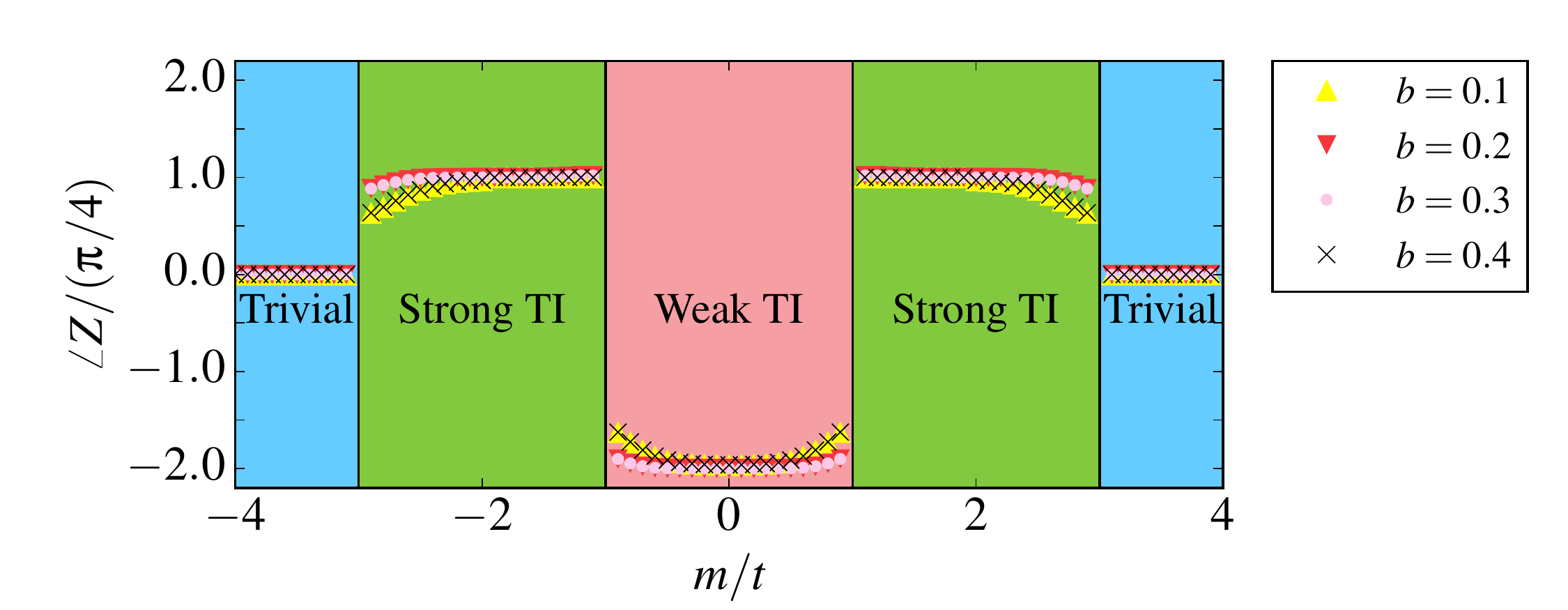}
  \includegraphics[scale=.5]{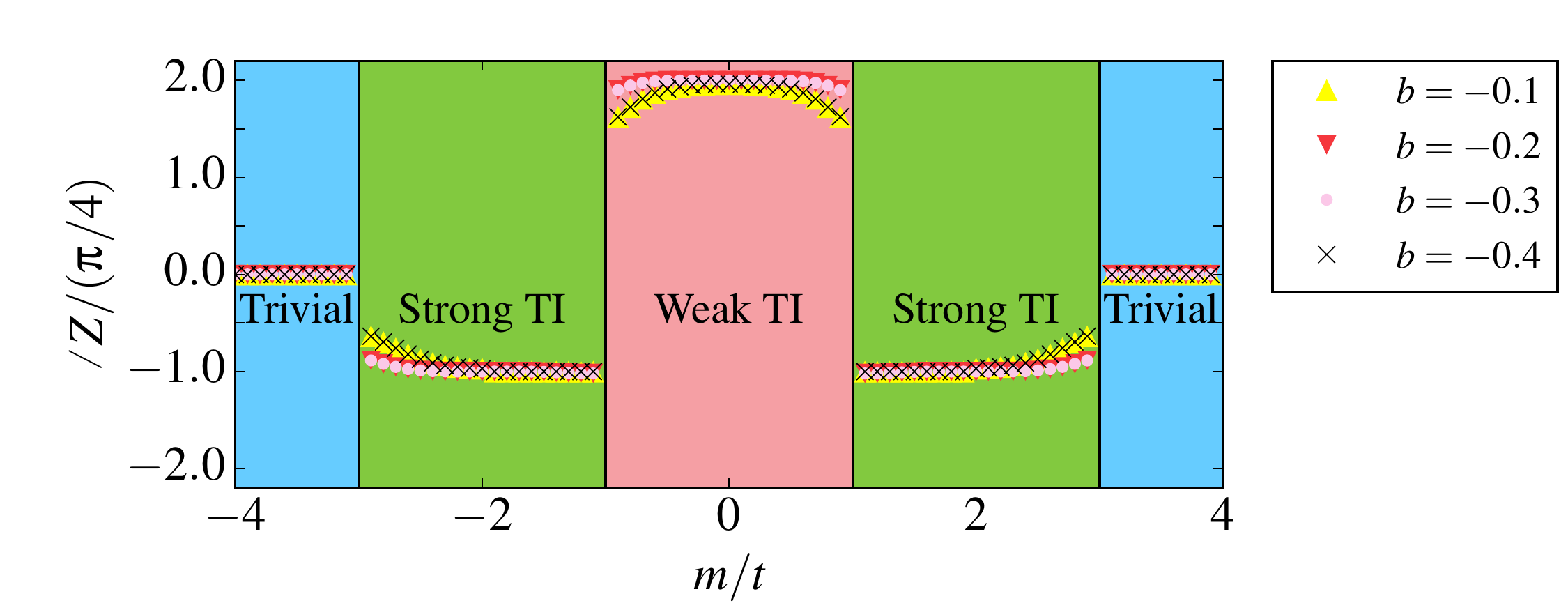}
\includegraphics[scale=.5]{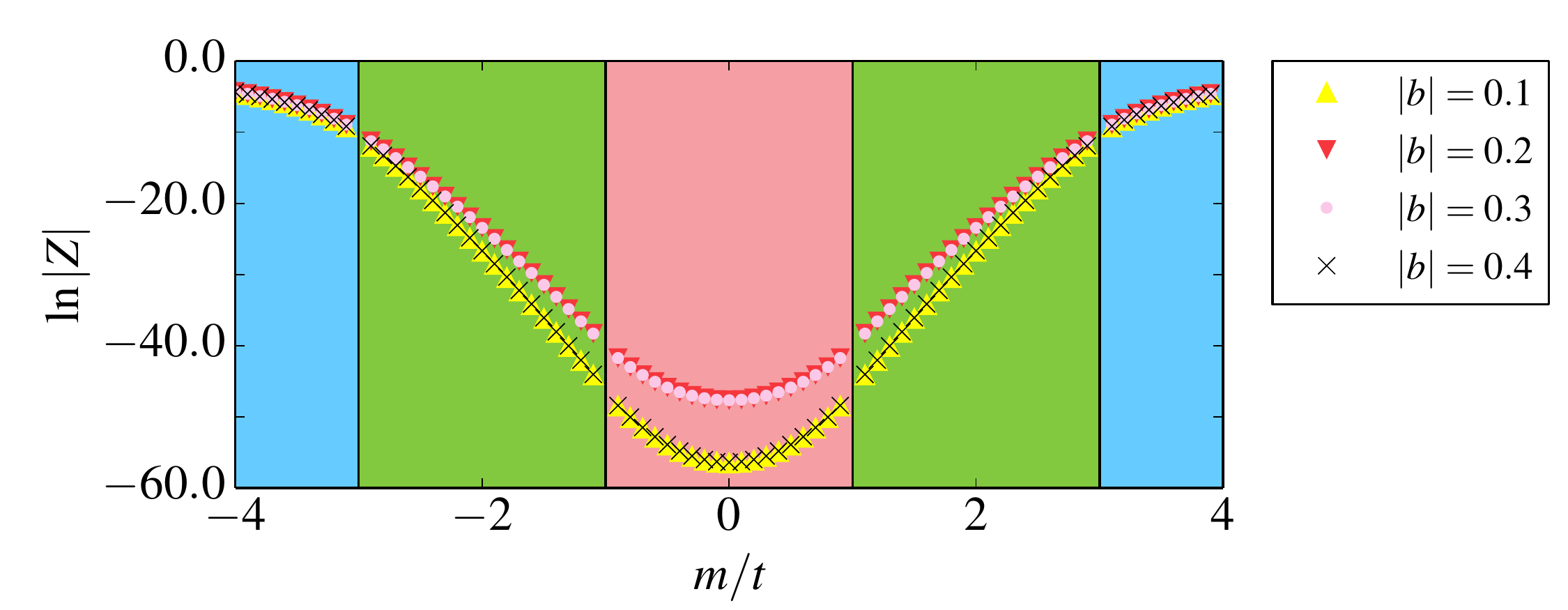}
\caption{\label{Fig:3DTI} (Color online)
  The phase ($\angle Z$) and modulus ($|Z|$)
  of the expectation value of the partial inversion,
  $Z = \braket{GS | I_D | GS}$,
  computed for the 3D inversion symmetric topological insulator (class A) as a function of
  the mass parameter $m$ for various values of
  the $U(1)$ phase transformation $b$ defined in Eq.~(\ref{Eq:U1}). Strong (weak) TI refers to the phase with odd (even) number of Dirac surface states. Here, we set $t=r$. The size of total system and  subsystem are $N=12^3$ and $N_\text{part}=6^3$, respectively.}
\end{figure}

In this section, we study the standard Wilson-Dirac Hamiltonian on a cubic lattice as a simple model of the three-dimensional inversion-symmetric TI~\cite{Wilson_3DTI,Qi_3DTI}
\begin{align} \label{eq:3DTI}
{H}=& \frac{1}{2} \sum_{\substack{\textbf{x}\\ s=1,2,3}} {\Big[} \psi_{\textbf{x}+\textbf{e}_s}^\dagger (i t\alpha_s - r \beta) \psi_\textbf{x} +{h.c.} {\Big]} 
+ m \sum_\textbf{x} \psi_\textbf{x}^\dagger \beta \psi^{\ }_\textbf{x}
\end{align}
where the Dirac matrices are given by
\begin{align*}
\alpha_s&= \tau_1\otimes \sigma_s=\left(\begin{array}{cc}
0 & \sigma_s \\ \sigma_s & 0
\end{array} \right), 
\quad 
\beta= \tau_3\otimes 1=\left(\begin{array}{cc}
\mathbb{I} & 0 \\ 0 & -\mathbb{I}
\end{array} \right). 
\end{align*}
In this convention the $\sigma$ and $\tau$ matrices act on the spin and orbital degrees of freedom respectively.
Transforming to reciprocal space, the Bloch Hamiltonian reads
\begin{align*}
h(\textbf{k})= \sum_{s=1,2,3}{\Big[} t \alpha_s \sin k_s - r\beta \cos k_s {\Big]}+ m\beta .
\end{align*}
This model can exhibit a non-trivial 3D TI phase protected by the inversion symmetry which is defined by Eq.~(\ref{Eq:invTI}).
In fact, as the mass parameter $m$ is varied, the Hamiltonian shows the following phases:
\begin{enumerate}
\item
$|m|<r$: weak TI with an even number of Dirac cones on each boundary surface.
\item
$r<|m|<3r$:  strong TI with a single Dirac cone on each boundary surface.
\item
$|m|>3r$: trivial phase equivalent to the atomic limit.
\end{enumerate}

\begin{figure}
\centering
\includegraphics[scale=.5]{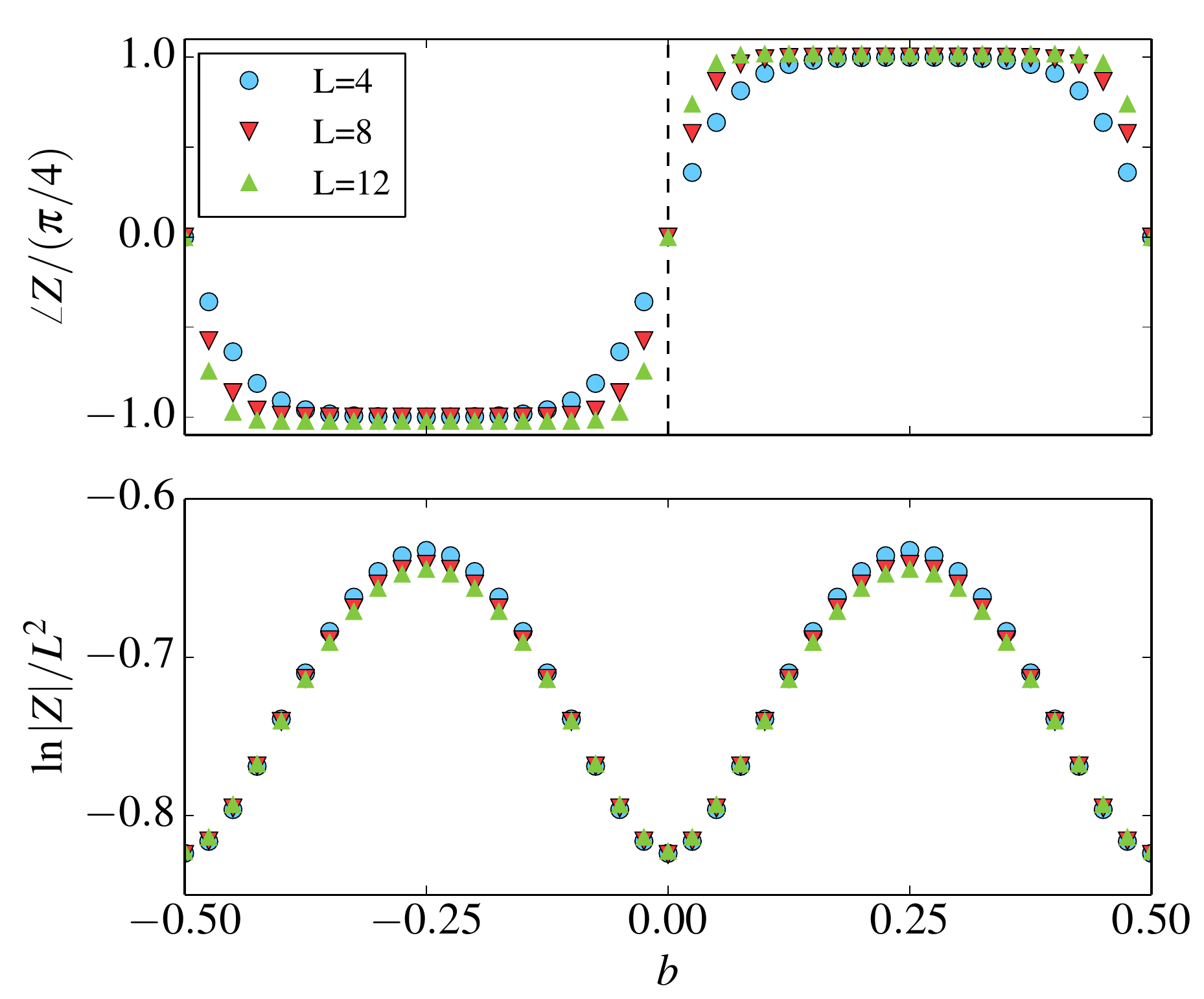}
\caption{\label{Fig:TI_vs_b}
  (Color online)
  The phase ($\angle Z$) and modulus ($|Z|$)
  of the expectation value
  of the partial inversion,
  $Z = \braket{GS | I_D | GS}$,
  computed for the 3D inversion symmetric topological insulator as a function of the $U(1)$ phase transformation $b$ defined in Eq.~(\ref{Eq:U1}). Here, we set $t=r$ and $m=2$. The sizes of the whole system and the subsystem are $N=16^3$ and $N_\text{part}=L^3$, respectively.}
\end{figure}

As shown in Fig.~\ref{Fig:3DTI}, we compute the complex phase 
$\angle Z = {\rm Im} \ln \braket{GS | U_b I | GS}$ 
of the partial inversion for various values of the $U(1)$ phase $e^{i2\pi b}$
for $t=r$. 
The calculation procedure here is very similar to the two-dimensional case
where in order to get $U_b I\ket{GS}$ we relocate the lattice points inside the
subsystem according to the inversion symmetry operator $I$ and multiply the
states by the $U(1)$ phase given by $U_b$
and finally the inner product $Z=\bra{GS}U_b I \ket{GS}$ is computed.
In particular, we observe that as $b$ changes from negative values to positive values the complex phase transitions from $-\pi/4$ to $\pi/4$ (see Fig.~\ref{Fig:TI_vs_b}). It is worth noting that as the subsystem is made larger, the transition becomes sharper and sharper indicating that this change will turn into a discontinuity in the thermodynamic limit. All these observations conform with our analytical results in the previous parts.

\subsection{General even spacetime dimensions}

The topological $U(1)$ phase emerging in the
expectation value of the partial inversion
for inversion symmetric topological superconductors 
can be generalized to any even spacetime dimensions. 
Let 
us consider the following BdG Hamiltonians 
which describe
odd parity topological superconductors in
$(2p+1)$-space dimensions:
\begin{align}
&H 
= \frac{1}{2} \sum_{\bk} \Psi^{\dag}(\bk) {\cal H}(\bk) \Psi(\bk), \quad 
\Psi(\bk) = (\psi(\bk), \psi^{\dag}(-\bk))^T, 
\nonumber \\
&{\cal H}(\bk) 
= \Delta \sum_{\mu=1}^{2p+1} k_{\mu} \Gamma_{\mu}  + m \Gamma_{2p+2}, 
\quad  
\{\Gamma_{\mu}, \Gamma_{\nu}\} = 2 \delta_{\mu \nu} \ \  (\mu, \nu = 1, \dots, 2p+2). 
\end{align}
Here, 
we consider the gamma matrices having 
the minimum possible dimension to satisfy the reality condition
\footnote{
These minimum dimensions (\ref{eq:minimum_dim_D+I}) can be confirmed by, 
for example, the suspension isomorphism of the $K$-theory 
in the presence of an additional order-two point group symmetry.\cite{Shiozaki2014} 
One can show that the topological superconductors with inversion symmetry 
with $I^2 = -1$ in $(2p+1)$ space dimensions are classified by 
$\pi_0({\cal R}_0) = \Z$ for $p \equiv 0, 1 \ (\mod \ 4)$ and $\pi_0({\cal R}_4) = 2 \Z$ for $p \equiv 2, 3 \ (\mod \ 4)$, 
where ${\cal R}_{p} (p=0, \dots, 7)$ are the classifying spaces of real Altland-Zirnbauer classes. 
}
\begin{align}
& p \equiv 0, 1 \ (\mod \ 4) \lto 2^{p+1} \times 2^{p+1}, 
\nonumber \\
& p \equiv 2, 3 \ (\mod \ 4) \lto 2^{p+2} \times 2^{p+2}. 
\label{eq:minimum_dim_D+I}
\end{align}
Inversion acts on the BdG Hamiltonian and 
the fermion operators as 
\begin{align}
I H I^{-1} = H, 
\quad 
I \psi^{\dag}(\bx) I^{-1} = i \psi^{\dag}(-\bx), 
\label{eq:any_dim_inversion_def}
\end{align}

Note that inversion (\ref{eq:any_dim_inversion_def}) 
is CPT dual of time-reversal
of class DIII for odd $p$ and class BDI for even $p$: 
The $\pi$ rotation of the real fermions is
associated with $\pm i$ phase
as shown in (\ref{eq:2d_px-ipy_rot_sym}), which implies 
that the inversion transformation with $I^2 = (-1)^F$ in $(2p+1)$-dimensions 
is equivalent to the reflection transformation with $R^2 = 1$ for odd $p$ and 
$R^2 = (-1)^F$ for even $p$. 
Also, $R^2 = 1$ and $R^2 = (-1)^F$ correspond to 
the TRS with $T^2 = (-1)^F$ and $T^2 = 1$, respectively, due to 
the Wick rotation.~\cite{Kapustin2015a}

We consider the reduced density matrix
for a $(2p+1)$-dimensional ball of radius $R$,
which can be described by 
the surface theory on the sphere $S^{2p}$
(the Dirac Hamiltonian on $S^{2p}$)
\cite{Camporesi1996}
\begin{align}
{\cal H}_{\rm surf} 
= \frac{\Delta}{R} \Big[ 
\big( -i \partial_{\theta} - \frac{(2p-1) i}{2} \cot \theta \big) \gamma_{2p} 
+ \frac{-i}{\sin \theta} \gamma_{i} (\tilde e_i - \frac{1}{2} \tilde \omega_{ijk} \Sigma^{jk})  \Big], 
\end{align}
where we have used the polar coordinates. 
The summations over $i,j,k$ are implicit.
$\theta$ is the angle from the north pole, 
$\{\tilde e_j\}_{j=1}^{2p-1}$ is a vielbein on $S^{2p-1}$, 
$\tilde \omega_{ijk} = \frac{1}{2} (\tilde C_{ijk} - \tilde C_{ikj} - \tilde C_{jki})$
with $[\tilde e_i, \tilde e_j] = \sum_{k=1}^{2p} \tilde C_{ijk} \tilde e_k$, 
and $\Sigma_{jk} = \frac{1}{4} [\Gamma_j, \Gamma_k]$. 
See Ref.~\onlinecite{Camporesi1996} for more details. 
The dimension of gamma matrices $\{\gamma_{i}\}_{i=1}^{2p}$ is 
a half of (\ref{eq:any_dim_inversion_def}). 
Eigenvalues of ${\cal H}_{\rm surf}$ are given by
\cite{Camporesi1996}
\begin{align}
E_{n,p,\pm} = \pm \frac{\Delta}{R} (n+p) 
\quad 
(n=0,1,\dots ), 
\label{eq:energy_eigenvalue_general_even_p}
\end{align}
with degeneracy 
\begin{align}
D_{2p}(n) = \frac{2^p (n+2p-1)! c_p}{n! (2p-1)!}, 
\quad 
c_p := \left\{\begin{array}{ll}
1, & p \equiv 0, 1 \ (\mod \ 4)  \\
2, & p \equiv 2, 3 \ (\mod \ 4)  \\
\end{array}\right.
\end{align}
Similar to our discussion on Sec.~\ref{Partial inversion}, 
the $U(1)$ phase of 
the expectation value of partial inversion is given by 
\begin{align}
&\im \big[ \ln \Braket{GS | I_D | GS} \big]
= \im \ I_{2p}(q = e^{- \frac{\xi}{R}}), 
\nonumber \\ 
&I_{2p}(q) 
= \ln \prod_{n=0}^{\infty} (1+i(-q)^{n+p})^{D_{2p}(n)} 
= \sum_{n=0}^{\infty} D_{2p}(n) \ln (1+i(-q)^{n+p}). 
\end{align}
One can show 
\begin{align}
\im \  I_{2p}(q=e^{- \delta})
= \frac{1}{2 \pi i} \int_{c-i \infty}^{c+i \infty} ds \frac{2^p c_p}{(2p-1)!} \delta^{-s} \Gamma(s) f_p(s-1) \beta(s+1), 
\end{align}
where 
\begin{align}
f_p(s-1) 
&:= \sum_{n=1}^{\infty} (n^2-1) (n^2-2^2) \cdots (n^2-(p-1)^2) (-1)^n n^{-s+1}, 
\end{align}
and $\beta(s) = \sum_{n=0}^{\infty} \frac{(-1)^n}{(2n+1)^s}$ is the Dirichlet beta function. 
All poles except for $s=0$ are canceled, and 
there is a single pole at $s=0$. 
We get a scale-independent exact result 
\begin{align}
\im \big[ \ln \Braket{GS | I_D | GS} \big] 
= (-1)^p (2 \pi) \times 
\left\{\begin{array}{ll}
2^{-(p+3)}, & p \equiv 0, 1 \ (\mod \ 4)  \\
2^{-(p+2)}, & p \equiv 2, 3 \ (\mod \ 4)  \\
\end{array}\right. 
\label{eq:any_dim_pi_sc}
\end{align}
This implies 
$\Z_{2^{p+3}}$ classification for $p \equiv 0, 1 \ (\mod \ 4)$ 
and 
$\Z_{2^{p+2}}$ classification for $p \equiv 2, 3 \ (\mod \ 4)$, 
which are consistent with 
Refs.~\onlinecite{you2014symmetry, Morimoto2015, Kapustin2015a}. 
Once again,
note that inversion symmetry (\ref{eq:any_dim_inversion_def}) 
is CPT dual of
time-reversal of
class DIII for odd $p$ and class BDI for even $p$. 

The partial inversion on class A insulators with inversion symmetry 
in even spacetime dimensions is given by adding the contribution from the 
negative energy eigenstates in (\ref{eq:energy_eigenvalue_general_even_p}), 
which leads to the twice of 
$p \equiv 0, 1 \ (\mod \ 4)$ cases in (\ref{eq:any_dim_pi_sc}), 
\begin{align}
\im \big[ \ln \Braket{GS | I_D | GS} \big] |_{{\rm class A} + I}
= (-1)^p (2 \pi) \times 2^{-(p+2)}. 
\end{align}
This is also consistent with 
Refs.~\onlinecite{you2014symmetry, Morimoto2015} and 
the free fermion part of $\pinc$ cobordism group~\cite{Bahri1987}
$\Z_{2^{p+2}} \subset \Omega^{\pinc}_{2p+2}(pt)$.

\section{Conclusion}
\label{Sec:6}

In this paper, 
we developed an approach to detect interacting fermionic SPT phases by 
introducing non-local order parameters, 
the expectation value of 
partial point group transformations
with respect to
a given ground state wave function on a closed space manifold. 
From the point of view of TQFTs, 
the interacting SPT invariants are 
defined as partition functions (path-integrals) 
on generating spacetime manifolds of cobordism group. 
In order to simulate the path integrals, in the operator formalism, on various generating manifolds such as 
real projective spaces and lens spaces, 
we showed that the partial point group transformations 
provide a unified way for this purpose: 
The SPT topological invariants emerge as 
the complex $U(1)$ phases of the expectation value of the partial point group transformations, 
if the subregion $D$ is sufficiently larger than the bulk correlation length. 
In addition to the topological $U(1)$ phases, 
we find that the amplitude part also includes scale-independent contributions, 
which is another signature of nontrivial SPT phases.

We confirm these results 
both by analytic calculations 
using gapless surface theories
and by numerics in lattice models. 
It is worth emphasizing that 
the definition of the partial point group transformation 
depends only on the symmetry of the problem,  
which contrasts with the modular transformation on the ground states on the 2-torus $T^2$.~\cite{Hung2014, HeMoradiWen2014, Moradi2014, You2015, HuangWei2015}

Let us close by mentioning a number of interesting future directions.

-- First, we focused in this paper on unitary symmetries. 
The definition of SPT invariants 
for SPT phases protected by time-reversal symmetry 
(and more general anti-unitary symmetries) 
is not fully understood. 
In $(1+1)d$ SPTs with time-reversal symmetry, 
it is known that the partial transpose plays the role of 
``gauging time-reversal symmetry'' and yields real projective plane 
and Klein bottle for bosons~\cite{Pollmann2012} and 
fermions.~\cite{Shapourian-Shiozaki-Ryu} 
The construction of SPT invariant for more general higher space dimensions is 
an open problem.

-- We note that our formula (\ref{Eq:PPGT}) can be applied to 
symmetry-enriched topological (SET) phases by point group symmetry. 
In topologically ordered phases where there are ground state degeneracies depending on 
the space manifold $M$, 
the ground state $\ket{GS}$ in the formula (\ref{Eq:PPGT}) is replaced by a linear 
combinations of degenerate ground states as 
$
\ket{GS} = \sum_{i} c_i \ket{GS_i}
$, 
$c_i \in \C$, 
$
\sum_{i} |c_i|^2 = 1, 
$
where $\ket{GS_i}$ is the ground state of the topological sector labeled by $i$. 
We leave the detailed studies of SET phase for the future. 

-- Throughout this paper, we assumed that the entanglement chemical potential 
$\mu_e$ associated with the reduced density matrix of the sub region $D$ is zero. 
However, in general, $\mu_e$ can be nonzero, that depends on the geometry of the 
region $D$ and other details. The agreement between TQFT partition functions and 
the numerical calculation of the partial point group transformations 
suggests that effect of a finite entanglement chemical potential can be 
neglected in the boundary theory. 

-- Due to the lattice translational symmetry, numerical 
calculations of partial rotations in this paper are limited into 
$C_2, C_3, C_4$, and $C_6$ rotations.  
It is an interesting problem to compute the partial rotations for rotation symmetries 
which can not be defined on translational symmetric lattice systems such as $C_5$
symmetry.

\acknowledgements

We thank useful discussions with 
Kiyonori Gomi, 
Hosho Katsura, 
Takahiro Morimoto, 
Masaya Nakagawa, 
Akinori Tanaka,
Apoorv Tiwari, 
Xueda Wen, 
and 
Peng Ye. 
Especially, K.S.\ is grateful to Chang-Tse Hsieh for introducing the 
spin cobordism group. 
Numerical computation of Pfaffians was carried out by the PFAPACK library.~\cite{Wimmer2012}
This work was supported in part 
by the National Science Foundation grant DMR-1455296,
by the U.S. Department  of  Energy,  Office  of  Science,  
Office  of  Advanced Scientific Computing Research, 
Scientific Discovery  through  Advanced  Computing  (SciDAC)  program
under Award Number FG02-12ER46875,  
and by Alfred P. Sloan foundation. 
K.S.\ is supported by JSPS Postdoctoral Fellowship for Research Abroad.

\appendix

\makeatletter
\renewcommand{\theequation}{%
\Alph{section}.\arabic{equation}}
\@addtoreset{equation}{section}
\makeatother

\section{Arf and Brown invariants}
\label{Arf and Brown invariant}

The purpose of this appendix is to introduce the Arf invariant~\cite{arf1941untersuchungen} $(-1)^{{\rm Arf}(\eta)}$
of $\spin$ structures $\eta \in \spin(\Sigma)$ on given 2-dimensional oriented manifolds $\Sigma$,~\cite{atiyah1971riemann, johnson1980spin}
and 
the Brown 
invariant~\cite{brown1971kervaire} for  
$\pinm$ structures on unoriented 2-manifolds.~\cite{Kirby, Cimasoni2009, Kapustin2015a}

\subsection{The Arf invariant}

The obstruction to give a $\spin$ structure on $\Sigma$ is measured by the 2nd Stiefel-Whitney class $w_2(T\Sigma) \in H^2(\Sigma,\Z_2)$. 
For oriented 2-manifolds, i.e., Riemann surfaces $\Sigma_g$ with genus $g$, $w_2(\Sigma_g)$ always disappears, hence one can define $\spin$ structures on $\Sigma_g$. 
The set of spin structures $\spin(\Sigma_g)$ on $\Sigma_g$ is equivalent to $H^1(\Sigma_g,\Z_2)$ as a set. 
Here, $H^1(\Sigma_g,Z_2) \cong \Hom(\pi_1(\Sigma_g),\Z_2)$ has $2^{2g}$ elements and can be thought of as the space of $\Z_2$-Wilson lines 
(background nontrivial $\Z_2$ gauge fields) on $\Sigma_g$. 
It is known that $H^1(\Sigma_g,\Z_2)$ acts on $\spin(\Sigma_g)$ freely and transitively (i.e. $\spin(\Sigma_g)$ is a $H^1(\Sigma_g,\Z_2)$-${\rm Torsor}$), 
which means any spin structures $\eta \in \spin(\Sigma_g)$ can be given by an action of a Wilson line $a \in H^1(\Sigma_g,\Z_2)$ on a some ``reference'' $\spin$ structure $\eta_0$. 
Note that there is no canonical choice of the reference spin structure $\eta_0$, which contrasts with $\Z_2$-Wilson lines where there is the zero flux in $H^1(\Sigma_g,\Z_2)$. 
The absence of any reference elements is a feature of ``${\rm Torsor}$''. 
See, for example, Ref.~\onlinecite{Cimasoni2007} for details, where the equivalence between the $\spin$ structures and the Kasteleyn orientations in the dimer model is also explained. 

The Arf invariant is defined in a pure algebraic manner. 
For a given vector space $V$ over the field
$\Z_2$ and a fixed bilinear form $\phi : V \times V \to \Z_2$, 
one can define the $\Z_2$-valued Arf invariant of the {\it quadratic forms} $Q_2(V,\phi)$, which will be described in Sec.~\ref{Quadratic form and Z2 Arf invariant}. 

It is known that the spin structures $\spin(\Sigma)$ on an oriented 2 manifold $\Sigma$ is equivalent to 
the quadratic forms $Q_2(H^1(\Sigma,\Z_2), \int x \cup y)$ on $H^1(\Sigma,\Z_2)$ with the intersection form $\int x \cup y$ as a 
$H^1(\Sigma,\Z_2)$-${\rm Torsor}$. 
(We do not explain this equivalence in the present paper. See Ref.~\onlinecite{Cimasoni2007} for details.) 
This implies that one can define the Arf invariant of the spin structures $\spin(\Sigma)$, which is nothing but the $\Z_2$ topological invariant of 
the $\spin$ TQFT in 2 spacetime dimensions.~\cite{Kirby, Moore2006, Kapustin2015a}

\subsubsection{Simplicial calculus}
To describe $\Z_2$-Wilson lines belonging to $H^1(\Sigma,\Z_2)$ and the intersection form $\int x \cup y$ in a rigorous way, 
we introduce the simplicial calculus, according to Appendix A in Ref.~\onlinecite{Kapustin2014bosonic}. 

Let $K$ be a triangulation of $\Sigma$ and $G$ be an Abelian group. 
A $p$-cochain $f$ is a function over $p$-simplices $\{(v_0, \dots, v_p)\}$ to $G$. 
We assume the vertices of $K$ are ordered in some way. 
The space of $p$-chains is denoted by $C^p(K,G)$. 
The 	differential operator 
$\delta : C^p(K,G) \to C^{p+1}(K,G)$ is defined by 
\begin{align}
(\delta f)(v_0, \dots, v_{p+1}) = \sum_{i=0}^{p+1} (-1)^i f(v_0, \dots, \hat v_i , \dots, v_{p+1}) 
\end{align}
for a $(p+1)$ simplex $(v_0, v_1, \dots, v_{p+1})$, where $\hat v_i$ means that $v_i$ is excluded.
One can show $\delta^2=0$ as 
\begin{align}
(\delta^2 f)(v_0, \dots, v_{p+2}) 
&= \sum_{i=0}^{p+2} (-1)^i (\delta f)(v_0, \dots, \hat v_i , \dots, v_{p+2}) 
\nonumber \\
&= \sum_{j<i}^{p+2} (-1)^{i+j} f(v_0, \dots, \hat v_j , \dots, \hat v_i, \dots, v_{p+2}) + \sum_{i<j}^{p+2} (-1)^{i+j-1} f(v_0, \dots, \hat v_i , \dots, \hat v_j, \dots, v_{p+2}) 
\nonumber \\
&=0. 
\end{align}
We define the coboundary $B^p(K,G) = {\rm Im}[\delta : C^{p-1}(K,G) \to C^p(K,G)]$ and 
the cocycle $Z^p(K,G) := {\rm Ker}[\delta : C^{p}(K,G) \to C^{p+1}(K,G)]$
and the cohomology $H^p(K,G) := Z^p(K,G)/B^p(K,G)$. 

If $G$ is a commutative ring (for example, a cyclic group $\Z_p$), we can define the cup product 
$\cup : C^p(K,G) \times C^q(K,G) \to C^{p+q}(K,G)$ by 
\begin{align}
(f \cup g)(v_0, \dots, v_{p+q}) = f(v_0, \dots, v_p) g(v_p,\dots, v_{p+q}). 
\end{align}
(Here, the r.h.s. is the product of the ring $G$.) 
The cup product satisfies the Leibniz rule 
\begin{align}
\delta(f \cup g) = \delta f \cup g + (-1)^p f \cup \delta g. 
\end{align}
Obviously, the cup product is well-defined in the cohomology $H^p(K,G)$ since 
$\delta (f \cup g) = 0$ for $\delta f = \delta g = 0$. 

\subsubsection{A quadratic form and the $\Z_2$ Arf invariant}
\label{Quadratic form and Z2 Arf invariant}
Let $V$ be a finite dimensional vector space over the field $\Z_2$, and let $\phi : V \times V \to \Z_2$ be a fixed bilinear form. 
A quadratic form $q$ on $(V,\phi)$ is a {\it function} $q : V \to \Z_2$ (not a linear form) which satisfies 
\begin{align}
q(x+y) = q(x) + q(y) + \phi(x,y). 
\end{align}
Note that the difference (= sum) of two quadratic forms $q_1 + q_2$ on $(V, \phi)$ is a linear form on $V$ because $2 \phi(x,y) = 0$. 
Therefore, the set $Q_2(V,\phi)$ of quadratic forms on $(V,\phi)$ is a $V^*$-torsor,
i.e.\ all quadratic forms $q$ is given by the action of a linear form $f : V \to \Z_2$ as $(f \cdot q)(x) := q(x) + f(x)$. 
The Arf invariant ${\rm Arf}(q) \in \{0,1\}$ on the quadratic forms $Q_2(V,\phi)$ is defined by 
\begin{align}
(-1)^{{\rm Arf}(q)} := \frac{1}{\sqrt{|V|}} \sum_{x \in V} (-1)^{q(x)}, 
\qquad 
q \in Q_2(V,\phi), 
\end{align}
where $|V|$ is the number of elements in $V$. 

\subsubsection{The Arf invariant for $\spin$ structures on $T^2$}
\begin{figure}[!]
 \begin{center}
  \includegraphics[width=\linewidth, trim=0cm 0cm 0cm 0cm]{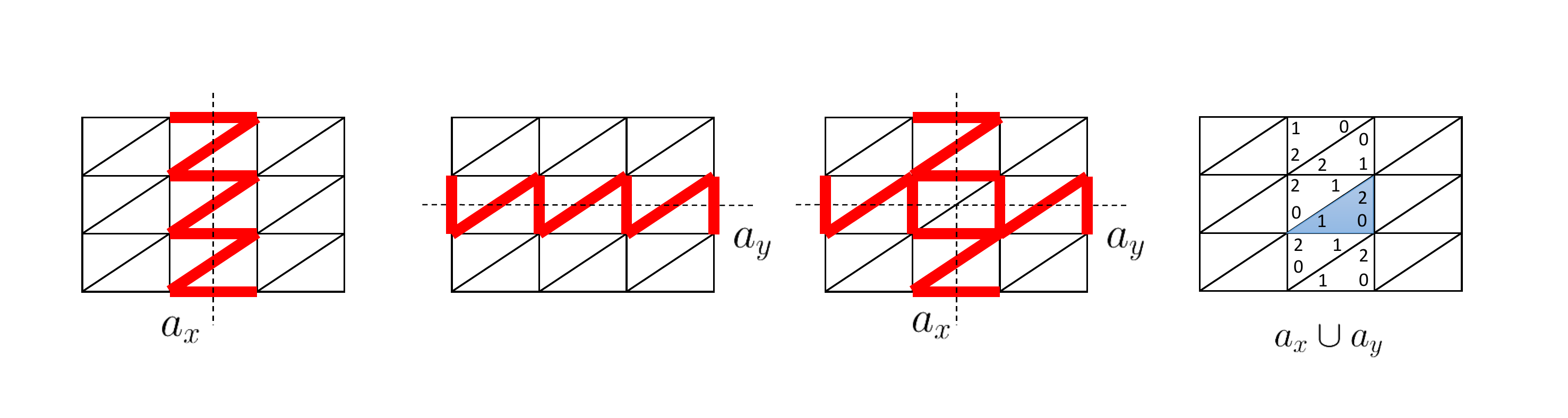}
 \end{center}
 \caption{(Color online) Wilson lines and the intersection forms.}
 \label{Fig:Arf}
\end{figure}

Let $T^2$ be the 2-torus.
Let us consider the quadratic forms $Q_2\big( H^1(T^2,\Z_2), \int_{T^2} x \cup y \big)$. 
Here $H^1(T^2,\Z_2) = \Z_2 \oplus \Z_2 = \{0, a_x, a_y, a_x+a_y\}$ is generated by two different Wilson lines $a_x$ and $a_y$ along the $x$ and $y$-directions. 
Intersection forms are given as 
\begin{align}
\int_{T^2} a_x \cup a_x = \int_{T^2} a_y \cup a_y=0, 
\qquad 
\int_{T^2} a_x \cup a_y = 1. 
\end{align}
This is the even/odd parity of the number of intersections between Wilson lines. 
See Fig.\ \ref{Fig:Arf}. 
There are $2^4 = 16$ kinds of functions $q : H^1(T^2,\Z_2) = \Z_2 \oplus \Z_2 \to \Z_2$. 
The quadratic forms have to obey 
\begin{align}
\left\{ \begin{array}{l}
q(0) = q(0) + q(0), \\
q(a_x) = q(a_x) + q(0), \\
q(a_y) = q(a_y) + q(0), \\
q(a_x+a_y) = q(a_x) + q(a_y) + 1. 
\end{array} \right. 
\end{align}
Then, $q(0)=0$ and $q(a_x+a_y) = q(a_x)+q(a_y)+1$. 
Solving these equations, we have $|H^1(T^2,\Z_2)| = 4$ distinct quadratic forms: 
\begin{align}
\left\{ 
\begin{array}{l}
q_{ns,ns}(a_x) = 0, \ q_{ns,ns}(a_y)=0 \\
q_{ns,r}(a_x) = 0, \ q_{ns,r}(a_y)=1 \\
q_{r,ns}(a_x) = 1, \ q_{r,ns}(a_y)=0 \\
q_{r,r}(a_x) = 1, \ q_{r,r}(a_y)=1 \\
\end{array} \right.
\end{align}
These four different quadratic forms correspond to spin structures $\{(ns,ns), (ns,r), (r,ns), (r,r)\}$, 
where $r$ ($ns$) represents the periodic (anti-periodic) boundary condition of the real fermion along the $x$ or $y$-directions. 
Through the bijection 
\begin{align}
\spin(T^2) \cong Q_2\big( H^1(T^2,\Z_2), \int_{T^2} x \cup y \big), 
\end{align}
we identify a quadratic form $q \in Q_2\big( H^1(T^2,\Z_2), \int_{T^2} x \cup y \big)$ with a spin structure $\eta \in \spin(T^2)$. 
Finally, we obtain the Arf invariant for each spin structures $\spin(T^2)$ as  
\begin{align}
&(-1)^{{\rm Arf}(ns,ns)} = (-1)^{{\rm Arf}(ns,r)} = (-1)^{{\rm Arf}(r,ns)} = 1,
\nonumber \\
&(-1)^{{\rm Arf}(r,r)} = -1. 
\label{eq:app_arf_torus}
\end{align}
This is the same as the partition function on $T^2$ of the Kitaev chain model (\ref{eq:z2inv_1d_sc}). 

\subsection{A Quadratic form and the $\Z_8$ Brown invariant}
\label{Quadratic form and Z8 Brown invariant}
In the same way, the Brown $\Z_8$ invariant \cite{brown1971kervaire} is constructed as follows. 
Let $V$ be a finite dimensional vector space over the field $\Z_2$ and let $\phi : V \times V \to \Z_2$ be a fixed bilinear form. 
We consider a quadratic form $q$ on $(V,\phi)$ is a function $q : V \to \Z_4$ which satisfies 
\begin{align}
q(x+y) = q(x) + q(y) + 2 \phi(x,y). 
\end{align}
Note that the difference (= sum) of two quadratic forms $q_1, q_2$ on $(V, \phi)$ is a linear form on $V$ since $4 \phi(x,y) = 0$. 
The set $Q_4(V,\phi)$ of quadratic forms $q : V \to \Z_4$ is an $V^*$-torsor by the action 
$(f \cdot q)(x) = q(x) + 2 f(x)$ of $f : V \to \Z_2$ on $q \in Q_4(V,\phi)$. 
The Brown invariant $\beta(q) \in \Z_8$ is defined by 
\begin{align}
e^{2 \pi i \beta(q)/8} := \frac{1}{\sqrt{|V|}} \sum_{x \in V} i^{q(x)}, 
\qquad 
q \in Q_4(V,\phi), 
\end{align}
where $|V|$ is the number of elements in $V$. 

\subsubsection{The Brown invariant for $\pinm$ structures on $\R P^2$}
\begin{figure}[!]
 \begin{center}
  \includegraphics[width=0.7\linewidth, trim=0cm 0cm 0cm 0cm]{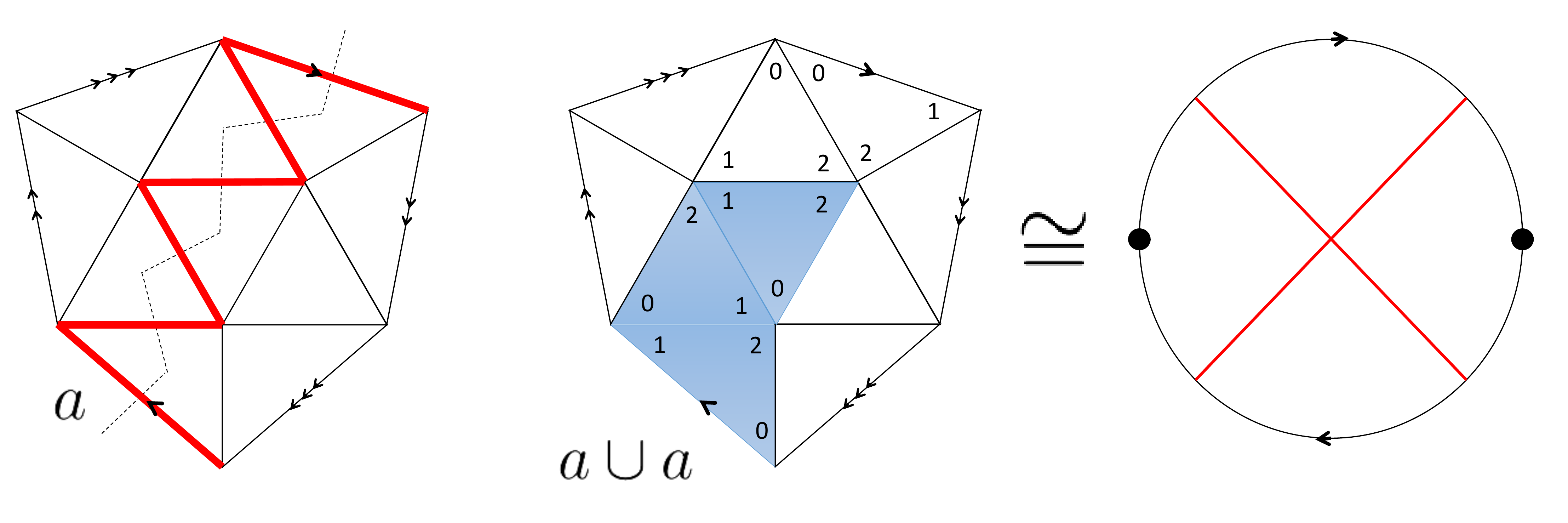}
 \end{center}
 \caption{(Color online) The Wilson line and the intersection form on $\R P^2$. }
 \label{Fig:Brown_RP2}
\end{figure}

There is a canonical 1 to 1 correspondence between $\pinm$ structures on an unoriented surface $\Sigma$ and quadratic forms $q : H^1(\Sigma,\Z_2) \to \Z_4$.~\cite{Kirby}
Let $\R P^2$ be the real projective plane.
Let us consider $\Z_4$-valued quadratic forms $q : H^1(\R P^2,\Z_2) \to \Z_4$ with the intersection form $\int_{\R P^2} x \cup y$. 
Here $H^1(\R P^2,\Z_2) = \Z_2= \{0, a\}$ is generated by the Wilson line $a$ for noncontractible loop of $\R P^2$. 
The intersection form is given by 
\begin{align}
\int_{\R P^2} a \cup a = 1
\end{align}
See Fig.\ \ref{Fig:Brown_RP2}. 
There are $4^2 = 16$ kinds of functions $q : H^1(\R P^2,\Z_2) \to \Z_4$. 
The quadratic form $q : H^1(\R P^2,\Z_2) \to \Z_4$ should satisfy
\begin{align}
q(0) = q(0) + q(0) , 
\qquad 
q(a+a) = q(a) + q(a) +2,  
\end{align}
which implies that there are two distinct quadratic forms: 
\begin{align}
q_+(a) = 1, \qquad q_-(a)=3. 
\end{align}
We identify these quadratic forms with two different $\pinm$ structures $\eta_+, \eta_-$ on $\R P^2$, respectively. 
Finally, we have the $\Z_8$ Brown invariant for each $\pinm$ structure as 
\begin{align}
e^{2 \pi i \beta(\eta_+)/8} = e^{\pi i/4}, 
\qquad 
e^{2 \pi i \beta(\eta_-)/8} = e^{- \pi i/4}. 
\end{align}
The former one is the $U(1)$ phase part which appeared in (\ref{eq:Brown_z8_inv_kitaev}).

\section{Lens space}
\label{App:1}
The lens space $L(p,q) = S^3/\mathbb{Z}_{p}$ is defined by the quotient 
\begin{align}
(z_1,z_2) \sim (e^{\frac{2 \pi i}{p}} z_1, e^{\frac{2 q \pi i}{p}} z_2), 
\end{align}
where $(z_1,z_2)$ with $|z_1|^2 + |z_2|^2=1$ represent the 3-sphere $S^3$, and 
$p$ and $q$ are coprime. 

The surgery construction of the lens space $L(p,q)$ is given by the following modular transformation~\cite{Jeffrey1992}
\begin{align}
S T^{m_{t-1}} S T^{m_{t-2}} \cdots S T^{m_1},   
\end{align}
on the boundary of a solid torus in $S^3$. 
Here, $(m_1, \dots, m_{t-1})$ is determined by the fraction expansion of $(p,q)$ 
\begin{align}
-\frac{p}{q} = m_{t-1} -\frac{1}{m_{t-2} - \frac{1}{\cdots - \frac{1}{m_1}}}. 
\end{align}
For example, the surgery by $S T^{n} S$ leads to $L(-n,1)$.

\section{Eta and theta functions}
In this section, we summarize the properties of the Dedekind eta function $\eta(\tau)$ 
and the generalized theta function $\theta_{a,b}(z|\tau)$ used in the main text. 
For $q = e^{2 \pi i \tau}$ ($\im \ \tau>0$), 
the Dedekind eta function is defined as
\begin{align}
\eta(\tau) 
:= q^{1/24} \prod_{n=1}^{\infty}(1-q^n).
\end{align}
The theta function is defined as 
\begin{align}
\theta_{a,b}(z|\tau) 
&:= \sum_{n \in Z} e^{i \pi \tau (n+a)^2 + 2 \pi i (n+a) (z+b)} 
\nonumber \\
&= e^{2 \pi i a (z+b) } q^{\frac{1}{2} a^2} 
\prod_{n=1}^{\infty} 
( 1-q^n )
\Big( 1 + e^{2\pi i (z+b)} q^{n+a-\frac{1}{2}} \Big) 
\Big( 1 + e^{-2\pi i (z+b)} q^{n-a-\frac{1}{2}} \Big),
\end{align}
where we noted the Jacobi's triple identity 
\begin{align}
\sum_{n \in \Z} e^{\pi i \tau n^2 + 2 \pi i n v} = \prod_{m=1}^{\infty} (1-e^{2 m \pi i \tau}) (1+e^{(2 m-1) \pi i \tau + 2 \pi i v}) (1+e^{(2 m-1) \pi i \tau - 2 \pi i v}). 
\end{align}
$\theta_{a,b}(z|\tau)$ obeys the following periodicities and modular transformations 
\begin{align}
&\theta_{a+1,b}(z|\tau) = \theta_{a,b}(z|\tau), && \theta_{a,b+1}(z|\tau) = e^{2 \pi i a} \theta_{a,b}(z|\tau), \\
&\theta_{a,b}(z|\tau+1) = e^{-\pi i (a^2+a)} \theta_{a,b+a+1/2}(z|\tau), && 
\theta_{a,b}(\frac{z}{\tau}|-\frac{1}{\tau}) = \sqrt{-i \tau} e^{2 \pi i a b} e^{\frac{i \pi}{\tau} z^2} \theta_{b,-a}(z|\tau), 
\label{eq:app_theta_modular} \\
&\eta(\tau+1) = e^{\pi i/12} \eta(\tau), && 
\eta(-\frac{1}{\tau}) = \sqrt{-i \tau} \eta(\tau). 
\end{align}

\section{Derivation of boundary theories}
\label{Derivation of surface theories}

Here, we summarize the derivation of gapless 
boundary theories 
in $(2+1)d$ and $(3+1)d$ free fermion topological phases. 
We employ the same approach as Ref.\ \onlinecite{imura2012spherical};
We first solve 
the radial part of the eigenvalue equation of single-particle Hamiltonians,
and then construct  
an effective single-particle boundary Hamiltonians
for directions tangential to the boundary. 

\subsection{Edge theory on the boundary of disc}
\label{Edge theory on the boundary of disc}

Let 
us consider the following single-particle Hamiltonian
describing 
a $(2+1)d$ bulk topological phase defined on the disc:
\begin{align}
{\cal H} = -i \sigma_x \partial_x - i \sigma_y \partial_y + m(r) \sigma_z.
\end{align}
Here, the profile of the position-dependent $m(r)$ ($r = \sqrt{x^2+y^2}$)
is chosen to create the disk geometry: 
$m(r)<0$ for $r<R$ and $m(r)>0$ for $r>R$, 
where $R$ is the radius of the disc. 
Introducing the polar coordinate $(x,y) = (r \cos \phi, r \sin \phi)$, 
${\cal H}$ can be written as 
\begin{align}
&{\cal H} 
= {\cal H}_{r} + {\cal H}_{\phi},
\nonumber \\
& {\cal H}_r = e^{-i \frac{\phi}{2} \sigma_z} \big[ -i \sigma_x \partial_r + m(r) \sigma_z \big] e^{i \frac{\phi}{2} \sigma_z}, 
\qquad 
{\cal H}_{\phi} = e^{-i \frac{\phi}{2} \sigma_z} \frac{1}{r} \sigma_y (-i \partial_{\phi} - \frac{\sigma_z}{2}) e^{i \frac{\phi}{2} \sigma_z}. 
\end{align}
In the following, we approximately derive the wave function of 
the edge state and effective edge Hamiltonian. 
Since gapless edge excitations are exponentially localized at $r \sim R \gg \xi$ 
($\xi$ is the correlation length of bulk which is determined by the gap $m(r)$), 
the edge state wave function is approximated by solving ${\cal H}_r$ as
\begin{align}
{\cal H}_r \phi_{\rm edge}(r,\phi) = 0, 
\quad
\phi_{\rm edge}(r,\phi) \sim u(\phi) \cdot 
e^{-i \frac{\phi}{2} \sigma_z}
\begin{pmatrix}
1 \\
i \\
\end{pmatrix}
e^{- \int^r m(r') dr'}, 
\quad
u(\phi+2 \pi) = - u(\phi). 
\label{Eq:Sol_2D_EdgeState}
\end{align}
Here $u(\phi)$ is a complex-valued function representing 
the $\phi$-dependence 
and obeys anti-periodic boundary condition to account for the factor 
$e^{i \frac{\phi}{2} \sigma_z}$. 
Since $\phi_{\rm edge}(r,\phi)$ is well localized at $r=R$, we can replace ${1}/{r}$ by 
${1}/{R}$. 
Then, the effective single-particle Hamiltonian 
for the edge excitations is given by 
\begin{align}
{\cal H}_{edge} = \int r d r [\phi_{\rm edge}(r,\phi)]^{\dag} {\cal H}_{\phi} \phi_{\rm edge}(r,\phi) \sim \frac{1}{R} u^*(\phi) (-i \partial_{\phi}) u(\phi).  
\end{align}

In the following, 
we use the above result for the single-particle bulk 
and the corresponding edge Hamiltonians to 
construct 
(second quantized) fermionic operators 
creating/annihilating edge excitations and 
the Hamiltonians. 
We will discuss topological insulators and 
superconductors separately. 

\subsubsection{Chern insulator}

Let $\psi(\bx) = (\psi_1(\bx), \psi_2(\bx))^T$ be a two-orbital complex fermion and consider a Chern insulator on the disk 
\begin{align}
H = \int d^2 \bx\, \psi^{\dag}(\bx) \Big[ -i \sigma_x \partial_x - i \sigma_y \partial_y + m(r) \sigma_z \Big] \psi(\bx). 
\end{align}
From (\ref{Eq:Sol_2D_EdgeState}), 
the complex fermion annihilation operator $\gamma(\phi)$ for edge excitations is given by 
\begin{align}
\gamma^{\dag}(\phi) \sim \big[ e^{-i \frac{\phi}{2}} \psi^{\dag}_1(r,\phi) + i e^{i \frac{\phi}{2}} \psi^{\dag}_2(r,\phi) \big] e^{- \int^r m(r') dr'}, 
\quad 
\gamma(\phi+2 \pi) = - \gamma(\phi) 
\end{align}
with the Hamiltonian 
\begin{align}
H = \frac{1}{R} \int d \phi\, \gamma^{\dag}(\phi) (- i \partial_{\phi}) \gamma(\phi).  
\end{align}

\subsubsection{$(p_x - i p_y)$ superconductor}
Let us consider the following model of a $(p_x - i p_y)$ superconductor, 
\begin{align}
H = \int d^2 \bx\, \Psi^{\dag}(\bx) \Big[ -i \sigma_x \partial_x - i \sigma_y \partial_y + m(r) \sigma_z \Big] \Psi(\bx), 
\quad 
\Psi(\bx) = (\psi(\bx), \psi^{\dag}(\bx))^T. 
\end{align}
From (\ref{Eq:Sol_2D_EdgeState}), 
the Majorana fermion operator  
$\gamma(\phi)$, 
$\gamma^{\dag}(\phi) = \gamma(\phi)$,  
for edge excitations is given by 
\begin{align}
\gamma(\phi) \sim \big[ e^{-i \frac{\phi}{2} - \frac{\pi}{4} i} \psi^{\dag}(r,\phi) + e^{i \frac{\phi}{2} + \frac{\pi}{4} i} \psi(r,\phi) \big] e^{- \int^r m(r') dr'}, 
\quad 
\gamma(\phi+2 \pi) = - \gamma(\phi) 
\end{align}
with the Hamiltonian 
\begin{align}
H = \frac{1}{R} \int d \phi\,  \gamma(\phi) (- i \partial_{\phi}) \gamma(\phi).  
\end{align}

\subsection{Surface theory on the boundary of ball}

Let 
us consider 
the following single-particle Hamiltonian 
\begin{align}
{\cal H} = -i \tau_x (\sigma_x \partial_x + \sigma_y \partial_y + \sigma_z \partial_z ) + m(r) \tau_z
\end{align}
representing a bulk topological phase on a 3-ball. 
We assume $m(r)<0$ for $r<R$ and $m(r)>0$ for $r>R$, 
where $R$ is the radius of the ball. 
By introducing polar coordinate $(x,y,z) = (r \sin \theta \cos \phi, r \sin \theta \sin \phi, r \cos \theta)$, 
${\cal H}$ can be written as 
\begin{align}
&{\cal H} = {\cal H}_r + {\cal H}_{\theta,\phi}, 
\nonumber \\
&{\cal H}_r = e^{-i \frac{\phi}{2} \sigma_z} e^{- i \frac{\theta}{2} \sigma_y} \big[ -i \tau_x \sigma_z \partial_r + m(r) \tau_z \big] e^{i \frac{\theta}{2} \sigma_y} e^{i \frac{\phi}{2} \sigma_z},
\nonumber \\
&{\cal H}_{\theta,\phi} 
= e^{-i \frac{\phi}{2} \sigma_z} e^{- i \frac{\theta}{2} \sigma_y} 
\Big[ \frac{1}{r} \tau_x \sigma_x (-i \partial_{\theta} - \frac{\sigma_y}{2}) + \frac{1}{r \sin \theta} \tau_x \sigma_y (-i \partial_{\phi} - \frac{\cos \theta}{2} \sigma_z - \frac{\sin \theta}{2} \sigma_x) \Big] 
e^{i \frac{\theta}{2} \sigma_y} e^{i \frac{\phi}{2} \sigma_z}.  
\end{align}
Two-component wave functions of boundary gapless excitations $\phi_{\rm surf}(r,\theta,\phi)$ are 
well approximated by solving the radial part as 
\begin{align}
&{\cal H}_r \phi_{\rm surf}(r,\theta,\phi) = 0, 
\nonumber \\
&\label{Eq:Sol_3D_Surf_State}
\phi_{\rm surf}(r,\theta,\phi) \sim 
e^{-i \frac{\phi}{2} \sigma_z} e^{- i \frac{\theta}{2} \sigma_y}  
\left( 
u_1(\theta,\phi)
\begin{pmatrix}
i \\
1 \\
\end{pmatrix}_{\tau} \otimes 
\begin{pmatrix}
0 \\
1 \\
\end{pmatrix}_{\sigma}, 
u_2(\theta,\phi)
\begin{pmatrix}
1 \\
i \\
\end{pmatrix}_{\tau} \otimes 
\begin{pmatrix}
1 \\
0 \\
\end{pmatrix}_{\sigma} \right)
e^{- \int^r m(r') dr'}, 
\nonumber \\
&u_i(\theta,\phi+2 \pi) = - u_i(\theta,\phi), \ \ (i=1,2). 
\end{align}
Here $u_i(\theta,\phi) \ (i=1,2)$ are scalar functions representing the $(\theta,\phi)$-dependence 
and obey anti-periodic boundary condition in $\phi$ to account for the factor $e^{i \frac{\phi}{2} \sigma_z}$. 
Since $\phi_{\rm surf}(r,\phi)$ is well localized at $r=R$, we can approximate ${1}/{r}$ by ${1}/{R}$. 
The effective single-particle Hamiltonian for the surface is given by 
\begin{align}
{\cal H}_{surf} 
&= \int r^2 d r [\phi_{\rm surf}(r,\phi)]^{\dag} {\cal H}_{\theta, \phi} \phi_{\rm surf}(r,\phi) 
\nonumber \\
&\sim ( u^*_1(\theta, \phi), u^*_2(\theta,\phi) ) 
\frac{1}{R} 
\begin{pmatrix}
0 & -i \partial_{\theta} - \frac{1}{\sin \theta} \partial_{\phi} - \frac{i \cot \theta}{2} \\
-i \partial_{\theta} + \frac{1}{\sin \theta} \partial_{\phi} - \frac{i \cot \theta}{2} & 0 \\
\end{pmatrix}
\begin{pmatrix}
u_1(\theta,\phi) \\
u_2(\theta,\phi). 
\end{pmatrix}
\end{align}

\subsubsection{$(3+1)d$ topological insulator}
\label{App:Der_Surf_(3+1)d_A}
Let $\psi(\bx) = \{ \psi_{\tau \sigma}(\bx) \}_{\tau,\sigma=1,2}$ be a four-orbital complex fermion. 
We consider a massive Dirac Hamiltonian defined on a ball of radius $R$, 
\begin{align}
H = \int d^3 \bx\, \psi^{\dag}(\bx) \Big[ -i \tau_x \bm{\sigma} \cdot \bm{\partial} + m(r) \tau_z \Big] \psi(\bx). 
\end{align}
From (\ref{Eq:Sol_3D_Surf_State}), gapless surface excitations $\gamma_i(\theta, \phi) (i=1,2)$ are given by 
\begin{align}
\gamma^{\dag}_1(\theta,\phi) 
& \sim \Big[ 
-e^{-i \frac{\phi}{2}} \sin \frac{\theta}{2} \big\{ i \psi^{\dag}_{\tau=1,\sigma=1}(r,\theta,\phi) + \psi^{\dag}_{\tau=2,\sigma=1}(r,\theta,\phi) \big\} 
\nonumber \\
& \ \ \ \ \ \ + e^{i \frac{\phi}{2}} \cos \frac{\theta}{2} \big\{ i \psi^{\dag}_{\tau=1,\sigma=2}(r,\theta,\phi) + \psi^{\dag}_{\tau=2,\sigma=2}(r,\theta,\phi) \big\}
\Big] e^{- \int^r m(r') dr'}, 
\nonumber \\
\gamma^{\dag}_2(\theta,\phi) 
& \sim \Big[ 
e^{-i \frac{\phi}{2}} \cos \frac{\theta}{2} \big\{ \psi^{\dag}_{\tau=1,\sigma=1}(r,\theta,\phi) + i \psi^{\dag}_{\tau=2,\sigma=1}(r,\theta,\phi) \big\} 
\nonumber \\
& \ \ \ \ \ \ +e^{i \frac{\phi}{2}} \sin \frac{\theta}{2} \big\{ \psi^{\dag}_{\tau=1,\sigma=2}(r,\theta,\phi) + i \psi^{\dag}_{\tau=2,\sigma=2}(r,\theta,\phi) \big\}
\Big] e^{- \int^r m(r') dr'}. 
\end{align}
and obey the boundary condition
\begin{align}
\gamma_i(\theta, \phi+2 \pi) = - \gamma_i(\theta, \phi) \quad (i=1,2) . 
\end{align}
The effective Hamiltonian on $S^2$ is given by
\begin{align}
&H = \frac{1}{R} \int \sin \theta d \theta d \phi\,  
(\gamma^{\dag}_1(\theta,\phi), \gamma^{\dag}_2(\theta,\phi) ) 
\begin{pmatrix}
0 & -i \partial_{\theta} - \frac{1}{\sin \theta} \partial_{\phi} - \frac{i \cot \theta}{2} \\
-i \partial_{\theta} + \frac{1}{\sin \theta} \partial_{\phi} - \frac{i \cot \theta}{2} & 0 \\
\end{pmatrix}
\begin{pmatrix}
\gamma_1(\theta,\phi) \\
\gamma_2(\theta,\phi) \\
\end{pmatrix}, 
\end{align}

\subsubsection{$(3+1)d$ topological superconductor}
\label{App:Der_Surf_(3+1)d_D}
Let $\psi(\bx) = \{ \psi_{\sigma}(\bx) \}_{\sigma=\uparrow,\downarrow}$ be a two-orbital complex fermion. 
We consider a topological superconductor on a ball of radius $R$, 
\begin{align}
H = \int d^3 \bx\, \Psi^{\dag}(\bx) \Big[ -i \tau_x \bm{\sigma} \cdot \bm{\partial} + m(r) \tau_z \Big] \Psi(\bx), 
\quad  
\Psi(\bx) = (\psi_{\uparrow}(\bx), \psi_{\downarrow}(\bx), \psi^{\dag}_{\downarrow}(\bx), -\psi^{\dag}_{\uparrow}(\bx))^T. 
\end{align}
From (\ref{Eq:Sol_3D_Surf_State}), 
complex fermion operators 
$\gamma^{\dag}_i(\theta, \phi) (i=1,2)$ for gapless surface excitations are given by 
\begin{align}
\gamma^{\dag}_1(\theta,\phi) 
& \sim \Big[ 
-e^{-i \frac{\phi}{2}} \sin \frac{\theta}{2} \big\{ i \psi^{\dag}_{\uparrow}(r,\theta,\phi) + \psi_{\downarrow}(r,\theta,\phi) \big\} 
+ e^{i \frac{\phi}{2}} \cos \frac{\theta}{2} \big\{ i \psi^{\dag}_{\downarrow}(r,\theta,\phi) - \psi_{\uparrow}(r,\theta,\phi) \big\}
\Big] e^{- \int^r m(r') dr'}, 
\nonumber \\
\gamma^{\dag}_2(\theta,\phi) 
& \sim \Big[ 
e^{-i \frac{\phi}{2}} \cos \frac{\theta}{2} \big\{ \psi^{\dag}_{\uparrow}(r,\theta,\phi) + i \psi_{\downarrow}(r,\theta,\phi) \big\} 
+e^{i \frac{\phi}{2}} \sin \frac{\theta}{2} \big\{ \psi^{\dag}_{\downarrow}(r,\theta,\phi) - i \psi_{\uparrow}(r,\theta,\phi) \big\}
\Big] e^{- \int^r m(r') dr'}. 
\end{align}
They
satisfy $\gamma^{\dag}_2(\theta,\phi) = -\gamma_1(\theta,\phi)$ and
the boundary condition,
\begin{align}
\gamma_1(\theta,\phi+2 \pi) = - \gamma_1(\theta,\phi). 
\end{align}
The effective Hamiltonian on $S^2$ is given by 
\begin{align}
&H 
= \frac{1}{R} \int \sin \theta d \theta d \phi 
\big( \gamma^{\dag}_1(\theta,\phi), -\gamma_1(\theta,\phi) \big) 
\begin{pmatrix}
0 & -i \partial_{\theta} - \frac{1}{\sin \theta} \partial_{\phi} - \frac{i \cot \theta}{2} \\
-i \partial_{\theta} + \frac{1}{\sin \theta} \partial_{\phi} - \frac{i \cot \theta}{2} & 0 \\
\end{pmatrix}
\begin{pmatrix}
\gamma_1(\theta,\phi) \\
-\gamma^{\dag}_1(\theta,\phi) \\
\end{pmatrix}, 
\end{align}

\section{${\rm Spin}^c$ cobordism group $\Omega^{{\rm Spin}^c}_3(B \Z_n)$}
\label{App:2}
${\rm Spin}^c$ cobordism groups $\Omega^{{\rm Spin}^c}_d(B G)$ for
a cyclic group 
$G = \Z_n$ for any dimensions are computed in Ref.\ \onlinecite{Bahri1987}. 
The result is
\begin{align}
\label{Eq:Spin^c_Bord_1}
&\widetilde \Omega_*^{{\rm Spin}^c}(B\Z_n) \cong \{A_*(n) \otimes \Z[\C P^2,\C P^4,\dots ]\} \oplus {\rm ker}_*(\eta,n), \\
&{\rm ker}_m(\eta,n) \cong \bigoplus_{j<m} {\rm Tor}(\Omega_j^{{\rm Spin}^c}), \ \ \ A_{2k+1} (n) \cong (1-t)/((1-t)^{k+2}), \ \ \ A_{2k}(n)=0, 
\label{Eq:Spin^c_Bord_2}
\end{align}
where $\widetilde \Omega_*^{{\rm Spin}^c}(B\Z_n)$ is the kernel of the symmetry forgetting functor of $\Z_n$ symmetry, says, 
\begin{align}
\Omega_*^{{\rm Spin}^c}(B\Z_n)
\cong \widetilde \Omega_*^{{\rm Spin}^c}(B\Z_n) \oplus \Omega_*^{{\rm Spin}^c}(pt). 
\end{align}
$(1-t)$ and $((1-t)^{k+2})$ are $R(\Z_n)$-modules defined by , 
\begin{align}
(1-t) = \{(1-t) f(t) | f(t) \in R(\Z_n)\}, 
\quad 
\big( (1-t)^{k+2}\big) = \{(1-t)^{k+2} f(t) | f(t) \in R(\Z_n)\}. 
\end{align}
$R(\Z_n)= \Z[t]/(1-t^n)$ is the representation ring of $\Z_n$. 
$\Z[\C P^2,\C P^4,\dots ]$ represents contributions from bosonic SPT phases. 
As will be explained in Sec.\ \ref{App:2-3}, 
the ideal $(1-t)$ means non-chiral SPT phases of free fermions and the quotient $(1-t)/((1-t)^k)$ is interpreted as the breakdown of 
free fermionic classification by interactions. 

For our purposes, only $\Omega^{{\rm Spin}^c}_3(B \Z_n)$ is needed, in which there is no 
contribution from bosonic SPT phases. 
We have 
\begin{align}
\Omega^{{\rm Spin}^c}_3(B \Z_n) \cong (1-t)/((1-t)^3). 
\end{align}
Calculations for some $n$ are illustrated in Sec.~\ref{Calculation of $(1-t)/((1-t)^3)$}.

\subsection{Smith homomorphism}
$\pinc$ cobordism groups $\Omega_{2k}^{{\rm Pin}^c}(pt)$ 
in even spacetime dimensions are isomorphic to 
the $\spinc$ cobordism group in one-higher spacetime dimensions with 
$\Z_2$ on-site symmetry. 
The Smith homomorphism provides an isomorphism \cite{Bahri1987}
\begin{align}
\widetilde \Omega_{2k+1}^{{\rm Spin}^c}(B\Z_2) 
\overset{\cong}{\longrightarrow} \widetilde \Omega_{2k}^{{\rm Pin}^c}(pt) 
\end{align}
in the present case. 
The l.h.s.\ means $(2k+1)d$ SPT phases of complex fermions with onsite 
$\Z_2$ symmetry, which can be computed by the formula 
(\ref{Eq:Spin^c_Bord_1}, \ref{Eq:Spin^c_Bord_2}). 
The r.h.s.\ represents $(2k) d$ SPT phases of complex fermions 
with orientation reversing symmetry (class A with reflection symmetry or class AIII, say).  
We show some examples in low dimensions 
\begin{align}
\wt \Omega^{\pinc}_2(pt) 
&\cong \wt \Omega_{3}^{{\rm Spin}^c}(B \Z_2) 
= \Z_4[A_3] = \Z_4, \\
\wt \Omega^{\pinc}_4(pt) 
&\cong \wt \Omega_{5}^{{\rm Spin}^c}(B \Z_2) 
= \Z_8[A_5] \oplus (\Z_2[A_1] \otimes \Z[\C P^2]) = \Z_8 \oplus \Z_2, \\
\wt \Omega^{\pinc}_6(pt) 
&\cong \wt \Omega_{7}^{{\rm Spin}^c}(B \Z_2) 
= \Z_{16}[A_7] \oplus (\Z_4[A_3] \otimes \Z[\C P^2]) = \Z_{16} \oplus \Z_4, \\
\wt \Omega^{\pinc}_8(pt) 
&\cong \wt \Omega_{9}^{{\rm Spin}^c}(B \Z_2) 
= \Z_{32}[A_9] \oplus (\Z_8[A_5] \otimes 
\Z[\mathbb{C}P^2]) 
\oplus (\Z_2[A_1] \otimes \Z[\mathbb{C}P^4]) \oplus (\Z_2[A_1] \otimes 
\Z[(\mathbb{C}P^2)^2]) 
\nonumber \\
&= \Z_{32} \oplus \Z_8 \oplus \Z_2 \oplus \Z_2. 
\end{align}
Here we used the relation 
\begin{align}
(1-t)/\big( (1-t)^{k+2} \big) = (1-t)/\big( 2^{k+1}(1-t) \big) = \Z_{2^{k+1}} 
\label{eq:app3/pinc}
\end{align}	
in the $R(\Z_2)$-module. 
(\ref{eq:app3/pinc}) is the breakdown formula of the free fermion topological phases in 
class A with reflection symmetry or class AIII, 
which is consistent with Refs.\ \onlinecite{you2014symmetry, Morimoto2015}.

\subsection{K-theory classification}
\label{App:2-3}
The $K$-theory classification of $(2d+1)$-dimensional class A topological insulators with 
on-site $G$ symmetry is given by the 
$G$-equivariant complex $K$-theory~\cite{Freed2013, segal1968equivariant, Shiozaki-Sato-Gomi}
\begin{align}
K_G(S^{2d}) \cong R(G) \otimes K(S^{2d}) = R(G) \otimes (\Z \oplus \Z) \cong R(G) \oplus R(G), 
\end{align}
where $G$ trivially acts on $S^{2d}$. 
We introduce the reduced $K$-theory to remove the trivial contribution from a point in $S^{2d}$ as 
\begin{align}
\widetilde K_G(S^{2d}) = R(G), 
\quad K_G(S^{2d}) = \widetilde K_G(S^{2d}) \oplus K_G(pt)
\end{align}
Here, $R(G)$ is the representation ring of $G$, which is generated by irreps.\ of $G$ and 
multiplicative structure is induced by the tensor product of representations. 

For our purposes, we consider $\widetilde K_{\Z_n}(S^{2}) \cong R(\Z_n)$ that represents 
the classification for $\Z_n$ symmetry in $2$ space dimensions.  
The representation ring is given by 
\begin{align}
R(\Z_n) = \Z[t]/(1-t^n) = \{m_0 + m_1 t + \dots + m_{n-1} t^{n-1} | m_p \in \Z \}. 
\end{align}
A base element $t^p \in R(\Z_n)$ corresponds to the following chiral Chern insulator,
in which the complex fermion operator
is in the the $e^{- 2 \pi i p/n}$ representation of $\Z_n$ symmetry:
\begin{align}
&H_{t^p} = \sum_{\bk} \psi^{\dag}(\bk) \big[ k_x \sigma_x + k_y \sigma_y + (m-k^2) \sigma_z \big] \psi(\bk), 
\quad (m>0), \\
&\psi(\bk) = (\psi_1(\bk), \psi_2(\bk))^T, 
\quad 
U \psi^{\dag}(\bk) U^{-1} = e^{- 2 \pi i p/n} \psi^{\dag}(\bk). 
\end{align}
For general insulators specified by 
$f(t) = m_0 + m_1 t + \dots + m_{n-1} t^{n-1} \in R(\Z_n)$, 
the 1st Chern number is given by the sum $\ch_1 \big( f(t) \big) 
= f(1) = m_0 + m_1 + \cdots + m_{n-1}$. 
Thus, the subgroup representing non-chiral phases is characterized by 
$f(1) = 0$, which is the ideal $(1-t)$, 
\begin{align}
(1-t) = \{(1-t) f(t) | f(t) \in R(\Z_n)\}, 
\end{align}
where the basis is spanned by 
\begin{align}
\{1-t, 1-t^2, \dots, 1-t^{n-1}\}. 
\end{align}
Here $1-t^p \in R(\Z_n)$ represent the following non-chiral topological insulators, 
\begin{align}
H_{1-t^p}
&= \sum_{\bk} \psi_{\uparrow}^{\dag}(\bk) \big[ k_x \sigma_x + k_y \sigma_y + (m-k^2) \sigma_z \big] \psi_{\uparrow}(\bk)
\nonumber \\
&\quad
+ \sum_{\bk} \psi_{\downarrow}^{\dag}(\bk) \big[ k_x \sigma_x - k_y \sigma_y + (m-k^2) \sigma_z \big] \psi_{\downarrow}(\bk), 
\quad 
(m>0),  
\end{align}
where
\begin{align}
&\psi_s(\bk) = (\psi_{s,1}(\bk), \psi_{s,2}(\bk))^T, \ \ (s= \uparrow, \downarrow),  
\nonumber \\
&U \psi_{\uparrow}^{\dag}(\bk) U^{-1} = \psi_{\uparrow}^{\dag}(\bk), \ \ 
U \psi_{\downarrow}^{\dag}(\bk) U^{-1} = e^{- 2 \pi i p/n} \psi_{\downarrow}^{\dag}(\bk). 
\end{align}

\subsection{Calculations of $(1-t)/((1-t)^3)$}
\label{Calculation of $(1-t)/((1-t)^3)$}
In this section, we illustrate the computation of the breakdown formula
\begin{align}
\Omega_3^{{\rm Spin}^c}(B \Z_n) \cong (1-t)/((1-t)^3). 
\end{align}
We show it for $n=2,3,4$. 

\subsubsection{$n=2$}
Since $(1-t)^3 = 4-4t$ in $R(\Z_2)$, we have 
\begin{align}
\Omega_3^{{\rm Spin}^c}(B \Z_2) \cong (1-t)/((1-t)^3) = (1-t)/(4-4t) \cong \Z_4[1-t]. 
\end{align}
Thus, the topological classification is given by $\Z_4$ and 
it is generated by the non-chiral topological insulator $H_{1-t}$.

\subsubsection{$n=3$}
In the representation ring $R(\Z_3) = \Z[t]/(1-t^3)$, 
$\big( (1-t)^3 \big)$ reads $\big( (1-t)^3 \big) = (-3t + 3 t^2)$, 
which is spanned by 
\begin{align}
\{-3t + 3t^2, t (-3t+3t^2) \} 
= \{3(1-t)-3(1-t^2), 3(1-t^2) \} 
\sim \{3(1-t), 3(1-t^2) \}. 
\end{align}
This leads to 
\begin{align}
\Omega_3^{{\rm Spin}^c}(B \Z_2) \cong (1-t)/((1-t)^3) = \Z_3[1-t] \oplus \Z_3[1-t^2]. 
\end{align}
$H_{1-t}$ and $H_{1-t^2}$ provide 
two independent generators of 
the $\Z_3$ classification.  

\subsubsection{$n=4$}
$((1-t)^3) = (1-3t+3t^2-t^3)$ is spanned by 
\begin{align}
&\{ 1-3t+3t^2-t^3. t(1-3t+3t^2-t^3), t^2(1-3t+3t^2-t^3) \} 
\nonumber \\
&= \{3(1-t) -3 (1-t^2) + (1-t^3), -(1-t) + 3(1-t^2) - 3 (1-t^3), (1-t) - (1-t^2) + 3(1-t^3)\}
\nonumber \\
& \sim \{3(1-t) -3 (1-t^2) + (1-t^3), 8(1-t), 2(1-t^2) \}, 
\end{align}
then, 
\begin{align}
(1-t)/((1-t)^3) = \Z_8[1-t] \oplus \Z_2[1-t^2]. 
\end{align}
The $\Z_8$ and $\Z_2$ direct summands are generated by 
the non-chiral topological phases,
$H_{1-t}$ and $H_{1-t^2}$,
respectively.

\section{Calculations of $\prod_{n=1}^{\infty} (1+e^{i \phi} (-q)^n)^n$}
\label{App:I(q,z)}
In this appendix, we compute 
\begin{align}
I(q=e^{- \delta},z) 
= \sum_{n=1}^{\infty} n \ln (1+z (-q)^n) 
= \sum_{n=1}^{\infty} (2n) \ln (1+z q^{2n}) + \sum_{n=1}^{\infty} (2n-1) \ln (1-z q^{2n-1}) 
\end{align}
with a pure phase $z = e^{i \phi}$. 
Using the Cahen-Mellin integral, 
\begin{align}
\sum_{n=1}^{\infty} (2n) \ln (1+z q^{2n}) 
&= -\sum_{n=1}^{\infty} (2n) \sum_{r=1}^{\infty} r^{-1} (-z)^r q^{2nr}
\nonumber \\
&= -\sum_{n=1}^{\infty} (2n) \sum_{r=1}^{\infty} r^{-1} (-z)^r \frac{1}{2 \pi i}
\int_{c- i \infty}^{c+i \infty} ds (2 \delta n r)^{-s} \Gamma(s)  
\nonumber \\
&= \frac{1}{2 \pi i} \int_{c- i \infty}^{c+i \infty} ds
\delta^{-s} \Big[ - 2^{1-s} \Gamma(s) \zeta(s-1) {\rm Li}_{s+1}(-z) \Big], 
\nonumber \\
\sum_{n=1}^{\infty} (2n-1) \ln (1-z q^{2n-1})
&= - \sum_{n=1}^{\infty} (2n-1) \sum_{r=1}^{\infty} r^{-1} z^r q^{(2n-1)r} 
\nonumber \\
&= - \sum_{n=1}^{\infty} (2n-1) \sum_{r=1}^{\infty} r^{-1} z^r \frac{1}{2 \pi i} \int_{c- i \infty}^{c+i \infty} ds
(\delta (2n-1) r)^{-s} \Gamma(s) 
\nonumber \\
&= \frac{1}{2 \pi i} \int_{c- i \infty}^{c+i \infty}ds
\delta^{-s} \Big[ - \Gamma(s) (1-2^{1-s}) \zeta(s-1) {\rm Li}_{s+1}(z) \Big]. 
\end{align}
We used 
\begin{align}
\zeta(s) = \frac{1}{1-2^{-s}} \sum_{n=1}^{\infty} \frac{1}{(2n-1)^s} \quad  ({\rm Re}(s) > 1), 
\end{align}
and introduced the polylogarithm function 
\begin{align}
{\rm Li}_s(z) := \sum_{n=1}^{\infty} \frac{z^n}{n^s}. 
\end{align}
Then, 
\begin{align}
I(q=e^{- \delta},z)  
= \frac{1}{2 \pi i} \int_{c- i \infty}^{c+i \infty} ds
\delta^{-s} \Big[ 
-2^{1-s} \Gamma(s) \zeta(s-1) {\rm Li}_{s+1}(-z)
- (1-2^{1-s}) \Gamma(s) \zeta(s-1) {\rm Li}_{s+1}(z)
\Big]. 
\end{align}
The integrand has a double pole at $s=0$ at $z=\pm 1$ 
since ${\rm Li}_{s+1}(1) = \zeta(s+1)$. 
From the contour integral, for small $\delta$, we have following expansions: 
\begin{align}
  \label{F6}
&
I(q=e^{- \delta},z=1) 
= - \frac{1}{8} \zeta(3) \delta^{-2} + \frac{1}{12} - \ln (A) + \frac{1}{12} \ln (\delta) + \frac{17}{2880} \delta^2 + \cdots, 
\nonumber \\
&
I(q=e^{- \delta},z=-1) 
= - \frac{1}{8} \zeta(3) \delta^{-2} - \frac{1}{6} - \frac{1}{12} \ln(2) + 2 \ln (A) - \frac{1}{6} \ln(\delta)- \frac{13}{2880} \delta^2 + \cdots, 
\end{align}
for $z=\pm 1$, 
and for $z \neq \pm 1$ 
\begin{align}
&I(q=e^{- \delta},z = e^{i \phi}) 
= - \frac{1}{8} {\rm Li}_3(e^{2 i \phi}) \delta^{-2} + \frac{1}{12} \big\{ \ln (1-e^{i \phi}) -2 \ln(1+e^{i \phi}) \big\} + \frac{1-15 \cos \phi}{480 \sin^2 \phi} \delta^2 + \cdots 
\nonumber \\
&= - \frac{1}{8} {\rm Li}_3(e^{2 i \phi}) \delta^{-2} + \frac{1}{12} \big\{ \ln| 2 \sin \frac{\phi}{2}| - 2 \ln |2 \cos \frac{\phi}{2}| \big\}
+ \frac{1}{12} \cdot \left\{\begin{array}{ll}
-\frac{\pi i}{2} - \frac{i \phi}{2} & (0 < \phi < \pi)  \\
\frac{\pi i}{2} - \frac{i \phi}{2} & (-\pi < \phi < 0) \\
\end{array}\right\}
+ \frac{1-15 \cos \phi}{480 \sin^2 \phi} \delta^2 + \cdots.
\end{align}
Here, $A \cong 1.2824\dots$ is the Glaisher-Kinkelin constant. 
Especially, 
\begin{align}
&I(q=e^{- \delta},z = \pm i) 
= \frac{3}{32} \zeta(3) \delta^{-2} - \frac{1}{24} \ln(2) \mp \frac{\pi i}{16} + \frac{\delta^2}{480} + \cdots 
\nonumber \\
&I(q=e^{- \delta},z = e^{i \phi}) + I(q=e^{- \delta},z = -e^{-i \phi}) 
\nonumber \\
&\qquad 
= - \frac{1}{8} \big\{ {\rm Li}_3(e^{2 i \phi}) + {\rm Li}_3(e^{-2 i \phi})  \big\} \delta^{-2} 
- \frac{1}{12} \ln| 2 \sin \phi| 
+ \left\{\begin{array}{ll}
-\frac{\pi i}{8} & (0 < \phi < \pi)  \\
\frac{\pi i}{8} & (-\pi < \phi < 0) \\
\end{array}\right\}
+ \frac{1}{240 \sin^2 \phi} \delta^2 + \cdots.
\end{align}

\subsection{The imaginary part of $I(q,e^{2 \pi i x})$}
If we look at the imaginary part of $I(q,e^{2 \pi i x})$, we obtain 
the following formula:
\begin{align}
{\rm Im} \big[ I(q=e^{- \delta},z = e^{2 \pi i x}) \big] 
&= \frac{1}{2 i} \big[ I(q=e^{- \delta},z = e^{2 \pi i x}) - I(q=e^{- \delta},z = e^{-2 \pi i x}) \big]
\nonumber\\
&= \frac{1}{2 \pi i} \int_{c- i \infty}^{c+i \infty} ds \delta^{-s} \Big[ 
-2^{1-s} \Gamma(s) \zeta(s-1) \frac{{\rm Li}_{s+1}(-e^{2 \pi i x}) - {\rm Li}_{s+1}(-e^{-2 \pi i x})}{2 i}
\nonumber \\
& \hspace{100pt} - (1-2^{1-s}) \Gamma(s) \zeta(s-1) \frac{{\rm Li}_{s+1}(e^{2 \pi i x}) - {\rm Li}_{s+1}(e^{-2 \pi i x})}{2i}
\Big].  
\end{align}
In this expression, poles of $\Gamma(s)$ at odd negative integers are canceled 
with zeros of $\zeta(s-1)$, 
and also poles of $\Gamma(s)$ at even negative integers are also canceled since 
\begin{align}
{\rm Li}_{-n} (z) + (-1)^n {\rm Li}_{-n}(z^{-1}) = 0 \ \ (n=1,2,3, \dots). 
\end{align}
The integrand has poles only at $s=0,2$ and we get an exact result 
\begin{align}
{\rm Im} \big[ I(q=e^{- \delta},z = e^{2 \pi i x}) \big]
&= \left\{\begin{array}{ll}
- \frac{\pi}{6} B_1(x + \frac{1}{2}) + \frac{\pi}{12} B_1(x+1) - \frac{\pi^3}{12} B_3(2x+1) \delta^{-2} & (-1/2 < x < 0) 
\nonumber \\
- \frac{\pi}{6} B_1(x + \frac{1}{2}) + \frac{\pi}{12} B_1(x) - \frac{\pi^3}{12} B_3(2x) \delta^{-2}  & (0 < x < 1/2) \\
\end{array}\right. \\
&= \left\{\begin{array}{ll}
-\frac{\pi}{12}(x-\frac{1}{2}) - \frac{2 \pi^3}{3} x (x+\frac{1}{4}) (x+\frac{1}{2}) \delta^{-2}  & (-1/2 < x < 0) \\
-\frac{\pi}{12}(x+\frac{1}{2}) - \frac{2 \pi^3}{3} x (x-\frac{1}{4}) (x-\frac{1}{2}) \delta^{-2}  & (0 < x < 1/2) \\
\end{array}\right. 
\end{align}
Here, $B_1(x) = x - \frac{1}{2}, B_3(x) = x(x-\frac{1}{2})(x-1)$ are the Bernoulli polynomials. 
At $x = \pm \frac{1}{4}$, ${\rm Im} \big[ I(q=e^{- \delta},z = e^{2 \pi i x}) \big]$ is 
$\delta$-independent, which is the origin of the very 
sharp plateau structure in 
the expectation value of partial inversion in  
$(3+1)d$ superconductors in Fig.\ \ref{fig:3Dsc}.

\bibliography{Draft.bib}

\end{document}